%% file: bare_jrnl.tex
\documentclass[journal]{IEEEtran}
\ifCLASSINFOpdf
\else
\fi

\usepackage[utf8]{inputenc}
\usepackage[margin=2.4cm]{geometry}

\usepackage{color}
\usepackage{url}
\usepackage{amssymb}
\usepackage{fancybox}
\usepackage[pdftex]{graphicx}
\usepackage{siunitx}
\usepackage{booktabs}
\newcommand{\rarray}[1]{\renewcommand{\arraystretch}{#1}} 

\usepackage[acronym, toc, nonumberlist]{glossaries}
\usepackage[nospace]{cite}
\usepackage{footnote}
\makesavenoteenv{tabular}
\makesavenoteenv{table}
\makesavenoteenv{table*}
\usepackage[colorlinks=false, pdfborder={0 0 0}, hyperfootnotes=true]{hyperref}

\usepackage[inline,marginclue,draft]{fixme}

\include{style/tikzset}
\include{style/macros}
\include{style/glossary}

\usetikzlibrary{external}
\begin{document}

%
\title{Far-Field Automatic Speech Recognition}
%
%
%

\author{Reinhold~Haeb-Umbach,
        Jahn~Heymann,
        Lukas~Drude,
        Shinji~Watanabe,
        Marc~Delcroix,
        and~Tomohiro~Nakatani
\thanks{J. Heymann and L. Drude were in part supported by a Google Faculty Research Award.}
}

%
%

\markboth{}%
{Shell \MakeLowercase{\textit{et al.}}: Far-Field Automatic Speech Recognition}
%



\maketitle

\begin{abstract}
The machine recognition of speech spoken at a distance from the microphones, known as far-field \gls{ASR}, has received a significant increase of attention in science and industry, which caused or was caused by an equally significant improvement in recognition accuracy. Meanwhile it has entered the consumer market with digital home assistants with a spoken language interface being its most prominent application. Speech recorded at a distance is affected by various acoustic distortions and, consequently, quite different processing pipelines have emerged compared to \gls{ASR} for close-talk speech. A signal enhancement front-end for dereverberation, source separation and acoustic beamforming is employed to clean up the speech, and the back-end ASR engine is robustified by multi-condition training and adaptation. We will also describe the so-called end-to-end approach to \gls{ASR}, which is a new promising architecture that has  recently been extended to the far-field scenario. This tutorial article gives an account of the algorithms used to enable accurate speech recognition from a distance, and it will be seen that, although deep learning has a significant share in the technological breakthroughs, a clever combination with traditional signal processing can lead to surprisingly effective solutions.
\end{abstract}


%
\IEEEpeerreviewmaketitle


\input{chapters/introduction}

\input{chapters/signal_model}

\input{chapters/enhancement}

\input{chapters/backend}

\input{chapters/summary_outlook}

\input{chapters/probe_further}


%

%


\ifCLASSOPTIONcaptionsoff
  \newpage
\fi



%

\bibliographystyle{IEEEtran}
\bibliography{references/IEEEfull,references/overview_refs}


%







\end{document}

%% file: style/tikzset.tex
\usepackage{tikz}
\usepackage{amssymb}
\usetikzlibrary{calc}
\usetikzlibrary{positioning}
\usetikzlibrary{chains}
\usetikzlibrary{fit}
\usetikzlibrary{arrows}
\usetikzlibrary{decorations.pathreplacing}
\usetikzlibrary{calligraphy}
\usetikzlibrary{arrows.meta}
\usetikzlibrary{backgrounds}
\usetikzlibrary{shapes.multipart}
\usetikzlibrary{shapes.geometric}

\tikzset{
    every text node part/.style={align=center},
    >=stealth,
    element/.style={
        draw=black!100,
        thick,
    },
    block/.style={
        element,
        rectangle,
        minimum width=4.5em,
        minimum height=2.5em,
        inner sep=0pt,
    },
    wide block/.style={
        block,
        minimum width=7.5em,
        minimum height=2.5em,
    },
    parameter/.style={
        element,
        rectangle,
        minimum width=2.5em,
        minimum height=2.5em,
        inner sep=0pt,
        text height=1.5ex,
        text depth=.25ex
    },
    random/.style={
        element,
        circle,
        minimum width=2.5em,
        minimum height=2.5em,
        inner sep=0pt,
        draw=black!100,
        text height=1.5ex,
        text depth=.25ex
    },
    observation/.style={
        random,
        double
    },
    branch/.style={
        circle,
        fill=black,
        draw=black,
        minimum size=0.2em,
        inner sep=0pt
    },
    apply/.style={
        circle,
        thick,
        draw=black!100,
        minimum size=1em,
        inner sep=0pt,
        label=center:{$\times$}
    },
    node distance=2em,
    arrow/.style={->,shorten >=0.1em},
    reverse arrow/.style={<-,shorten <=0.1em},
}

%% file: style/macros.tex
\usepackage{amsmath,amssymb,amsfonts}
\usepackage{bbm}       
\usepackage{siunitx}   
\usepackage{trfsigns}  

\DeclareMathOperator*{\argmax}{argmax}
\DeclareMathOperator*{\argmin}{argmin}

\DeclareMathOperator*{\CE}{CE}
\newcommand{\vect}[1]{\boldsymbol{\mathbf{#1}}}
\newcommand{\vectTilde}[1]{\boldsymbol{\mathbf{\tilde{#1}}}}
\newcommand{\vectBar}[1]{\boldsymbol{\mathbf{\bar{#1}}}}
\newcommand{\vectHat}[1]{\boldsymbol{\mathbf{\hat{#1}}}}

\renewcommand{\H}{^\mathrm{H}}
\newcommand{\T}{^\mathrm{T}}

\newcommand{\watanabeStyleC}[1]{\rarray{1}\begin{tabular}[c]{@{}c@{}}#1\end{tabular}}
\newcommand{\watanabeStyleL}[1]{\rarray{1}\begin{tabular}[c]{@{}l@{}}#1\end{tabular}}

%% file: style/glossary.tex
\newacronym{AIR}{AIR}{acoustic impulse response}
\newacronym{AM}{AM}{acoustic model}
\newacronym{ASR}{ASR}{automatic speech recognition}
\newacronym{ATF}{ATF}{acoustic transfer function}
\newacronym{BLSTM}{BLSTM}{bi-directional LSTM}
\newacronym{BSS}{BSS}{blind source separation}
\newacronym{CACGMM}{CACGMM}{Complex Angular Central GMM}
\newacronym{CD}{CD}{Cepstral Distortion}
\newacronym{CE}{CE}{Cross Entropy}
\newacronym{CNN}{CNN}{Convolutional Neural Network}
\newacronym{CTC}{CTC}{Connectionist Temporal Classification}
\newacronym{MVN}{MVN}{mean and variance normalization}

\newacronym{DAN}{DAN}{Deep Atractor Network}
\newacronym{DC}{DC}{Deep Clustering}
\newacronym{DER}{DER}{Diarization Error Rate}
\newacronym{DFT}{DFT}{Discrete Fourier Transform}
\newacronym{DL}{DL}{Deep Learning}
\newacronym{DNN}{DNN}{Deep Neural Network}
\newacronym{DOA}{DOA}{Direction-Of-Arrival}
\newacronym{DSP}{DSP}{Digital Signal Processing}
\newacronym{EM}{EM}{Expectation-Maximization}
\newacronym{FF}{FF}{Feed Forward}
\newacronym{FWSSNR}{FWSNR}{Frequency-Weighted Segmental SNR}
\newacronym{GEV}{GEV}{Generalized Eigenvalue Decomposition}
\newacronym{GMM}{GMM}{Gaussian Mixture Model}
\newacronym{HMM}{HMM}{Hidden Markov Model}
\newacronym{IBM}{IBM}{ideal binary mask}
\newacronym{ICA}{ICA}{Independent Component Analysis}
\newacronym{IVA}{IVA}{Independent Vector Analysis}
\newacronym{ILRMA}{ILRMA}{Independent Low-Rank Matrix Analysis}
\newacronym{LM}{LM}{language model}
\newacronym{LP}{LP}{linear prediction}
\newacronym{LSTM}{LSTM}{Long-Short Term Memory}
\newacronym{LMF}{LMF}{log-Mel filterbank}
\newacronym{MINT}{MINT}{multiple-input/output inverse theorem}
\newacronym{MIMO}{MIMO}{multiple input multiple output}
\newacronym{ML}{ML}{Maximum Likelihood}
\newacronym{MMI}{MMI}{maximum mutual information}
\newacronym{MMSE}{MMSE}{Minimum Mean Squared Error}
\newacronym{MCLP}{MCLP}{multi-channel linear prediction}
\newacronym{MPDR}{MPDR}{Minimum Power Distortionless Response}
\newacronym{MSE}{MSE}{Mean Squared Error}
\newacronym{MVDR}{MVDR}{Minimum Variance Distortionless Response}
\newacronym{MWF}{MWF}{Multichannel Wiener Filter}
\newacronym{NLP}{NLP}{natural language processing}
\newacronym{NMF}{NMF}{Nonnegative Matrix Factorization}
\newacronym{NN}{NN}{neural network}
\newacronym{PESQ}{PESQ}{Perceptual Evaluation of Speech Quality}
\newacronym{PIT}{PIT}{Permutation Invariant Training}
\newacronym{PLDA}{PLDA}{Probabilistic Linear Discriminant Analysis}
\newacronym{PSD}{PSD}{Power Spectral Density}
\newacronym{RNN}{RNN}{Recurrent Neural Network}
\newacronym{RSAN}{RSAN}{Recursive Selective Attention Network}
\newacronym{RTF}{RTF}{relative transfer function}
\newacronym{SAD}{SAD}{Speech Activity Detection}
\newacronym{SCER}{SCER}{Speaker Confusion Error Rate}
\newacronym{SDW}{SDW}{Speech Distortion Weighted}
\newacronym{SDR}{SDR}{Signal-to-Distortion Ratio}
\newacronym{SDW-MWF}{SDW-MWF}{Speech Distortion Weighted MWF}
\newacronym{sMBR}{sMBR}{segmental Minimum Bayes-Risk}
\newacronym{SNR}{SNR}{Signal-to-Noise Ratio}
\newacronym{SPP}{SPP}{Speech Presense Probability}
\newacronym{STFT}{STFT}{Short-Time Fourier Transformation}
\newacronym{STOI}{STOI}{Short-Time Objective Intelligibility}
\newacronym{TasNet}{TasNet}{Time Domain Audio Separation Network}
\newacronym{TF}{TF}{time-frequency}
\newacronym{TDOA}{TDOA}{Time Difference Of Arrival}
\newacronym{TDNN}{TDNN}{Time-Delay Neural Network}
\newacronym{VAD}{VAD}{Voice Activity Detection}
\newacronym{WER}{WER}{Word Error Rate}
\newacronym{WPE}{WPE}{Weighted Prediction Error}
\newacronym{WSJ}{WSJ}{Wall Street Journal}

%% file: chapters/introduction.tex
\section{Introduction}

\IEEEPARstart{F}{ar}-field, also called distant \gls{ASR} is concerned with the machine recognition of speech spoken at a distance from the microphone. Such recording conditions are common for applications like voice-control of digital home assistants, the automatic transcription of meetings, human-to-robot communication, and several other more. In recent years far-field \gls{ASR} has witnessed a great increase of attention in the speech research community.
This popularity can be attributed to several factors. There is first the large gains in recognition performance enabled by \gls{DL}, which made the more challenging task of accurate far-field \gls{ASR} come within reach. A second reason is  the  commercial success  of speech enabled digital home assistants, which has become possible through progress in various fields, including signal processing, \gls{ASR} and \gls{NLP}.  Finally, scientific challenges related to far-field noise and reverberation robust \gls{ASR}, such as the REVERB challenge \cite{REVERB2016}, the series of CHiME challenges \cite{chime3CS, Vincent2016Chime4, Barker2018CHiME5,Watanabe2020},  and the ASpIRE challenge \cite{harper2015automatic} exposed the task to a wide research audience and met with a lot of publicity. Conversely, those challenges have also helped to get a clearer picture as to which techniques and algorithms are helpful for far-field \gls{ASR}.

The reason why far-field \gls{ASR} is more challenging than \gls{ASR} of speech recorded by a close-talking microphone is the degraded signal quality.
First, the speech signal is attenuated when propagating from the speaker to the microphones, resulting in low signal power and often also low \gls{SNR}. Second, in an enclosure, such as the living or a meeting room, the source signal is repeatedly reflected by walls and objects in the room, resulting in multi-path propagation, which causes a temporal smearing of the source signal called reverberation, much like multi-path propagation does in wireless communications. Third, it is likely that the microphone will capture other interfering sounds, in addition to the desired speech signal, such as the television or HVAC equipment. These sources of acoustic interference can be diverse, hard to predict, and often nonstationary in nature and thus difficult to compensate for. All these factors have a detrimental impact on \gls{ASR} recognition performance.

Given these  signal degradations, it is not surprising that quite different processing pipelines have emerged compared to \gls{ASR} for close-talk speech. There is, foremostly, the use of a microphone array instead of a single microphone for sound capture. This allows for multi-channel speech enhancement, which has proven very successful in noisy reverberant environments. Second, the speech recognition engine is trained with data which represents the typical signal degradations the recognizer is exposed to in a far-field scenario. This robustifies the \gls{AM}, which is the component of the recognizer which translates the speech signal into linguistic units. The following examples demonstrate the power of enhancement and acoustic modeling:
\begin{itemize}
	\item The REVERB challenge data  consists of recordings of the text prompts of the \gls{WSJ} data set, respoken and rerecorded in a far-field scenario with a distance of 2-3 m between speaker and microphone array \cite{REVERB2016}. The challenge baseline \gls{ASR} system, defined in 2014, which operates on a single channel microphone signal, achieved a \gls{WER} of 49\%. Using a strong \gls{AM} based on \gls{DL}, the \gls{WER} could be reduced to 22.2\% \cite{Delcroix2015Strategies,McDonough2014REVERB,Weninger2014REVERB}, while the addition of a multi-microphone front-end and strong dereverberation brought the error rate down to 6.14\% \cite{Wang2020}.
	\item The data set of the CHiME-3 challenge consists of recordings of the \gls{WSJ} sentences in four different noise environments (bus, street, cafe, pedestrian area) \cite{spcom2017_CHiME3}. The data was recorded using a tablet computer with six microphones mounted around the frame of the device. The baseline system reached a \gls{WER} of 33\%, while a robust back-end speech recognizer achieved 11.4\% \cite{Chen2018CHiME4}. Finally, the multi-microphone front-end processing brought the error rate down to 2.7\% \cite{Chen2018CHiME4}.
	\item CHiME-5/6 consists of recordings of casual conversations among friends during a dinner party. The spontaneous speech, reverberation, and the large portion of times where more than one speaker is speaking simultaneously results in a \gls{WER} of barely below 80\% achieved by the baseline system. Using a strong back-end, approximately 60\% \gls{WER} is achieved \cite{Catalin2019Single}, while the addition of multi-microphone source separation and dereverberation results in a \gls{WER} of 43.2\% \cite{Catalin2019Single}. Improvements in both front-end and back-end resulted in 30.5\% WER in the follow-up CHiME-6 campaign \cite{USTC_chime6}.
\end{itemize}
The progress in \gls{ASR} brought about by \gls{DL} is well documented in the literature \cite{6296526, Yu2015, LiDeHaGo2015}. In this contribution we therefore concentrate on those aspects of acoustic modeling that are typical of far-field \gls{ASR}.  But those aspects, although improving the error rate a lot, proved to be insufficient to cope with high reverberation, low \gls{SNR} and concurrent speech, as is typical of far-field \gls{ASR}. This is because common \gls{ASR} feature representations are agnostic to phase (a.k.a. spatial) information and are vulnerable to reverberation, i.e., the temporal dispersion of the signal over multiple analysis frames, and because it is difficult for a  single \gls{AM} to decide which speech source to decode, if multiple are present.  Therefore, front-end processing for cleaning up the signals has been developed, including  techniques for acoustic beamforming \cite{Souden2010MVDR,VanCompernolle1990}, dereverberation \cite{Naylor2010,Nakatani2010WPE}, and source separation/extraction \cite{Araki2016Spatial}. All of those have been shown to significantly improve speech recognition performance, as can be seen in the examples above. 

In the last years, \glspl{NN} have challenged the traditional signal processing based solutions for speech enhancement \cite{Wu:2017:RAS:3068681.3068688,Hershey2016DeepClustering,7952155}, and achieved excellent performance on a number of tasks. However, those advances come at a price. The networks are notorious for their computational and memory demands, often require large sets of parallel data (clean and distorted version of the same utterance)  for training, which have to be matched to the test scenario, and are ``black box'' systems, lacking interpretability by a human. In multi-channel scenarios, it is furthermore not obvious how to handle phase information. As a consequence researchers tried to combine the best of both worlds, i.e., to blend classic multi-channel signal processing with deep learning. 

\begin{figure}
\centering
\footnotesize 
\tikzsetnextfilename{fig1}
\includegraphics[width=0.5\textwidth]{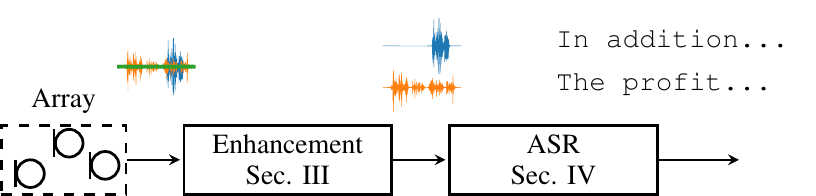}
\vspace{-1em}
\caption{Typical far-field \gls{ASR} system. Here, exemplarily with $M=3$ sensors, $I=2$ sources and additive noise.}
\label{fig:block_diagram_system}
\end{figure}

The purpose of this tutorial article is to describe the specific challenges of far-field \gls{ASR} and how they are approached. We will discuss the general components of an \gls{ASR} system only as much as is necessary to understand the modifications introduced in the far-field scenario. The organization of the paper is oriented along the processing pipeline of a typical far-field \gls{ASR} as shown in Fig.~\ref{fig:block_diagram_system}. First, the signal is captured by an array of $M$ microphones. The signal model, which describes the typical distortions encountered, is given in Section~\ref{sec:signal_model_performance_metric}.  Although recently good single-channel dereverberation \cite{Nakatani2010WPE} and source separation techniques have been developed \cite{Kolbaek2017PIT, Hershey2016DeepClustering, Yi19ConvTasNet}, the use of an array of microphones instead of a single one has the clear advantage that spatial information can be exploited, which often leads to a much more effective suppression of noise and competing audio sources, as well as to better dereverberation performance. Dereverberation, acoustic beamforming and source separation/extraction techniques will be discussed in Section~\ref{sec:se}.

Once the signal is cleaned up it is forwarded to the \gls{ASR} back-end, whose task it is to transcribe the audio in a machine readable form. In far-field \gls{ASR} it is particularly important to make the acoustic model robust against remaining signal degradations. We will explain in Section~\ref{sec:asr} how this can be achieved by so-called multi-style training and by adaptation techniques. Section~\ref{sec:e2e} discusses end-to-end approaches to ASR. In this rather new approach, the recognizer consists of a monolithic neural network, which directly models the posterior distribution of linguistic units given the audio. This paradigm has recently been extended to the far-field scenario, as we explain in that section. 

We conclude this tutorial article with a summary and discuss remaining challenges in Section~\ref{sec:summary}. We further provide pointers to software and databases in Section~\ref{sec:probe_further} for  those who want to gain some hands-on experience.


This article  primarily focuses on speech recognition accuracy, a.k.a. word error rate (WER), in far-field conditions as a criterion for success. Factors such as algorithmic latency or computational efficiency are only touched on in passing, although they are certainly of pivotal importance for the success of a technology in the market. 


%% file: chapters/signal_model.tex
\section{Signal model and performance metrics}
\label{sec:signal_model_performance_metric}

\subsection{Signal model}
\label{sec:signal_model}

In a typical far-field scenario the signal of interest is degraded due to room reverberation, competing speakers, and ambient noise. Assuming an array of $M$ microphones, the signal at the $m$th microphone can be written as follows:
\begin{align}
	\label{eq:signal_model_time_domain}
	y_m[\ell] = \sum_{i=1}^I\left(a_m^{(i)} \ast s^{(i)} \right)[\ell] + n_m[\ell],
\end{align}
where $\ast$ is a convolution operation, $a_m^{(i)}[\ell]$ is the \gls{AIR} from the origin of the $i$th speech signal $s^{(i)}[\ell]$ to the $m$th microphone, and $n_m[\ell]$ is the additive noise.
Depending on the application, we might only be interested in one of the $I$ signals, say $s^{(i)}[\ell]$, while the remaining ones are considered unwanted competing audio signals.

In the following we assume that the \gls{AIR} is time invariant, although it is well-known that movements of the speaker or changes in the environment, and even room temperature changes, cause a change of the \gls{AIR}. Nevertheless, time invariance is a common assumption in \gls{ASR} applications, justified by the fact, that  an utterance, for which the \gls{AIR} is assumed to be constant, is only a few seconds long. 

However, the nonstationarity of the speech and noise signals has to be taken into account. When moving to a frequency domain representation we therefore have to use the \gls{STFT}, i.e., apply the DFT to windowed segments of the signal. Typical window, also called frame, lengths are \SIrange{25}{128}{ms} and frame advances are \SIrange{10}{32}{ms}.

When expressing the signal model of Eq.~\eqref{eq:signal_model_time_domain} in the \gls{STFT} domain, it is important to note that, in a common setup, the \gls{AIR} is much longer than the length of the analysis window. In a typical living room environment it takes \SIrange{0.3}{0.7}{s} for the \gls{AIR} to decay to \SI{-60}{dB} of its initial value, which is considerably longer than the aforementioned window length. Then the convolution in Eq.~\eqref{eq:signal_model_time_domain} no longer corresponds to a multiplication in the \gls{STFT} domain, but instead to a convolution over the frame index. To a good approximation \cite{Avargel2007OnMT, Gilloire1992},  Eq.~\eqref{eq:signal_model_time_domain} can be expressed in the STFT domain as
\begin{align}
  \label{eq:signal_model_stft_domain_m}
	y_{m,t,f} = \sum_{i=1}^I\sum_{\tau=0}^{L-1}a_{m,\tau,f}^{(i)}s_{t-\tau,f}^{(i)} + n_{m,t,f}, 
\end{align}
where $a_{m,t,f}^{(i)}$ is a time-frequency representation of the \gls{AIR}, called \gls{ATF}; $s_{t,f}^{(i)}$ and $n_{m,t,f}$ are the \glspl{STFT} of  the $i$th source speech signal and of the noise at microphone index $m$, frame index $t$, and frequency bin index $f$. Furthermore, $L$ denotes the length of the \gls{ATF} in number of frames. Note that we used, in an abuse of notation, the same symbols for the time domain and frequency domain representations. This should not lead to confusion, because time domain signals have an argument, as in $y[\ell]$, while frequency domain variables have an index, as in  $y_{t,f}$.
The model of Eq.~\eqref{eq:signal_model_stft_domain_m} strongly contrasts with the model for a close-talking situation, where $y_{t,f} = s_{t,f} + n_{t,f}$, or where even the noise term can be neglected. 

When trying to extract $s_{t,f}^{(i)}$ from $y_{m,t,f}$, it comes to our help that multi-channel input is available, i.e., $m\in[1,\ldots , M]$. Defining the vector of microphone signals $\vect{y}_{t,f} = \begin{bmatrix} y_{1,t,f} & \ldots , y_{M,t,f}\end{bmatrix}\T$, we can write
\begin{align}
  \label{eq:signal_model_stft_domain}
	\vect{y}_{t,f} = \sum_{i=1}^I\sum_{\tau=0}^{L-1}\vect{a}_{\tau,f}^{(i)}s_{t-\tau,f}^{(i)} + \vect{n}_{t,f}, 
\end{align}
where $\vect{a}_{t,f}^{(i)}$ and $\vect{n}_{t,f}$ are similarly defined as $\vect{y}_{t,f}$.

\begin{figure}
	\centering
	\footnotesize 
	\hspace{-0.5em}\includegraphics[width=0.5\textwidth]{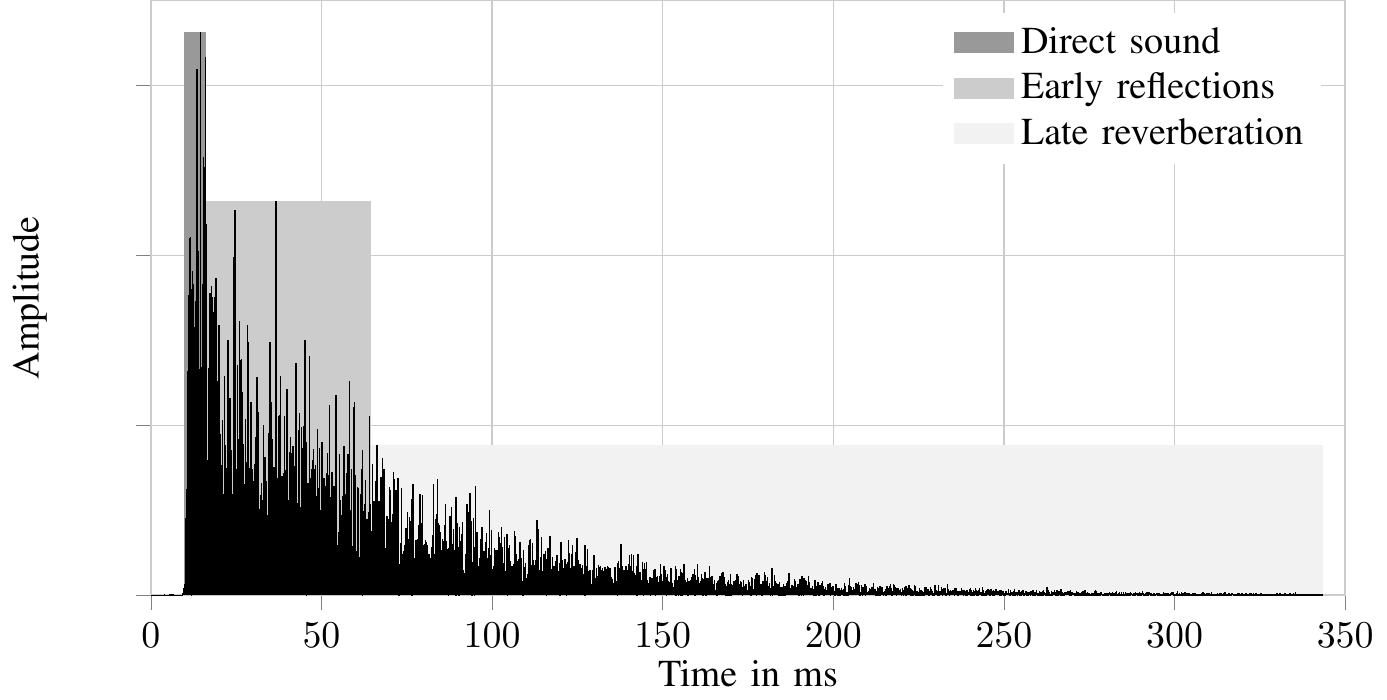}
	\caption{An acoustic impulse response consists of the direct sound, early reflections and late reverberation.}
	\label{fig:air}
\end{figure}

Fig.\ref{fig:air} displays a typical \gls{AIR}: it consists of three parts, the direct signal, early reflections and late reverberation caused by multiple reflections off walls and objects in the room.
The early reflections are actually beneficial both for human listeners and for \gls{ASR}. Its intelligibility is even better than that of the ``dry'' line-of-sight signal.
After the mixing time, which is in the order of  \SI{50}{ms}, the diffuse reverberation tail begins.
This late reverberation degrades human intelligibility and also leads to a significant loss in recognition accuracy of a speech recognizer.
Thus, we split the \gls{ATF} in an early and a late part: 
\begin{align}
  \label{eq:signal_model_stft_domain_separated}
	\vect{y}_{t,f}
	&= \sum_{i=1}^I \vect{d}_{t,f}^{(i)}
	+ \sum_{i=1}^I \vect{r}_{t,f}^{(i)}
	+ \vect{n}_{t,f}, 
\end{align}
where the early-arriving speech signals are given by
\begin{align}
  \label{eq:direct_signal}
	\vect{d}_{t,f}^{(i)}
	&= \sum_{\tau=0}^{\Delta-1}\vect{a}_{\tau,f}^{(i)}s_{t-\tau,f}^{(i)}\approx\vect{h}_f^{(i)}s_{t,f}^{(i)},
\end{align}
and the late-arriving speech signals are given by
\begin{align}
  \label{eq:late_reverberation_signal}
	\vect{r}_{t,f}^{(i)} &= \sum_{\tau=\Delta}^{L-1}\vect{a}_{\tau,f}^{(i)}s_{t-\tau,f}^{(i)}.
\end{align}
Here, $\Delta$ is the temporal extent of the direct signal and early reflections, which is typically set to correspond to the mixing time. For example, $\Delta$ is set at $3$ when a frame advance is set at 16~ms.
In Eq.~(\ref{eq:direct_signal}), the desired signal is approximated by the product of a time-invariant (non-convolutive) \gls{ATF} vector $\vect{h}_f^{(i)}$ with the clean speech $s_{t,f}^{(i)}$, disregarding the spread of the desired signal over multiple analysis frames. Other works have tried to overcome this approximation by employing a convolutive transfer function model for the desired signal \cite{Talmon2009, Li2017a}.

Considering Eq.~\eqref{eq:signal_model_stft_domain_separated}, the tasks of the enhancement stage can be defined as follows:
\begin{itemize}
	\item Dereverberation (also known as deconvolution) aims at removing the late reverberation component from the observed mixture signal.
	\item The goal of source separation is to disentangle the mixture into its $I$ speech components,\footnote{where in some approaches, the noise is treated like an additional, $(I+1)$st component.} while
	\item Beamforming aims at extracting a target speech signal, which can be any of the $I$ sources, by projecting the input signal to the one-dimensional subspace spanned by the target signal, thereby diminishing signal components in other subspaces.
\end{itemize}
We will discuss each of the above tasks in detail in Section~\ref{sec:se}.

\subsection{Performance metrics}
\label{sec:performance_metrics}

Clearly, the ultimate performance measure depends on the application. For a transcription task it is the word error rate, while it is the success rate (high precision and recall) for an information retrieval task. 
However, when developing the speech enhancement front-end it is very helpful to be able to assess the quality of the enhancement with an instrumental measure which is independent of the \gls{ASR} or a downstream \gls{NLP} component.
This will give not only smaller turnaround times in system development, but also gives more insight in how to improve front-end performance. 

Clearly, speech quality and intelligibility is most informatively assessed by human listening experiments. But because these are too expensive and time consuming there is a whole body of literature devoted to how to measure speech quality or intelligibility by an ``instrumental'' measure. Measures, which have been originally developed to evaluate speech communication systems and which have found widespread use in speech enhancement are \gls{PESQ} \cite{PESQ} for speech quality and \gls{STOI} for speech intelligibility \cite{Taal2010}. Note that both measures are ``intrusive'', which means that a clean reference signal is required. 
Please do also note that those measures are only moderately correlated with ASR performance, as has been empirically observed, e.g., in \cite{Chen2018CHiME4}. They are still useful in system development, but for a definite assessment of the benefits of an enhancement system for ASR, recognition experiments  are indispensable.

For the evaluation of source separation performance the most common measure is the \gls{SDR} \cite{Vincent2006Performance}. It measures the ratio of the power of the signal of interest to the power of the difference between the signal of interest and its prediction (obtained by the source separation algorithm). Today, values of more than \SI{10}{dB} are not uncommon.

%% file: chapters/enhancement.tex
\section{Multi-channel speech enhancement}
\label{sec:se}

We now discuss enhancement techniques to address the aforementioned signal degradations. While linear and non-linear filtering approaches are developed for speech enhancement, the linear filtering has empirically been shown to be advantageous to estimate the desired signal $\vect{d}_{t,f}^{(i)}$ in Eq.~(\ref{eq:signal_model_stft_domain}) from the observation $\vect{y}_{t,f}$ in terms of \gls{WER} reduction of far-field speech recognition \cite{REVERB2016,chime3CS,Catalin2019Single}.
This linear filtering leverages information the \gls{AM} typically does not have access to, while not introducing time-dependent artifacts such as musical tones. On the other hand, the non-linear filtering approach has been shown to be useful for estimating statistics of signals, such as time-frequency dependent variances and masks of signals \cite{Wu:2017:RAS:3068681.3068688,DeliangWang2018}, which are effectively used for estimating a linear filter. 

A very general form of a (causal time-invariant) linear filter can be represented by a convolutional beamformer \cite{Buchner2004icassp,Nakatani2019ML,Li2017a,Talmon2009}.
It is defined as
\begin{align}
\label{eq:convolutional_beamforming}
    \hat{d}_{1,t,f}^{(i)} &=\sum_{\tau=0}^{L_w-1}\left(\vect{w}_{\tau,f}^{(i)}\right)\H\vect{y}_{t-\tau,f},
\end{align}
where $\hat{d}_{1,t,f}^{(i)}$ is an estimate of $\vect{d}_{t,f}^{(i)}$ at the $1$st microphone\footnote{Without loss of generality, we here declare the first microphone as the reference microphone.}, $\vect{w}_{\tau,f}^{(i)}=[w_{1,\tau,f}^{(i)},\ldots,w_{M,\tau,f}^{(i)}]\T\in\mathbb{C}^{M\times 1}$ is a coefficient vector of the convolutional beamformer to be optimized for the estimation of $\hat{d}_{1,t,f}^{(i)}$, $L_w$ is the length of the convolutional beamformer, and $\smash{(\cdot)\H}$ denotes transposition and complex conjugation. 
While many techniques have been developed for optimizing a convolutional beamformer \cite{Buchner2004icassp,Li2017a,Talmon2009}, an approach decomposing it into a \gls{MCLP} filter and a beamformer is widely used as a frontend for the far-field ASR.
With $\Delta = 1$ and by applying the distributive property, Eq.~(\ref{eq:convolutional_beamforming}) can be rewritten as
\begin{align}
\label{eq:factorized_beamformer}
    \hat{d}_{1,t,f}^{(i)} &=\underbrace{\left(\!\vect{w}_{0,f}^{(i)}\!\right)\H\vphantom{\Bigg|}}_{\text{Beamformer}}
    \underbrace{\left(\!\vect{y}_{t,f}-\sum_{\tau=\Delta}^{L_w-1}\left(\!\vect{C}_{\tau,f}^{(i)}\!\right)\H\vect{y}_{t-\tau,f}\!\right)\vphantom{\Bigg|}}_{\text{\acrshort{MCLP} filter}}\!,\!\!
\end{align}
where $\vect{C}_{\tau,f}^{(i)}\in\mathbb{C}^{M\times M}$ is a \gls{MCLP} coefficient matrix satisfying $\vect{C}_{\tau,f}^{(i)}\vect{w}_{0,f}^{(i)}=-\vect{w}_{\tau,f}^{(i)}$.
Equation~(\ref{eq:factorized_beamformer}) highlights that a convolutional beamformer that estimates $\hat{d}_{1,t,f}^{(i)}$ can be decomposed into two consecutive linear filters: A \gls{MCLP} filter \cite{Slock1994mclp} corresponding to the terms in the parentheses, and a (non-convolutional) beamformer $\vect{w}_{0,f}^{(i)}$ \cite{VanVeen1988Beamforming,Sharon2017taslp}. 
As will be discussed later, the \gls{MCLP} filter can perform reduction of late reverberation, namely dereverberation.
The beamformer, on the other hand, can perform reduction of noise, i.e., denoising, and extraction of a desired source from other competing sources, i.e., source separation.

The factorization in Eq.~(\ref{eq:factorized_beamformer}) 
allows us to use a cascade configuration for speech enhancement, i.e., dereverberation followed by denoising and source separation.
This is advantageous because we can decompose the complicated enhancement problem into sub-problems that are easier to handle. Furthermore, it is
shown that, under  certain moderate conditions, even when we separately optimize dereverberation and beamforming, the estimate obtained by the cascade configuration is equivalent to (or can be even better than) that obtained by direct optimization of the convolutional beamformer in Eq.~(\ref{eq:convolutional_beamforming}) \cite{boeddeker2020icassp}.

Although both dereverberation and beamforming are well-known concepts from antenna arrays \cite{VanVeen1988Beamforming,Abed-Meraim1997pe}, acoustic signal processing in a non-stationary acoustic environment requires additional efforts, such as estimation of time-varying statistics of temporally-correlated desired sources and noise, and ``broadband'' processing in the time-frequency domain \cite{Vincent2018, Makino2018}. 
For this purpose, many techniques have been developed:
\begin{itemize}
	\item For dereverberation, estimation and subtraction of the spectrum of the late reverberation has been employed, e.g., \cite{Xiong2015}. Also, \gls{MCLP} filtering with delayed prediction and a time-varying Gaussian source assumption have been developed and shown effective for both single and multiple desired source scenarios \cite{Yoshioka2012GeneralWPE,caroselli2017adaptive}.
	\item For denoising, techniques for effectively estimating the time-varying statistics of the desired signal and the noise have been developed based on estimation of a time-frequency dependent mask. \cite{Souden2010MVDR,7472671}. 
  \item For source separation, sophisticated techniques for estimating masks of multiple competing sources have been developed. Modern techniques are even able to handle multiple sources in single-channel input \cite{Isik2016DeepClustering, 7952155, Kolbaek2017PIT}.
\end{itemize}
While these techniques are well established in classical signal processing areas \cite{1142739, Makino2007BSS, Naik2014, Naylor2010}, recently purely deep learning based solutions have challenged those solutions, e.g. \cite{DeliangWang2018, Wu:2017:RAS:3068681.3068688}. The advantage of the deep learning-based solutions is their powerful capability of modeling source magnitude spectral patterns over wide time frequency ranges, which were very difficult to handle by classical signal processing approaches. The deep learning approaches, however, are also notorious for being resource hungry and hard to interpret. Their training for speech enhancement tasks requires parallel data, i.e., a database which contains each speech utterance in two versions, distorted and clean, one serving as input to the network, and the other as training target.
Reasonably, this can only be obtained by artificially adding the distortions to a clean speech utterance, leaving an unavoidable mismatch between artificially degraded speech in training and real recordings in noisy reverberant environments during test. Classical signal processing solutions are typically much more resource efficient and do not have this parallel data training problem. We will show for each of the three enhancement tasks how ``neural network-supported signal processing'' or ``signal processing supported neural networks'' can combine the advantages of both worlds, achieving high enhancement performance, being resource efficient and rendering parallel data unnecessary \cite{Hey2016,7472778, Drude2017Integration,Kinoshita2017Neural,Heymann2019Joint, HeymannIWAENC2018, Heymann2017Beamnet}.

\begin{figure*}
\centering
\footnotesize 
\includegraphics[width=1.0\textwidth]{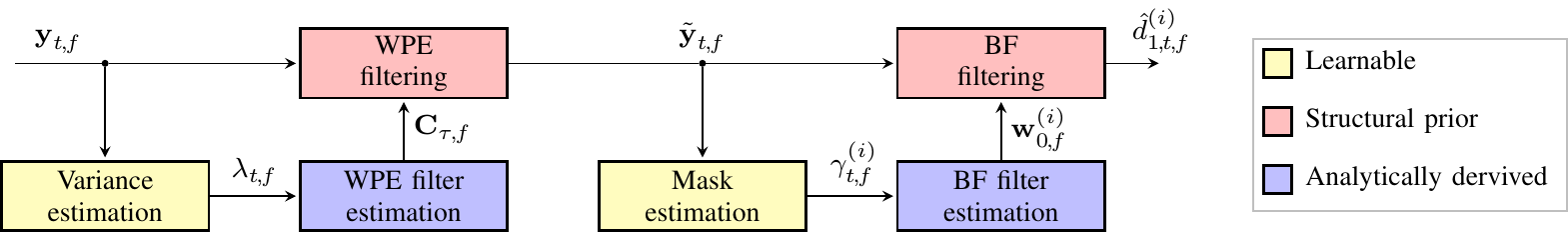}
\vspace{-1em}
\caption{Overview of the enhancement system consisting of a neural network supported dereverberation module and a neural network supported or spatial clustering model supported beamforming module. The \gls{MCLP} coefficient matrix $\vect C_{\tau,f}$ as well as the time-varying variance $\lambda_{t,f}$ are speaker-independent as argued in the last paragraph of Sec.~\ref{subsec:dereverb}. The BF filtering block may contain additional postfiltering to compensate for potential artifacts the beamformer may have produced.}
\label{fig:enhance}
\end{figure*}

A typical processing pipeline for dereverberation, separation, and extraction is illustrated in Fig.~\ref{fig:enhance}.

\subsection{Dereverberation}
\label{subsec:dereverb}
The goal of dereverberation is to reduce the late reverberation $\vect{r}_{t,f}^{(i)}$ from the observation $\vect{y}_{t,f}$ in Eq.~(\ref{eq:signal_model_stft_domain_separated}) while keeping the desired signal $\vect{d}_{t,f}^{(i)}$ unchanged. 
Based on the decomposition in Eq.~(\ref{eq:factorized_beamformer}), we here highlight a technique based on \gls{MCLP} filtering, referred to as \gls{WPE} dereverberation \cite{Nakatani2010WPE,Yoshioka2012GeneralWPE}. %
In the following, we first explain \gls{WPE} dereverberation in the noiseless single source case, i.e., assuming $\vect{y}_{t,f}=\vect{d}_{t,f}^{(i)}+\vect{r}_{t,f}^{(i)}$, and then explain its applicability to the noisy multiple source case at the very end of this section.

\begin{figure}
\centering
\footnotesize 
\includegraphics[width=0.5\textwidth]{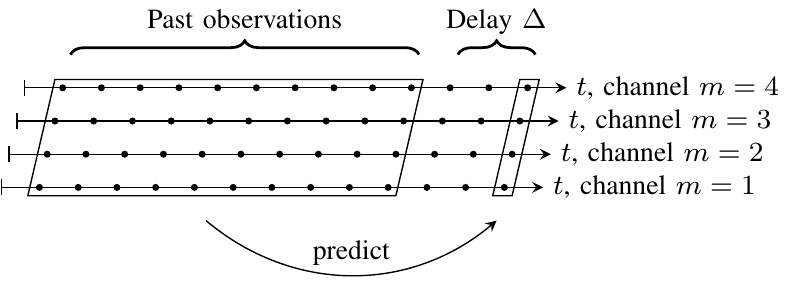}
\vspace{-1em}
\caption{\Gls{WPE} estimates a filter to predict the late reverberation in the current observation from the past observations (skipping $\Delta-1$ frames).
The late reverberation is then subtracted from the current observation.}
\label{fig:mimo}
\end{figure}

The core idea of \gls{WPE} dereverberation is to predict the late reverberation of the desired signal from past observations.
This late reverberation is then subtracted from the observed signal to obtain an estimate of the desired signal.
Just as Eq.~(\ref{eq:factorized_beamformer}) indicates which past observations are used for prediction, Fig.~\ref{fig:mimo} visualizes the past observations, the prediction delay and which frame of late reverberation is predicted.
A unique characteristic of \gls{WPE} is the introduction of the prediction delay $\Delta$, which corresponds to the duration of the direct signal and early reflections in Eq.~(\ref{eq:direct_signal}). It avoids the desired signal being predicted from the immediately past observations, because this would destroy the short-time correlation typical of a speech signal.  Thanks to this, the \gls{WPE} can only predict the late reverberation and keep the desired signal unchanged. 

To deal with the time-varying characteristics of speech in the \gls{MCLP} framework, \Gls{WPE} estimates the coefficient matrix $\vect{C}_{\tau,f}^{(i)}$ based on maximum likelihood estimation.
It is assumed that the desired signal $\vect{d}_{t,f}^{(i)}$ follows a zero-mean circularly-symmetric complex Gaussian distribution with the unknown channel-independent time-varying variance $\lambda_{t,f}^{(i)}$ of the early-arriving speech signal:
\begin{align}
p\left(\vect{d}_{t,f}^{(i)}\right) = \mathcal N_{\mathbb C}\left(\vect{d}_{t,f}^{(i)};\vect 0,\lambda_{t,f}^{(i)}\vect{I}_M\right),
\end{align}
where $\vect{d}_{t,f}^{(i)}$ is obtained from \gls{MCLP} filtering in Eq.~(\ref{eq:factorized_beamformer}) and $\vect{I}_M$ is an $M\times M$-dimensional identity matrix.
With this model, the objective to minimize becomes
\begin{align}
{\cal L}(\psi_f)
&= -\sum_t\log p\left(\vect{d}_{t,f}^{(i)};\psi_f\right) \nonumber \\
&= \sum_t \frac{||\vect{y}_{t,f}-\sum_{\tau=\Delta}^{L_w-1}\left(\vect{C}_{\tau,f}^{(i)}\right)\H\vect{y}_{t-\tau,f}||_2^2}{\lambda_{t,f}}\nonumber\\
&\phantom{=}\quad+\sum_t M\log\lambda_{t,f}+\text{const.}
\label{eq:wpe_cost}
\end{align}
where $\psi_f$ is a set of parameters to be estimated at frequency $f$, composed of $\vect{C}_{\tau,f}$ and $\lambda_{t,f}$ for all $\tau$ and $t$, and $||\cdot||_2$ denotes the Euclidean norm. Variations of the objective have also been proposed for better dereverberation performance by introducing sparse source priors \cite{Ante2015taslp,Chetupalli2019taslp}.

The minimization of the above objective leads to an iterative algorithm which alternates between estimating the time-varying variance $\lambda_{t,f}^{(i)}$ and the coefficient matrix $\vect C_{\tau,f}^{(i)}$.
The steps can be summarized as follows:
{
\savebox\strutbox{$\vphantom{\dfrac11}$}
\begin{align}
\text{Step 1)}\!\!\!\!\!\!\phantom{\mathbf R_f^{(i)}}
\lambda_{t,f}^{(i)} &= \frac{1}{\left(\delta + 1 + \delta\right)M} \sum_{\tau=t-\delta}^{t+\delta} \sum_{m} | d_{m,\tau,f}^{(i)} |^2\!,\hspace{-20em} \label{eq:wpestep2} \\
\text{Step 2)}\!\!\!\!\!\!\phantom{\lambda_{t,f}^{(i)}}
\mathbf R_f^{(i)} &= \sum_t \frac{
\vectBar y_{t-\Delta,f}
\vectBar y_{t-\Delta,f}\H
}{
\lambda_{t,f}^{(i)}
} &&\!\!\!\!\!\!\!\in \mathbb C^{MK \times MK}, \label{eq:wpestep1} \\
\mathbf P_{f}^{(i)} &= \sum_t \frac{
\vectBar y_{t-\Delta,f}
\vect y_{t,f}\H
}{
\lambda_{t,f}^{(i)}
} &&\!\!\!\!\!\!\!\in \mathbb C^{MK \times M}, \label{eq:wpestep1p} \\
\vectBar C_{f}^{(i)} &= \left(\vect R_f^{(i)}\right)^{-1} \mathbf P_{f}^{(i)} &&\!\!\!\!\!\!\!\in \mathbb C^{MK \times M}, \label{eq:wpestep1end}
\end{align}
}%
where $\vectBar y_{t-\Delta,f} \in \mathbb C^{MK \times 1}$ is the stacked observation vector as depicted by the box on the left hand side of Fig.~\ref{fig:mimo} and $\delta$ defines a temporal context.

In the variance estimation step in Eq.~(\ref{eq:wpestep2}), $\lambda_{t,f}^{(i)}$ is updated dependent on the previous estimate of $\vect{C}_{\tau,f}^{(i)}$, i.e., it is estimated as the variance of the signal dereverberated with $\vect{C}_{\tau,f}^{(i)}$ according to the \gls{MCLP} filter in Eq.~\eqref{eq:factorized_beamformer}.
Often, smoothing by averaging over neighboring frames with a left context of $\delta$ and a right context of $\delta$ is introduced to reduce the variance of this variance estimate.

In the filter matrix estimation step in Eqs.~(\ref{eq:wpestep1})--(\ref{eq:wpestep1end}), fixing $\lambda_{t,f}^{(i)}$ at its value estimated in the previous step makes Eq.~(\ref{eq:wpe_cost}) a simple quadratic form, and thus we can reach a global minimum  by a closed-form update.  
Here, $\vect R_f^{(i)}$ can be interpreted as an auto-correlation matrix of normalized stacked observation vectors.
Further, $K = L_{w} - \Delta$ is the number of filter taps.
Finally, Eq.~\eqref{eq:wpestep1end} computes the stacked filter matrix
\begin{align}
\vectBar C_{f}^{(i)} = \left[
\left(\vect C_{\Delta,f}^{(i)}\right)\T,
\dots,
\left(\vect C_{L_w - 1,f}^{(i)}\right)\T
\right]\T
\end{align}
using the Wiener-Hopf equation.

This iterative algorithm may be started by initializing the time-varying variance $\lambda_{t,f}^{(i)}$ with that of the observation. Although this is a rather crude approximation, it typically converges within three iterations.

The use of a neural network further allows to estimate the time varying variance $\lambda_{t,f}^{(i)}$ within a single step avoiding the iterative estimation, and eases the transition towards online  processing~\cite{Kinoshita2017Neural, HeymannIWAENC2018}.
In \cite{Kinoshita2017Neural} a neural network is trained with a \gls{MSE} loss to predict (the logarithm of) the time-varying variance $\lambda_{t,f}^{(i)}$ and applied to offline and block-online processing, while \cite{HeymannIWAENC2018} extends this to frame-online processing. 

In order to handle noisy multi-source cases, we slightly revise the goal of the \gls{WPE} dereverberation to estimate a single set of coefficient matrices $\vect{C}_{\tau,f}$ that can reduce the late reverberation $\vect{r}_{t,f}^{(i)}$ for all $i$ at the same time, rather than estimating a different set of matrices $\vect{C}_{t,f}^{(i)}$ separately for dereverberation of each source $i$. 
Existence of such a set of coefficient matrices is guaranteed by the \gls{MINT} \cite{miyoshi1988mint} when $M\ge I$, $\vect{n}_{t,f}=0$, and the acoustic transfer functions share no common zeros.
The coefficient matrices can be estimated based on the objective of the \gls{WPE} in Eq.~(\ref{eq:wpe_cost}), by setting $\lambda_{t,f}$ to represent the variance of the mixture of all $\vect{d}_{t,f}^{(i)}$.
Although $\vect{n}_{t,f}=0$ is usually not satisfied within the far-field setting, due to the inherent robustness of the \gls{MCLP} filtering, \gls{WPE} works well with such additive noise.

While we discussed here WPE in some detail, because it has found widespread use in the ASR community, this is by no means the only approach to dereverberation. Instead of estimating the direct signal and early reflections, one can estimate the power spectral density of the late reverberation and subtract it from the observed signal, thereby achieving a dereverberating  effect \cite{Xiong2015, Braun2018a}. Also, neural networks trained to estimate the nonreverberant signal from the observed reverberant one are very successful \cite{Wang2020}.

\subsection{Beamforming}
\label{ssec:Beamforming}
Beamforming aims at reducing additive noise and residual reverberation from the observation.
As in the decomposition in Eq.~(\ref{eq:factorized_beamformer}), a spatial filter $\vect w_{0,f}^{(i)}$ (commonly referred to as beamformer) is used to obtain an estimate of the desired signal from the output of the \gls{WPE} dereverberation.
Consequently, we here define new variables which describe the signal components after \gls{WPE} processing.
Let us define the input of the beamformer as 
\begin{align}
\vectTilde y_{t,f}
&=\vect{d}_{t,f}^{(1)}+ \dots + \vect{d}_{t,f}^{(I)} + \vectTilde n_{t,f}
=\vect{d}_{t,f}^{(i)}+\vectTilde n_{t,f}^{(i)},
\label{eq:bf_in_decomposition}
\end{align}
where $\vectTilde n_{t,f}$ contains all residual reverberation and noise, and where $\vectTilde n_{t,f}^{(i)}$ collectively represents all the interference signal components from the viewpoint of speaker $i$:
these are the remaining reverberation, the source signals other than the desired signal, ambient noise, and possible other deviations from $\vect{d}_{t,f}^{(i)}$.
In other words, Eq.~(\ref{eq:bf_in_decomposition}) shows the decomposition from the perspective of speaker $i$ and not for all speakers.
Then, the beamforming step is meant to remove all interferences $\vectTilde n_{t,f}^{(i)}$ from $\vectTilde y_{t,f}$ while keeping $\vect{d}_{t,f}^{(i)}$ unchanged.

Most statistical beamforming approaches rely on estimated second order statistics, namely the spatial covariance matrices of the desired signal $\vect \Phi_{\vect{d}\vect{d},f}^{(i)}$ and that of the interference ${\Phi}_{\vectTilde n\vectTilde n,f}^{(i)}$.
A beamforming algorithm is derived by defining an optimization criterion.
A widely used approach is \gls{MVDR} beamforming
which minimizes the expected variance of the resultant interference subject to a distortionless constraint involving the \gls{ATF} vector $\vect{h}_{f}^{(i)}$ in Eq.~(\ref{eq:direct_signal}).
It is defined as
\begin{align}
\label{eq:def_mvdr}
    \vect{w}_{0,f}^{(i)}=\argmin_{\vect{w}}\vect{w}\H \vect \Phi_{\vectTilde n\vectTilde n,f}^{(i)}\vect{w}~~\text{s.t.}~~\vect{w}\H\vect{h}_f^{(i)}=h_{1,f}^{(i)},
\end{align}
where $\vect \Phi_{\vectTilde n\vectTilde n,f}^{(i)}$ is the spatial covariance matrix of all interferences, assumed to be time-invariant, and $\smash{h_{1,f}^{(i)}}$ is the $1$st microphone element of $\smash{\vect{h}_{f}^{(i)}}$.
Thanks to the distortionless constraint, the beamformer keeps the desired signal unchanged, while reducing the additive distortions.
The optimization problem in Eq.~(\ref{eq:def_mvdr}) results in
\begin{align}
\vect w_{0,f}^{(i)} = \frac{
\left(\vect \Phi_{\vectTilde n\vectTilde n,f}^{(i)}\right)^{-1} \tilde{\vect{h}}_{f}^{(i)}
}{
\left(\tilde{\vect{h}}_{f}^{(i)}\right)\H \left(\vect \Phi_{\vectTilde n\vectTilde n,f}^{(i)}\right)^{-1} \tilde{\vect{h}}_{f}^{(i)}
},
\end{align}
where $\tilde{\vect{h}}_{f}^{(i)}$ is a relative transfer function (RTF)  \cite{Gannot2001SP,Schwarts2015TASLP} defined as the ATF vector normalized by its $1$st microphone component, i.e., $\tilde{\vect{h}}_{f}^{(i)}={\vect{h}}_{f}^{(i)}/h_{1,f}^{(i)}$. The RTF is a  widely used representation to avoid scale ambiguity of ATF vector estimation.

Techniques for estimating the RTF vector $\tilde{\vect{h}}_{f}^{(i)}$ have been developed, which in general require an estimate of a spatial covariance matrix $\vect\Phi_{\vect d \vect d,f}^{(i)}$ of the desired signal $\vect{d}_{t,f}^{(i)}$ \cite{Souden2010MVDR,Sharon2017taslp}.
%
Alternative objectives can also be used for beamforming, such as likelihood maximization with a time-varying Gaussian source assumption, similar to WPE, resulting in the weighted Minimum Power Distortionless response (wMPDR) beamformer \cite{boeddeker2020icassp}, and maximization of expected output \gls{SNR} resulting in maximum \gls{SNR} beamformer (also called Generalized Eigenvalue Decomposition (GEV) beamformer) \cite{WaKrHa08}.

One way to estimate these covariance matrices is to select time frames in which just one signal component is active, e.g., the beginning of a recording where only noise is active.
This approach is appropriate under the assumption that the corresponding signals are stationary.
However, a better and more fine-grained approach is to use a time-frequency mask, $\gamma_{t,f}^{(i)}$, to decide for each \gls{TF} bin how well it corresponds to the target speaker or the interference.
This leads to a covariance matrix calculation with time- and frequency-dependent masks $\gamma_{t,f}^{(i)}$:
\begin{align}
\vectHat\Phi_{\vect{d}\vect{d},f}^{(i)}
&= {\sum_{t} \gamma_{t, f}^{(i)} \vectTilde y_{t, f} \vectTilde y_{t, f}\H} \bigg/ {\sum_{t} \gamma_{t, f}^{(i)}}, \\
\vectHat\Phi_{\vectTilde n\vectTilde n,f}^{(i)} &= {\sum_{t}\sum_{i'\neq i}\gamma_{t, f}^{(i')} \vectTilde y_{t, f} \vectTilde y_{t, f}\H} \bigg/ {\sum_{t} \sum_{i'\neq i}\gamma_{t, f}^{(i')}}.
\label{eq:cov_estimate}
\end{align}
Conceptually, assuming that the selected \gls{TF} bins indeed only contain the desired signal, $\vectHat\Phi_{\vect{d}\vect{d},f}^{(i)} \in\mathbb C^{M\times M}$ approximates the covariance matrix of $\vect d_{t,f}^{(i)}$, and on a similar assumption, $\vectHat\Phi_{\vectTilde n\vectTilde n,f}^{(i)} \in\mathbb C^{M\times M}$ approximates the covariance matrix of all interferences $\vectTilde n_{t,f}^{(i)}$.

Depending on the acoustic environment, the a priori knowledge for the given utterance, the number of speakers in the recording, and the available training data different ways to estimate the masks for each speaker are possible.
The two predominant approaches for mask estimation are unsupervised spatial clustering and neural network-based mask estimation and are explained in the following section.

\subsection{Mask estimation for denoising, single source extraction, and source separation}
\label{sec:spp}
The goal of a mask estimator is to estimate a presence probability mask for each speaker and for noise.
This section first describes unsupervised spatial clustering approaches for single- and multi-speaker scenarios and then continues with neural network-based approaches again for single- and multi-speaker scenarios.

\subsubsection{Unsupervised spatial clustering}
Unsupervised spatial clustering is a technique used to assign each \gls{TF} bin to a particular class based solely on spatial cues, i.e., phase and level differences between microphone channels that provide information about the direction of sound with respect to the microphone array.
A class then models the different speakers characterized by different locations or noise with more diffuse characteristics.
Assuming that the speakers speak from different locations, it is possible to separate the microphone signals into speech signals of the different speakers by clustering the spatial cues~\cite{Mandel2007EM}.

To do so, one typically formulates a statistical model which consists of a class-dependent distribution for each source $i$ and an additional noise class which is here indexed by $i = I+1$:
\begin{align}
p(\vectTilde y_{t,f}) = \sum_{i=1}^{I+1} p(\vectTilde y_{t,f} | \vect\theta^{(i)}) p(z_{t,f} {=} i),
\label{eq:clustering_model}
\end{align}
where $z_{t,f}$ is the hidden class affiliation variable, and $\vect\theta^{(i)}$ summarizes the class-dependent parameters.
Typical class-dependent distributions are the complex Watson distribution \cite{TranVu2010cW}, complex Bingham distribution \cite{Ito2016cBMM}, or the complex angular central Gaussian distribution \cite{Ito2016cACGMM}.
The parameters and the masks are then obtained through an \gls{EM} algorithm in which the E-step and the M-step alternate.
In the M-step, the class-dependent parameters are updated.
In the E-step, the masks $\smash{\gamma_{t,f}^{(i)} = p(z_{t,f} {=} i | \vectTilde y_{t,f})}$, which here correspond to posterior probabilities, are obtained using Bayes' rule:
\begin{align}
\gamma_{t,f}^{(i)}
= \frac{
p(z_{t,f} {=} i) \, p(\vectTilde y_{t,f} | \vect\theta^{(i)})
}{
\sum_{i' = 1}^{I+1} p(z_{t,f} {=} i') \, p(\vectTilde y_{t,f} | \vect\theta^{(i')})
}.
\label{eq:spatial_estep}
\end{align}
In a single-speaker scenario, where one just wishes to distinguish between target speaker and noise, one can use spatial clustering with $I=1$.
To name an example, the winning system of the CHiME~3 robust speech recognition challenge employed such an unsupervised clustering approach with $I=1$ successfully to single-speaker recordings~\cite{Yoshioka2015CHiME3}.
In case of a multi-source scenario, $I$ has to be set to the number of speakers in the mixtures, which either has to be known a-priory, or estimated separately.
The consecutive steps, e.g., beamforming as in Fig.~\ref{fig:enhance} are then repeated for each speaker.

\subsubsection{Neural network-based mask estimation}
In contrast,  mask estimation networks are trained with a supervision signal.
To discuss neural network-based approaches, we first introduce a neural mask estimator as used in neural network-based beamforming in the following.
We then introduce SpeakerBeam as a speaker-informed mask estimator.
Lastly, we introduce neural network-based blind source separation approaches.

\begin{figure}
\centering
\footnotesize 
\includegraphics[width=0.5\textwidth]{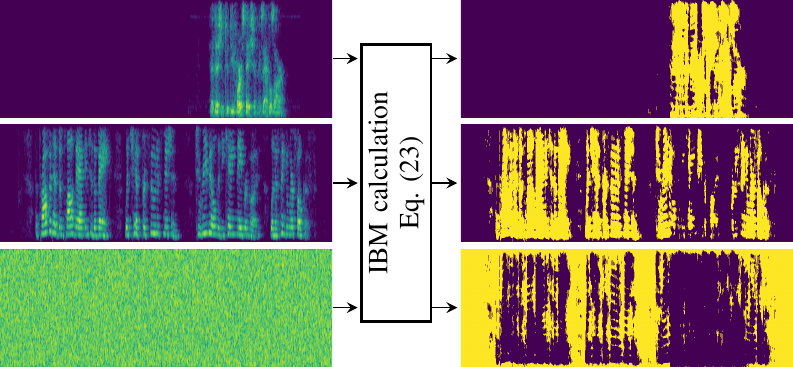}
\vspace{-1em}
\caption{Visualization of the spectrograms of the underlying images $\vect d_{t,f}^{(1)}$, $\vect d_{t,f}^{(2)}$, and $\vectTilde n_{t,f}$ on the left and the ideal binary masks $\text{IBM}_{t,f}^{(1)}$, $\text{IBM}_{t,f}^{(2)}$, and $\text{IBM}_{t,f}^{(3)}$ on the right. Bright colors indicate higher values.}
\label{fig:ibm_process}
\end{figure}

For neural network-based mask estimation, a supervision signal such as an \gls{IBM} \cite{Li2009idealBinaryMask} is first extracted on each training mixture.
To do so, one needs access to the speech images and the noise image, i.e., each individual speech component and the noise component at the microphones, separately:
\begin{align}
\text{IBM}_{t,f}^{(i)}
&= \begin{cases}
1, & \text{for } \big\Vert \vect d_{t,f}^{(i)} \big\Vert_2^2 > \big\Vert \vect d_{t,f}^{(i')} \big\Vert_2^2 \,\forall\, i' \neq i \\
0, & \text{otherwise},
\end{cases}
\label{eq:ibm}
\end{align}
where $i$ corresponds to the source index.
This definition can be extended to an additional noise class by treating the oracle noise signal as $\vect d^{(I+1)} := \vectTilde n_{t,f}$.
Fig.~\ref{fig:ibm_process} illustrates the underlyinging signal components and the corresponding \gls{IBM} with an additional noise class.
Further definitions of oracle masks suitable for supervision can be found in, e.g., \cite{Erdogan2015Masks, hey_asru_2015}.

\begin{figure}
\centering
\footnotesize 
\includegraphics[width=0.5\textwidth]{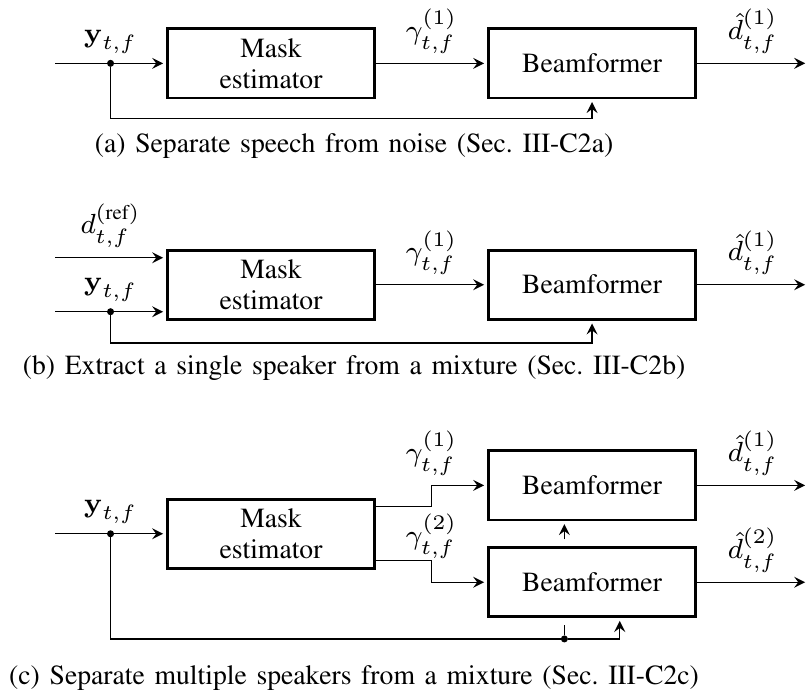}
\vspace{-1em}
\caption{Processing flow for different use-cases of a mask estimator. The corresponding interference mask to calculate the interference covariance matrix in Eq.~(\ref{eq:cov_estimate}) is not shown for brevity.}
\label{fig:mask_flow}
\end{figure}

Then, depending on the particular use-case, a neural network can be trained with such a supervision signal.
The different use-cases are illustrated in Fig.~\ref{fig:mask_flow}.

\paragraph{Separate speech from noise}
\label{paragraph:speech_vs_noise}
One can now train a neural network with, e.g., log-amplitude spectrogram features as input, to predict a speech mask and a noise mask by providing noisy speech training data from various speaker and the corresponding \glspl{IBM} or clean speech signals.
Since speech and noise have different spectro-temporal characteristics, a neural network can distinguish between these signals very well.
An exemplary training criterion is the binary cross entropy between the estimated mask $\gamma_{t,f}^{i}$ and the corresponding oracle $\text{IBM}_{t,f}^{(i)}$.

This mask estimation procedure with a subsequent beamforming step led to dramatic \gls{WER} reductions, e.g., in the CHiME~3/4 challenges~\cite{hey_asru_2015, Erdogan2016MVDR}.
Often, these masks are estimated independently for each channel and then pooled over all channels such that a single mask can be used, e.g., in Eq.~(\ref{eq:cov_estimate}).

\paragraph{Extract a single speaker from a mixture}
\label{paragraph:extract_single}
In many practical applications one is interested in one target speaker in a mixture, e.g., the speaker who is actually interacting with the digital home assistant.
While dealing with speech mixtures, simply training a neural network to extract the target speaker is not possible because both the target speech and interference signal have similar spectro-temporal characteristics.
However, if additional information about the target speaker is available, a neural mask estimator can be informed about speaker-dependent characteristics.
These characteristics may stem from a separate adaptation utterance or from the wake-up keyword.
In the SpeakerBeam framework~\cite{Zmolikova2017Speaker, Zmolikova2019SpeakerBeam} a sequence-summarizing neural network~\cite{Vesely2016Summarizing} which captures the speaker-dependent characteristics is jointly trained with a mask estimation network which uses these characteristics as additional features to estimate a target speaker mask and an interference mask.
VoiceFilter implements this approach with a \gls{CNN} architecture~\cite{Wang2018VoiceFilter}.

\paragraph{Separate multiple speakers and noise}
\label{paragraph:separate}
While, in a single-speaker scenario, a mask estimator only needs to distinguish between speech and non-speech time frequency bins (compare Sec.~\ref{paragraph:speech_vs_noise}), source separation approaches have to solve the following problem: given the observation the algorithm should yield a mask for each speaker as well as an additional noise mask.
For quite some time it has been complicated to do this with neural networks due to the permutation problem: while the order in which the speakers appear at the different output channels of the system is unpredictable, a loss function which assumes a particular order can result in misleading gradients.
While the spatial clustering model in Eq.~(\ref{eq:clustering_model}) is naturally permutation invariant (switching speaker indices does not change the likelihood), permutation invariant losses for neural networks appeared just recently.

Kolbaek et al. formulated a way to turn any loss function, e.g., \gls{CE}, into a permutation invariant loss function~\cite{Kolbaek2017PIT}: the original loss is calculated for every possible permutation.
Then, only the minimal loss is used for back-propagation, e.g.:
\begin{align}
J
&= \argmin_{\Pi} \sum_{i=1}^{I+1} \CE_{t,f}\left(\gamma_{t,f}^{(\Pi(i))}, \text{IBM}_{t,f}^{(i)}\right),
\end{align}
where $\Pi$ is a permutation of $(1, \dots, I+1)$.
A neural network with $I+1$ mask outputs can now be trained with such a \gls{PIT} loss.
The estimated masks $\gamma_{t,f}^{(i)}$ can then be used, e.g., for beamforming.
In its original formulation the network architecture of a \gls{PIT} system depends on the maximum number of speakers expected in a mixture.
The system can be trained in such a way that some output channels are empty when there are less speakers.

Fundamentally differently, Deep clustering, while pioneering this area, used a neural network to calculate embedding vectors for each time frequency bin~\cite{Hershey2016DeepClustering}.
The loss, as any typical embedding loss, is designed in such a way that the embedding vectors belonging to the same class move closer together while the embedding vectors of different classes move further apart.
Naturally, such a formulation is permutation invariant in itself.
The embedding vectors can then be used for clustering yielding masks in a similar way as explained in the clustering approach before.
Interestingly, at least the embedding network is then independent of the number of speakers in a mixture~\cite{Hershey2016DeepClustering}.

\subsubsection{Comparison of spatial and spectral approaches and integrations thereof}
The main advantage of spatial clustering models over neural network-based mask estimation is the interpretability of the underlying stochastic dependencies.
Closely related, this interpretability allows to incorporate a priori knowledge by modifying the parameter updates, e.g., \cite{Boeddeker2018CHiME5} uses externally provided time annotations for the CHiME~5 database.
Due to the spatial features, it exploits spatial selectivity and, as long as the spatial properties of each source are distinct enough, is able to produce meaningful separation results.
Since no training phase is involved, this unsupervised clustering approach naturally generalizes well to unseen conditions.

One drawback of the spatial clustering approaches is, that it is most suited for offline processing.
Although quite a few online or block-online clustering approaches had been proposed, these did not find a lot of application in far-field \gls{ASR} challenges yet.
Moving sources, if no online algorithm is used, can only be handled to some extent: small head movements can still be captured in the class dependent parameters.
Larger movements, however, invalidate the underlying model assumptions.
Further, since clustering is often performed independently across frequency bins, a frequency permutation problem arises~\cite{Sawada2004Robust}:
from one frequency bin to another the spatial clustering solution may have resulted in switched speaker indices.
This frequency permutation problem is independent of the aforementioned global permutation problem when discussing \gls{PIT}.

In contrast to the spatial clustering approaches, neural network-based approaches rely on spectral cues and process all frequency bins jointly.
Therefore, a frequency permutation problem does not occur.
Quite remarkably, the neural network-based separation models learn relations from training databases and tend to perform better with an ever increasing amount of training data.

However, alongside this comes their biggest limitation: depending on the variability of the training data, the models have limited generalizability to unseen conditions, e.g., Yu et al. demonstrated that the performance already degrades significantly when switching from English to Danish~\cite{Yu2016Invariant}.
The training corpus needs to contain the mixed speech as well as access to the clean sources to be able to compute gradients.
A notable exception are unsupervised approaches to train a neural network-based source separator~\cite{Drude2019Unsupervised, Seetharaman2018Bootstrapping, Tzinis2018Unsupervised}.
Further, most neural network-based approaches are single-channel.
Even when multi-channel features are employed~\cite{Wang2018MCDC}, in which way those contribute to better separation performance is far from understood.

By no means these approaches are mutually exclusive.
Judging by the aforementioned advantages and disadvantages, both methods are highly complementary, e.g., \cite{Nakatani2017Integrating} proposed to combine neural network-based mask estimation with spatial clustering for speech enhancement, while \cite{Drude2017Integration, Drude2019Integration} proposed an integration of Deep Clustering and spatial clustering for multi-talker scenarios.

\subsection{Front-end overview}
The entire front-end system is now composed of dereverberation, mask estimation, and beamforming.
An established configuration is depicted in Fig.~\ref{fig:enhance}.
The optimal processing order, as demonstrated in \cite{Delcroix2015Strategies} for conventional beamforming and in \cite{Drude2018WPE} for neural network supported beamforming turns out to be applying \gls{WPE} on the multi-channel signal first and then applying the beamforming step on the dereverberated signal.

Spatial clustering based source separation approaches profit in particular from a preceding \gls{WPE} dereverberation (experimental results in~\cite{Boeddeker2018CHiME5}) since the sparseness assumption, which implies that different speakers populate different \gls{TF} bins, is much better fulfilled for less reverberant speech.
Further experiments also report improved separation performance with neural network-based separation methods~\cite{Yoshioka2018Recognizing}.
However, it is worth to acknowledge that a publication which clearly tracks down the gains of better source separation due to better dereverberation is still missing.

In Fig.~\ref{fig:enhance} the variance estimation network and the mask estimation network conceptually perform a similar task (at least in the single-speaker scenario).
Thus, it might be worth investigating if both models can be fused into a single model with two different outputs. 
Further, for practical reasons, the mask estimation network often operates on the observation signal $\vect y_{t,f}$ to avoid needing to train on dereverberation results.

From a machine learning perspective, it is worth highlighting that the building blocks in Fig.~\ref{fig:enhance} are very differently motivated: the filtering blocks can be seen as structural priors motivated by an a priori understanding of field experts.
The filter coefficient estimation blocks are derived analytically from separate optimization criteria, and the variance estimation neural network as well as the mask estimation neural network are trained independently with gradient descent on a separate training database.
More recently, it has been demonstrated that the neural networks can also be trained with gradients from a downstream task~\cite{Heymann2017Beamnet, Heymann2019Joint} (compare Sec.~\ref{sec:e2e}).

%% file: chapters/backend.tex
\section{ASR back-end}
\label{sec:asr}
To achieve high \gls{ASR} performance in a far-field scenario, we need not only employ a powerful 
speech enhancement front-end but also design carefully the \gls{ASR} back-end.
The \gls{ASR} back-end used for far-field ASR has essentially the same structure as a general back-end
used for recognition of clean speech. 
Those interested can find an overview of legacy ASR systems in \cite{Rabiner2008}, while \cite{6296526, Yu2015, LiDeHaGo2015} describe general ASR in the era of deep learning. 
However, several elements need careful 
consideration when dealing with far-field \gls{ASR}.
In this section, we will first briefly review a general \gls{ASR} back-end 
and then emphasize the key components and design choices that are most relevant for far-field \gls{ASR}.

\subsection{Overview of a general ASR back-end}
\label{sec:overview_asr}
The goal of the \gls{ASR} back-end is to find the most likely 
word sequence, $\vectHat v$,  given a sequence of observed speech features $\vect{O}$.
Here for generality, the speech features can be derived either from clean speech, microphone observations or enhanced speech, as described in Section \ref{sec:se}.
The task of \gls{ASR} is formulated with the Bayes decision theory as
\begin{equation}
 \vectHat v  = \argmax _{\vect{v} \in \mathcal{V}} p\left(\vect{v}|\vect{O}\right),
 \label{eq:asr_bayes_decision}
\end{equation}
where $\vect{O}=(\vect{o}_1, \ldots, \vect{o}_t, \ldots, \vect{o}_T )$ 
is a sequence of speech features, $\vect{o}_t \in \mathbb{R}^D$, 
is a feature vector for frame $t$, 
$\vect{v} = (v_1, \ldots, v_j, \ldots, v_J  )$ is a $J$-length word sequence, 
$v_j \in \mathcal{V}$ is a word at position $j$,
and  $\mathcal{V}$ is the set of possible words, called vocabulary.
Since it is complex to deal with $p(\vect{v}|\vect{O})$ directly, 
the problem is usually rewritten using the Bayes theorem as,
\begin{equation}
 \vectHat v  = \argmax _{\vect{v} \in \mathcal{V}} p\left(\vect{O}|\vect{v}\right) p(\vect{v}),
 \label{eq:asr_dec}
\end{equation}
where the likelihood function $p(\vect{O}|\vect{v})$ is called the \glsfirst{AM} 
and the prior distribution $p(\vect{v})$ is the \glsfirst{LM} \cite{rabiner1989tutorial}.
Note that some recent end-to-end \gls{ASR} systems described in Section \ref{sec:e2e} 
aim at directly modeling $p(\vect{v}|\vect{O})$. 

\begin{figure}[t]
\centering
\footnotesize 
\includegraphics[width=0.5\textwidth]{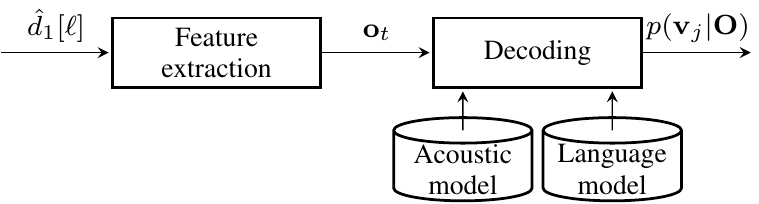}
\vspace{-1em}
\caption{Schematic diagram of a general \gls{ASR} back-end.}
\label{fig:asr_backend}
\end{figure}

Fig.~\ref{fig:asr_backend} depicts a general \gls{ASR} back-end with its main components, i.e., the feature extraction module, the \gls{AM} and the language model, which are briefly described below.

\subsubsection{Feature extraction}
\label{ssec:feat_extract}
The first component of an \gls{ASR} back-end is a feature extraction module that converts the time domain signal $\hat{d}_{1}[l]$ into speech feature $\vect{o}_t$ more suitable for \gls{ASR}.
There has been a lot of research on designing robust features for \gls{ASR}.
However, the simple \gls{LMF} coefficients are widely used both for general and far-field \gls{ASR}.
\gls{LMF} coefficients are obtained by computing the power spectrum of the time-domain signal using a \gls{STFT}, then applying a Mel filter to emphasize low-frequency components of the spectrum.
Finally, the dynamic range is compressed using the logarithm operation as,
\begin{align}
o_{t,\nu} &= \text{Feat}\left([\hat{d}_{t,1},\ldots ,\hat{d}_{t,f}, \ldots,\hat{d}_{t,F}]\T\right), \nonumber\\
        &= \text{MVN}\Bigg(\log\Big( \sum_{f} b_{\nu,f} | \hat{d}_{t,f} |^2 \Big) \Bigg),    
\end{align}
where Feat denotes the feature extraction process. Further, $\hat{d}_{t,f}$ is the \gls{STFT} coefficient of the enhanced speech, $F$ is the number of frequency bins, $b_{\nu,f} $ represents the Mel filterbank associated with the $\nu$-th channel, and $\text{MVN}(\cdot)$ represents the \gls{MVN} operation. Note that in general the parameters of the \gls{STFT} (window type, length and overlap) used for speech enhancement and recognition differ. Therefore, the speech enhancement front-end usually converts the signals back to the time domain before doing feature extraction for \gls{ASR}.
The features are often normalized with \gls{MVN} to have zero-mean and unit variance using statistics computed either for each utterance or over the whole training data set.

\subsubsection{Acoustic model} 
\label{sec:am}
The \gls{AM} employs phonemes as a basic unit of speech sounds.
In this section, we focus our discussion on \gls{HMM} based \glspl{AM}, where each phoneme is associated with a \gls{HMM} that models the dynamic evolution of speech within that phoneme~\cite{rabiner1989tutorial,Huang:2001}.\footnote{Note that other types of \glspl{AM} such as \gls{CTC}-based \gls{AM} are also becoming widely used~\cite{graves2006connectionist,Li2017}.}
An \gls{HMM} representing the whole word sequence is constructed from several phoneme \gls{HMM}s using a pronunciation dictionary to map each word to a phoneme sequence.
\gls{HMM} based \glspl{AM} make the conditional independence assumption, according to which an observed feature vector only depends on the current state and is independent of neighboring HMM states. This gives the following expression for the likelihood,
\begin{align}
    p\left(\vect{O}|\vect{v}\right) = a_{\sigma_0,\sigma_1} p(\vect{o}_1| \sigma_1) \prod_{t=2}^T p(\vect{o}_t| \sigma_t) a_{\sigma_{t-1},\sigma_{t}},
    \label{eq:hmm}
\end{align}
where $\sigma_t$ is an HMM state at time $t$, $a_{\sigma_t,\sigma_{t+1}}$ is the transition probability between state $\sigma_t$ and $\sigma_{t+1}$, $a_{\sigma_0,\sigma_1}$ is the initial state probability, and $p(\vect{o}_t| \sigma_t)$ is the emission probability. 

In legacy systems, the emission probability was modeled with a \gls{GMM}. 
More recent systems use a \gls{DNN} instead and are called \gls{DNN}-\gls{HMM} hybrid systems.
Let $\vect{g}(\vect{o}_t;\vect\theta)$ be the $\Sigma$-dimensional softmax output vector of a \gls{DNN} \gls{AM} with parameters $\vect\theta$, where $\Sigma$ is the total number of \gls{HMM} states, 
and $g_{\sigma}(\vect{o}_t;\vect\theta)$ is the output associated with \gls{HMM} state $\sigma$.
$g_{\sigma}(\vect{o}_t;\vect\theta)$ can be interpreted as a posterior probability $p(\sigma|\vect{o}_t)$, which can be be converted into a pseudo likelihood using Bayes rule as \cite{6296526}
\begin{align}
    p(\vect{o}_t|\sigma) &\propto \frac{p(\sigma|\vect{o}_t)}{p(\sigma)}, \nonumber \\
                        &= \frac{g_{\sigma}(\vect{o}_t;\vect\theta)}{p(\sigma)} ,
\end{align}
where a prior probability $p(\sigma)$ is derived from the statistics of the training data set. 
    
There has been much research on designing appropriate network architectures for $g_{\sigma}(\vect{o}_t;\vect\theta)$.
The choice for a specific architecture foremostly depends on latency constraints during inference time and the amount of available training data. It is a fast-evolving research field with new results claiming state-of-the-art performance due to often only slight modifications of the architecture being published almost on a weekly basis. Equally important is the choice of training hyperparameters and schemes. Both need extensive tuning for a fair comparison among architectures but this is often not possible due to a limited compute budget. In general, a solid baseline architecture are time delay neural networks (TDNNs)~\cite{21701} or convolutional neural networks in general
(e.g.~\cite{7178920}) possibly followed by some (bi-directional) \gls{LSTM} layers~\cite{7178838}. Variants of this architecture have been employed successfully in the latest CHiME challenges. Recently, architectures with self-attention~\cite{vaswani2017attention}, often referred to as transformers, have shown competitive results on several benchmark tasks~\cite{karita2019comparative, 9054345}.


\subsubsection{Language model}
The \gls{LM} provides the prior probability of a word sequence. There exist  N-gram \glspl{LM} and neural \glspl{LM} such as \gls{RNN} \gls{LM}~\cite{Mikolov}.
The \gls{LM} is trained on a large text corpus, and, unlike the other components of the \gls{ASR} back-end, it is not affected by the acoustic conditions such as noise or reverberation. It can thus be very effective to improve the performance of far-field \gls{ASR} when the language is well constrained such as for read speech tasks~\cite{Yoshioka2015CHiME3,duChime4}. However, for conversational situations, it is more difficult to model the speech content and thus the \gls{LM} appears less effective \cite{DuChime52018}.

\subsubsection{Training procedure}
Building an \gls{ASR} back-end requires training the \gls{AM} with speech training data and the associated transcriptions. 
The goal of the training is finding the \gls{DNN} parameters, $\vect\theta$, which optimize a training criterion as,
\begin{equation}
    \label{eq:optimize_am}
    \vectHat\theta = \argmax_{\vect\theta} \sum_u \mathcal{C}\left(\vect{g}(\vect{O}_u; \vect\theta), \vect{v}_u\right),
\end{equation}
where $\mathcal{C}(\cdot)$ is an objective function, $\vect{O}_u$ and $\vect{v}_u$ are the
sequence of feature vectors and words associated with the $u$th utterance of the training set, respectively.
By abuse of notation, $\vect{g}(\vect{O}_u; \vect\theta)$ refers to the sequence of output vectors
of the \gls{DNN} \gls{AM} with $\vect{O}_u$ at its input.
The model parameters $\vect\theta$ are learned by backpropagation.

Various criteria can be used for training the \gls{AM}.
The most basic criterion is the \gls{CE}, which is given as\cite{Yu2015},
\begin{align}
    \mathcal{C}^{\text{CE}} =&  \sum_u \sum_t \sum_{\sigma=1}^\Sigma p(\sigma) \log(g_{\sigma}(\vect{o}_t;\vect\theta))  \nonumber \\
    = & \sum_u \sum_t \log( g_{\tilde{\sigma}_{u,t}}(\vect{o}_t;\vect\theta)),
    \label{eq:ce}
\end{align}
where $(\tilde{\sigma}_{u,\tau})_{\tau=1}^T$ is the \gls{HMM}-state label sequence associated with the reference word sequence $\vect{v}_u$.
Because we use hard \gls{HMM}-state labels, $p(\sigma) = \delta_{\sigma,\tilde{\sigma}_{u,t}}$ where $\delta_{i,j}$ is the Kronecker Delta.
Thus, the \gls{CE} takes the expression of the log-likelihood in Eq.~(\ref{eq:ce})~\cite{Yu2015}.
Besides, the sign of $\mathcal{C}^{\text{CE}}$ is opposite to the \gls{CE} loss~\cite{Yu2015} because we defined the training as a maximization problem in Eq.~(\ref{eq:optimize_am}).
The HMM-state label sequence can be obtained from the transcription using forced alignment (see section \ref{sec:force_alignment}).
\gls{CE} is a frame level criterion, that does not consider the whole context of the sequence in the loss computation and thus differs from what is performed by the \gls{ASR} decoding in Eq.~(\ref{eq:asr_dec}).

Alternatively, sequence-level criteria have been proposed to better match the \gls{ASR} decoding scheme, such as 
\gls{MMI} or \gls{sMBR}~\cite{vesely2013sequence,Yu2015}. For example \gls{MMI} aims at directly maximizing the posterior probability,
\begin{align}
    \mathcal{C}^{\text{MMI}}&=\sum_u \log(p(\vect{v}_u|\vect{O}_u; \vect\theta)) \nonumber \\
        & =  \sum_u \log\left( \frac{p(\vect{O}_u|\vect{v}_u; \vect\theta) p(\vect{v}_u)}
        {\sum_{\vect{v}'} p(\vect{O}_u|\vect{v}'; \vect\theta) p(\vect{v}')} \right).
\end{align}
The numerator represents the likelihood of the observed speech given the correct word sequence. 
It can be obtained from forced alignment as for \gls{CE}.
The denominator represents the total likelihood of the observed speech features obtained over all possible word sequences (i.e. all word sequences that could be obtained by recognizing the training utterance using the acoustic and language models). 
\gls{MMI} is a sequence discriminative criterion that offers the possibility to make correct word sequences more likely by maximizing the numerator, while making all other word sequences less likely by minimizing the denominator. 
\gls{MMI} and other sequence discriminative criteria have shown to improve performance over \gls{CE}~\cite{vesely2013sequence}. 
However, the summation in the denominator makes \gls{MMI} computationally complex. Recently, an efficient way to implement \gls{MMI} called lattice-free \gls{MMI} has been proposed~\cite{DBLP:conf/interspeech/PoveyPGGMNWK16}. It has become the standard for \gls{ASR} and is also widely used for far-field \gls{ASR}~\cite{kanda2018hitachi, DuChime52018}.

\subsection{Practical considerations for far-field ASR}
\label{sec:practical_asr}
\subsubsection{Multi-condition training data}
\label{sec:mc_train}
To train the \gls{ASR} back-end, we need training speech data and their corresponding word transcriptions.
Training the \gls{ASR} back-end on clean speech would expose it to too little variation of the acoustic conditions,
which may severely affect its performance when exposed to far-field conditions. Indeed, the speech enhancement front-end cannot completely remove acoustic distortions caused by the environment.
Therefore, to make the \gls{ASR} back-end robust, 
it is usually trained with multi-condition data that cover many acoustic conditions, 
including various types and levels of noise, reverberation, etc.

It is very costly to collect and transcribe a large amount of speech data in various real environments. 
Consequently, it is common to resort to simulation to create far-field speech data. 
If we have access to a clean speech training corpus, creating far-field speech signals can be easily done by convolving clean speech signals with acoustic impulse responses and adding noise, as shown in the signal model of Eq.~(\ref{eq:signal_model_time_domain}).
The procedure to create multi-condition data is thus as follows:
\begin{enumerate}
    \item Prepare a set of clean speech training data $\mathcal{S}^{\text{Train}}$, noise samples $\mathcal{N}$ and \glspl{AIR} $\mathcal{A}$,
    \item For each clean training speech signal $s^{\text{Train}} \in \mathcal{S}^{\text{Train}}$, create noisy and reverberant speech as, 
    \begin{align}
    y^{\text{Train}}_m[\ell] &= \left(a_m \ast s^{\text{Train}} \right)[\ell] + n_m[\ell], \nonumber \\
    \text{where } & (a_1, \ldots, a_m, \ldots, a_M) \sim \mathcal{A}, \nonumber\\
    & (n_1, \ldots, n_m, \ldots, n_M) \sim \mathcal{N}.
    \label{eq:mc_training}
\end{align}
\end{enumerate}
It is thus possible to create any amount of distant speech data by varying the \glspl{AIR} and the type and level of noise. 

The \glspl{AIR} can be obtained from databases of \glspl{AIR} measured in real environments\cite{nakamura-etal-2000-acoustical,jeub2009,Szoke2019} or artificially generated using the image method which is a simple model of sound propagation in an enclosure~\cite{Allen1979Image, Habets2010}.
With the image method, it is simple to generate far-field speech data in various rooms with different reverberation time and microphone/speaker positions.
To add background noise, we can use several noise recordings datasets~\cite{musan2015},
and increase the acoustic variations by changing the \gls{SNR}.

The above data augmentation techniques affect only the acoustic environment.
It is also possible to modify the speech signal itself by, e.g., modifying the speed of the audio signal~\cite{KoIS15}.

Although simulation data can be used to create various acoustic conditions, some aspects cannot be well simulated such as, e.g., head movements, the Lombard effect\footnote{The Lombard effect describes the phenomenon that speech is articulated differently when uttered in heavy noise.} etc. 
It is thus usually beneficial to augment the training data with some amount of real recordings.
Moreover, if multi-microphone recordings are available, using each microphone recording as separate training samples can also help increase the acoustic variation~\cite{Yoshioka2015CHiME3}.

Besides these data augmentation techniques that rely on physical models of speech or the room acoustics, there have been a number of approaches proposed recently to artificially augment training data without relying on physical models by e.g. generating adversarial training examples~\cite{Hsu_2017,adversarial_examples,Hu_2018,Qian_2019}.
Moreover, the recently proposed Spectral Augmentation (SpecAug) technique~\cite{SpecAugment} has also been employed to increase the robustness of acoustic models for far-field ASR tasks~\cite{STC_chime6,USTC_chime6}. It can also be combined with physically motivated augmentation yielding significant improvements even for large scale data sets~\cite{SpecAugment}.

The usefulness of multi-condition training data covering various acoustic conditions has been demonstrated in various tasks and challenges~\cite{Delcroix2015Strategies,Yoshioka2015CHiME3,DuChime52018}, and in the development of commercial products~\cite{Li2017}. Note, however, that using simulation to create such data can only increase the acoustic context seen during training but not the actual speech content (spoken words), which can be a limitation if the clean speech training corpus used as a basis for simulation is too small.

In theory, the impact of noise and reverberation on ASR could be largely mitigated by training acoustic models with a very large amount of training data that would cover the acoustic variety seen during application. In such a case, the speech enhancement front-end could eventually become unnecessary. However, in many scenarios, the acoustic conditions can be so diverse that it would require a prohibitively large amount of transcribed training data. This is especially true if multiple microphones are available. There are a few studies that investigate the impact of data augmentation on far-field ASR with and without any front-end, but currently it remains unclear how much data would be sufficient to address a general far-field scenario \cite{Delcroix2015Strategies,Xue_2018,Tang_2018,Manohar_2019,Kanda_2019}. Most studies suggest that an ASR back-end trained with data augmentation techniques alone cannot solve the far-field ASR problem even when using a large amount of training data. For example, for the CHiME 5 challenge, a system trained with 4500 hours of training data
~\cite{kanda2018hitachi} was outperformed by systems using 10 times less data~\cite{DuChime52018,Catalin2019Single}. Moreover, even when using a large amount of data to train the ASR back-end, higher performance is usually achieved when is it combined with a  SE front-end, although for some systems the impact of the front-end may become small~\cite{Li2017,Xue_2018}.

\subsubsection{HMM-state alignments}
\label{sec:force_alignment}
As mentioned in the description of the training procedure, 
training the \gls{AM} requires the HMM-state labels, $(\tilde{\sigma}_{u,\tau})_{\tau=1}^T$. 
Such labels can be obtained by Viterbi forced alignment, which performs Viterbi decoding on the HMM model constructed from the reference word sequence to obtain for each observed speech feature in the utterance the most likely HMM-state, thus performing time-alignment of the input speech and the HMM states~\cite{Huang:2001}.

Viterbi forced alignment can provide accurate alignments when using clean speech. 
However, when the observed speech is corrupted by noise, 
reverberation or other persons' voices, there may be alignment errors. 
For example, when the observed speech also contains speech of an interfering speaker,
that speaker's speech may be mapped to HMM-states of the utterance of the target speaker, which distorts the alignments~\cite{Peddinti+2016}. 
Reverberation and noise also make it harder to correctly identify phoneme boundaries.

These problems can be mitigated if clean speech is available to compute the alignments, leading to more accurate HMM-state labels. 
For example, when using simulated far-field data, we can use the clean speech signals used to generate the training data to perform the alignment. With real recordings, it is sometimes possible to use a headset or lapel microphone synchronized with the distant microphone to obtain a cleaner version of the target speaker's speech that can provide more reliable HMM-state labels.
The training procedure is thus as follows,
\begin{enumerate}
\item For each training utterance,
\begin{enumerate}
    \item construct the utterance HMM from the word labels and the pronunciation dictionary,
    \item compute the HMM-state alignments $(\tilde{\sigma}^{\text{clean}}_{u,\tau})_{\tau=1}^T$ from clean speech and utterance HMM using Viterbi decoding.
 
\end{enumerate}
    \item Train the \gls{AM} using e.g. cross entropy criterion as defined in Eq.~(\ref{eq:ce}),
    \begin{equation}
    \mathcal{C}^{\text{CE}} = \sum_u \sum_t \log( g_{\tilde{\sigma}^{\text{clean}}_{u,t}}(\vect{o}^{\text{noisy}}_t;\vect\theta)),
    \label{ce:am}
\end{equation}
    where $\vect{o}^{\text{noisy}}_t$ is the noisy speech training sample and 
    $\tilde{\sigma}^{\text{clean}}_{u,t}$ is computed from the clean training utterances.
\end{enumerate}

Simply using clean speech for computing the alignments instead of the microphone signals can improve ASR performance by up to 10\% when using \gls{CE} for training~\cite{Peddinti+2016,decroix_IS13}.  Besides,  using heuristics to filter out training utterances that could not be properly aligned can also be important~\cite{Peddinti+2016}.
Lattice-free \gls{MMI} is less sensitive than \gls{CE} to alignment errors. 
Moreover, the state alignment issue may not occur with other types of \gls{AM} such as \gls{CTC}-based \gls{AM} because they do not require HMM-state labels for their training.

\subsubsection{Adaptation of the ASR back-end to the speech enhancement front-end}

The speech enhancement front-end does not fully remove the acoustic interference and may introduce artifacts, 
which causes a mismatch between the input speech signal and the \gls{AM} that is trained using multi-condition training data.
Several approaches can be used to mitigate such a mismatch.
For example, we can process the far-field training data with the enhancement front-end and add this processed speech data to the unprocessed multi-condition training dataset, 
so that the \gls{AM} is exposed to some enhanced speech during training.
Note that in general using only enhanced speech for training the \gls{AM} may reduce the 
acoustic variation observed during training and generate a weaker \gls{AM}~\cite{vincent:hal-01588876,Gonzalez:2015:MTT:2812295.2812300}.

Alternatively, we can use the enhanced speech to adapt an already trained \gls{AM}.
For example, we can obtain an \gls{AM} matched to the test conditions by retraining its parameters with adaptation data  that is similar to the test conditions as
\begin{align}
 \vect\theta^{\text{adapt}} = \argmax_{\vect\theta} \sum_u \mathcal{C}\left(\vect{g}(\vect{O}^{\text{adapt}}_u; \vect\theta), \vectHat{v}_u \right),
\end{align}
where $\vect{O}^{\text{adapt}}_u$ is the sequence of  feature vectors of the $u$-th adaptation utterance, 
and $\widehat{\mathbf{v}}_u$ is the word sequence associated with the adaption utterance.
We can use the training data processed with the speech enhancement front-end as adaptation data, 
in which case $\widehat{\mathbf{v}}_u$ simply corresponds to the transcriptions.
Alternatively, if the adaptation data has no transcriptions (as is the case in unsupervised adaptation), $\widehat{\mathbf{v}}_u$ can be obtained by a first \gls{ASR} decoding pass.

There may be much fewer adaptation data than training data, which makes the process prone to overfitting.
In practice, overfitting can be mitigated by regularization techniques, 
early stopping, or only updating some parameters of the \gls{AM} such as the input layer~\cite{liao2013speaker,yu2013kl}.
Adaptation has been shown to consistently improve the performance of top systems in recent challenges by \SIrange{5}{10}{\percent} relative word error rate reduction~\cite{Yoshioka2015CHiME3,erdogan2016multi}.

\begin{figure*}[t]
\centering
\footnotesize 
\includegraphics[width=1.0\textwidth]{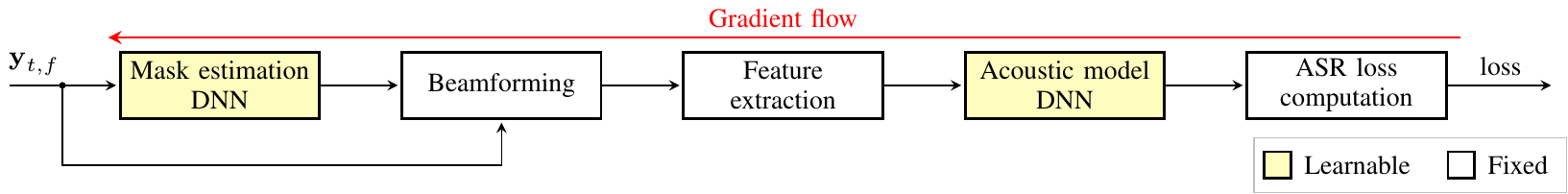}
\vspace{-1em}
\caption{Schematic diagram of the joint training of the speech enhancement front-end and \gls{ASR} back-end.}
\label{fig:asr_joint}
\end{figure*}

\subsubsection{Joint-training}
\label{sec:joint_training}
The above adaptation technique only adjusts the \gls{AM} of the \gls{ASR} back-end to the speech enhancement front-end. However, the speech enhancement front-end is usually optimized for a criterion that is not directly related to \gls{ASR}. Recent works have explored a tighter integration of the speech enhancement front-end and \gls{ASR} back-end, enabling optimization of the front-end for the \gls{ASR} criterion~\cite{Heymann2017Beamnet,xiao2017time,Heymann_ICASSP18}. This is relatively easy to realize because both the front-end and the back-end use neural networks, and therefore it is possible to combine them into a single neural network with learnable and fixed computational nodes. Both systems can then be jointly optimized with backpropagation as,
\begin{equation}
    \vectHat\theta = \argmax_{\vect\theta} 
    \sum_u \mathcal{C}\left(
    \vect{g}\!\left( \text{Feat}\!\left( \text{Enh}(\mathcal{Y}_u; \vect\theta^{\text{enh}})\!\right); \vect\theta^{\text{am}}\!\right),\vect{v}_u\!\right),
    \label{eq:joint_obj}
\end{equation}
where $\text{Enh}(\cdot)$ represents the processing of the enhancement front-end,
$\mathcal{Y}_u$ represents the multi-channel STFT coefficients for a training utterance,
and $\vect\theta = \{\vect\theta^{\text{am}}, \vect\theta^{\text{enh}}\}$ are the model parameters of the \gls{AM} and front-end, respectively.

Fig.~\ref{fig:asr_joint} shows an example of a joint-training scheme that combines 
a beamforming based front-end with the \gls{AM} of the \gls{ASR} back-end~\cite{Heymann2017Beamnet,xiao2017time,Heymann_ICASSP18}.
The mask estimation \gls{DNN} of the front-end and the \gls{DNN} of the \gls{AM} 
are the learnable components of the system.
They are interconnected with fixed computational 
blocks that consist of the beamformer computation (see Sec.~\ref{ssec:Beamforming}) and the feature extraction (see Sec.~\ref{ssec:feat_extract}).
The gradient can flow from the \gls{AM} to the speech enhancement front-end,
which enables optimization of the front-end for \gls{ASR}. 

We have discussed the joint-training scheme with a beamforming front-end,
but joint training has also been used for dereverberation\cite{Heymann2019Joint} and source separation/extraction\cite{Zmolikova2019SpeakerBeam}.
Significant \gls{ASR} gains have been reported on several tasks with joint training schemes.
However, joint-training can sometimes lead to a performance drop because it may
 weaken the \gls{AM}~\cite{Heymann_ICASSP18}. 

One advantage of joint-training is that the whole system can be optimized using only far-field speech and the associated word transcriptions. 
Therefore, it alleviates the need for parallel clean and far-field speech data to train the speech enhancement front-end, which may be an advantage when training or adapting systems with real recordings.

%% file: chapters/summary_outlook.tex
\section{Toward Far-Field End-to-End ASR}
\label{sec:e2e}

This section describes the recent efforts towards end-to-end solutions which allow to optimize all components of the front-end speech enhancement and back-end speech recognizer jointly.
This optimization is performed with respect to our final objective, the Bayes decision rule, as introduced in Eq.~(\ref{eq:asr_bayes_decision}).

\subsection{End-to-End ASR}
End-to-end \gls{ASR} approaches directly model the output distribution $p(\vect{v}|\vect{O})$ over the character, subword, or word sequence $\vect{v} = (v_1, \dots, v_{J})$,  given the speech feature sequence $\vect{O} = (\vect{o}_1, \dots, \vect{o}_{T})$.
This is quite different from conventional approaches to \gls{ASR}~\cite{rabiner1989tutorial} composed of the acoustic model $p(\vect{O}|\vect{v})$ and language model $p(\vect{v})$, as we discussed in \ref{sec:overview_asr}.
End-to-end models subsume all of these components in a single neural network, which greatly simplifies the model building process and also enables joint training of the whole system.
The end-to-end neural speech processing has become a popular alternative to conventional \gls{ASR}, and several approaches have been proposed including \gls{CTC} ~\cite{graves2006connectionist}, attention-based encoder-decoder models~\cite{chorowski2015attention,chan2016listen}, and their variants~\cite{graves2013speech,kim2017joint}.


For example, attention-based methods start from the Bayes decision theory, similar to Section \ref{sec:asr}, but do not use any conditional independence assumption, and simply factorize the posterior probability $p(\vect{v}| \vect{O})$ based on the probabilistic chain rule and the attention mechanism, as follows:
\begin{align}
	p(\vect{v}| \vect{O}) & = \prod _{j} p(v_j|\vect{v}_{1:j-1}, \vect{O})  \nonumber \\
	& = \prod _{j} p(v_j|\vect{v}_{1:j-1}, \vect{c}_j; \vect\theta^{\text{dec}}),
	\label{eq:e2e_asr}
\end{align}
where $\vect{v}_{1:j-1} = (v_1, \dots, v_{j-1})$ is a subsequence of $\vect{v}$ representing the word history before word $v_j$.
$\vect{c}_j$ is called a context vector obtained at each token position $j$, and is extracted from the input speech feature $\vect{O}$ based on the attention mechanism, which we will  explain below.
$p(v_j|\vect{v}_{1:j-1}, \vect{c}_j; \vect\theta^{\text{dec}})$ is computed with a neural network called a decoder network with its set of model parameters $\vect\theta^{\text{dec}}$, which can generate a token sequence $v_j$ given the history $\vect{v}_{1:j-1}$ and a context vector $\vect{c}_j$.
The decoder network is often represented as an \gls{LSTM} model with hidden state vector $\vect{z}_{j}$ for each token position $j$.

To obtain context vector $\vect{c}_j$ in Eq.~\eqref{eq:e2e_asr}, we first focus on an input feature conversion based on an encoder network.
The encoder network takes the original speech feature sequence $\vect{O}$ as input and converts it to high-level hidden vector sequence $\vect{O} ^{\text{enc}} = (\vect{o}_1 ^{\text{enc}}, \dots, \vect{o}_{T'} ^{\text{enc}})$, as follows:\footnote{In general, the length of the encoder output sequence $T'$ is shorter than the length of the original sequence $T$, i.e., $T' < T$ due to subsampling.}
\begin{align}
	\vect{O} ^{\text{enc}} = \text{Enc}(\vect{O}; \vect\theta^{\text{enc}}),
	\label{eq:enc}
\end{align}
where $\vect\theta^{\text{enc}}$ is a set of model parameters in the encoder network.

We often use \gls{BLSTM} or self-attention models as an encoder network.

\begin{figure}[t]
	\centering
    \footnotesize 
		\includegraphics[width=0.5\textwidth]{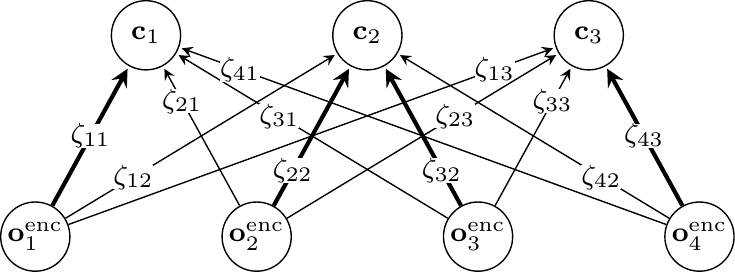}
	\caption{The attention mechanism to compute the alignment between input encoder vector $\vect{o}_t ^{\text{enc}}$ at frame $t$ and output context vector $\vect{c}_j$ at token $j$. $\zeta _{jt}$ denotes the attention weight. The bold lines correspond to the higher attention weights and the attention mechanism obtains the soft alignment between these input and output vectors.}
	\label{fig:att}
\end{figure}

Given $\vect{O} ^{\text{enc}}$, an attention mechanism produces context vector $\vect{c}_j$ for each token $v_j$ as follows~\cite{chorowski2015attention}:
\begin{align}
	\vect{c}_j= \text{Att} \left(\vect{O} ^{\text{enc}}, \vect{z}_{j-1}; \vect\theta^{\text{att}}\right),
\end{align}
where $\vect{z}_{j-1}$ is a hidden state vector introduced in the decoder network.
$\text{Att} (\cdot)$ is an attention network with a set of model parameters $\vect\theta^{\text{att}}$, which first computes the attention weight $\zeta _{jt} \in [0, 1]$ given the encoder output vector $\vect{o}_t ^{\text{enc}}$ and the decoder hidden vector $\vect{z}_{j-1}$ obtained in the previous output time step \cite{chorowski2015attention}, as follows:
\begin{equation}
\zeta _{jt} = f^{\text{att}}(\vect{o}_t ^{\text{enc}}, \vect{z}_{j-1}),
\label{eq:attention_weight}
\end{equation}
where $f^{\text{att}}(\cdot)$ is a function to produce the attention weight, which can be a dot product or neural network-based operations with trainable parameters.
$\zeta _{jt}$ satisfies the sum-to-one condition across the input frames, i.e., $\sum _{t=1} ^{T'} \zeta _{jt} = 1$.
Given the attention weight $\zeta _{jt}$ in Eq.~\eqref{eq:attention_weight}, the context vector $\vect{c}_j$ is obtained as a weighted summation of encoder output sequence $\vect{O} ^{\text{enc}}$, i.e., 
\begin{align}
	\vect{c}_j = \sum _{t=1} ^{T'} \zeta _{jt} \vect{o}_t ^{\text{enc}}.
	\label{eq:context_vec}
\end{align}
Note that Eq.~\eqref{eq:context_vec} can perform a conversion between two values with different time scales (input time $t$ and output time $j$) through the soft alignment based on the weighted summation.
For example, Fig.~\ref{fig:att} depicts the attention mechanism based on Eq.~\eqref{eq:context_vec}.
The bold lines correspond to the higher attention weights and the attention mechanism obtains the soft alignment between these input and output vectors.
This is different from the alignment process in conventional \gls{ASR}, which is based on \glspl{HMM},  as discussed in Section \ref{sec:am}.

\begin{figure}[t]
	\centering
	\footnotesize
		\includegraphics[width=0.5\textwidth]{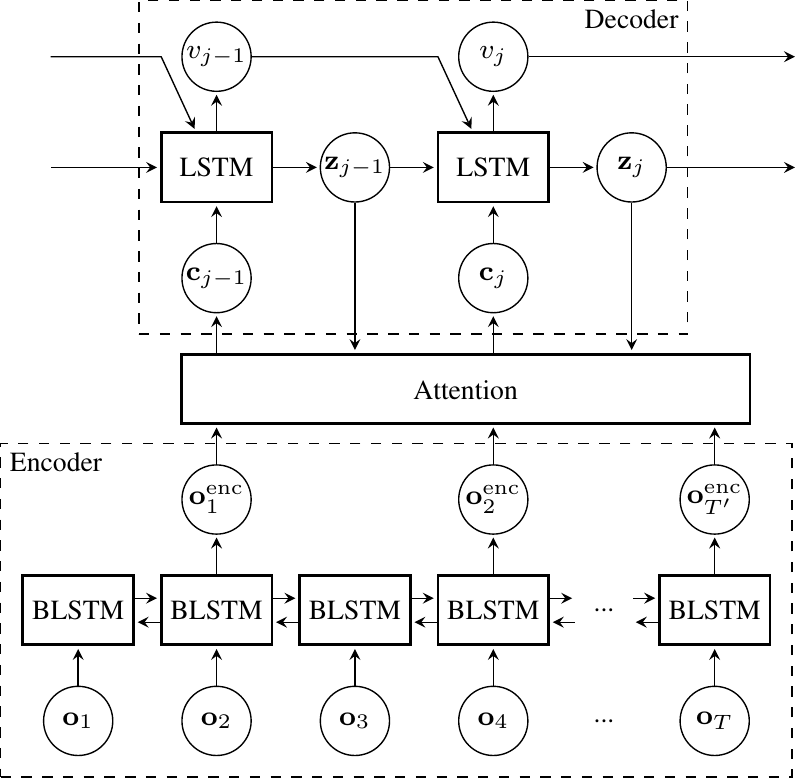}
	\vspace{-1em}
	\caption{Attention-based encoder decoder network. The attention is controlled by the decoder LSTM state.}
	\label{fig:encdec}
\end{figure}

The forward computation of the attention-based end-to-end ASR is processed as follows:
\begin{enumerate}
	\item Encoder processing: $\vect{O} ^{\text{enc}} = \text{Enc}(\vect{O}; \vect\theta^{\text{enc}})$
	\item For each $j$
	\begin{enumerate}
		\item compute $\vect{c}_j = \text{Att} (\vect{O} ^{\text{enc}}, \vect{z}_{j-1}; \vect\theta^{\text{att}})$
		\item obtain $p(v_j|\vect{v}_{1:j-1}, \vect{c}_j; \vect\theta^{\text{dec}})$.
	\end{enumerate}
\end{enumerate}
	Figure \ref{fig:encdec} shows an entire encoder-decoder neural network with an attention mechanism.
Note that the history subsequence $\vect{v}_{1:j-1}$ can be obtained from the reference transcription during training and from prediction results during decoding.
All of these steps are differentiable, and we can estimate the model parameters $\vect\theta = \{ \vect\theta^{\text{enc}},\vect \theta^{\text{att}}, \vect\theta^{\text{dec}} \}$ by maximizing the following log-likelihood, similar to Eq.~\eqref{eq:optimize_am},
\begin{equation}
\vectHat \theta = \argmax _{\vect \theta} \sum_u \log( p(\vect{v} _u| \vect{O} _u; \vect \theta) ).
\label{ce:e2e}
\end{equation}
Thus, the attention-based encoder decoder network represents an entire ASR process with a single neural network, and can be trained in an end-to-end manner unlike the conventional HMM-based ASR system.
Alternatively, a transformer architecture, which is originally proposed in neural machine translation \cite{vaswani2017attention} to replace RNNs with self-attention networks, has been used as a variant of attention based methods for ASR \cite{karita2019comparative}.

\subsection{Multi-Channel End-to-End ASR}
The straightforward extension of this methodology to far-field speech recognition is to combine all speech enhancement modules and \gls{ASR} with a single neural network to enable joint optimization  \cite{Ochiai2017MultichannelES,ochiai2017unified}.
This method can be regarded as an extension of the joint-training methods~\cite{Heymann2017Beamnet,xiao2017time,Heymann_ICASSP18} of multi-channel speech enhancement and acoustic modeling as discussed in Section \ref{sec:joint_training}.

By following Eq.~\eqref{eq:e2e_asr}, multi-channel end-to-end \gls{ASR} directly models the posterior distribution $p(\vect{v}| \mathcal{Y})$, given the sequence of multi-channel (STFT) signals $\mathcal{Y} = ([\vect{y}_{1,1}\T, \dots, \vect{y}_{1,F}\T], \dots, [\vect{y}_{t,1}\T, \dots, \vect{y}_{t,F}\T], \dots)$:
\begin{align}
	p(\vect{v}| \mathcal{Y}) & = \prod _{j} p(v_j|\vect{v}_{1:j-1}, \mathcal{Y}) = \prod _{j} p(v_j|\vect{v}_{1:j-1}, \hat{\vect{O}}),
	\label{eq:me2e_asr}
\end{align}
where 
\begin{equation}
\hat{\vect{O}} = \text{Feat}(\text{Enh}(\mathcal{Y}; \theta ^{\text{enh}})).
\end{equation}
$\text{Enh}(\cdot)$ corresponds to the multi-channel enhancement with a set of parameters $\theta ^{\text{enh}}$ and $\text{Feat }(\cdot)$ denotes  the  standard speech feature extraction to produce an enhanced speech feature sequence, $\hat{\vect{O}}$.
Both are introduced in the joint training of speech enhancement and recognition in Eq.~\eqref{eq:joint_obj}.

As an instance of the multi-channel enhancement function, \cite{Ochiai2017MultichannelES} uses \gls{BLSTM} mask-based beamforming \cite{Souden2010MVDR,Erdogan2016MVDR}, as described in Section \ref{sec:se}.
This model is trained with an end-to-end \gls{ASR} objective (cross entropy given the reference transcriptions $\vect{v} _u$ for utterance $u$) as follows:
\begin{equation}
\vectHat\theta = \argmax _{\vect\theta} \sum_u \log( p(\vect{v} _u| \mathcal{Y}_u; \vect\theta) ),
\label{ce:me2e}
\end{equation}
where the model parameters $\vect\theta$ consist of the parameters of the enhancement, encoder, attention and decoder networks as,
\begin{equation}
\vect\theta = \{\vect\theta^{\text{enh}}, \vect\theta^{\text{enc}}, \vect\theta^{\text{att}}, \vect\theta^{\text{dec}}\}.
\end{equation}
Compared with the standard end-to-end ASR training in Eq.~\eqref{ce:e2e}, the multi-channel extension can jointly estimate both ASR model parameters and the enhancement parameters $\vect\theta^{\text{enh}}$ in an end-to-end manner.
Note that this model can be trained  without requiring any  parallel data (pairs of clean and noisy speech data), as described in Section \ref{sec:se} or any other intermediate HMM state/phoneme alignments compared with standard acoustic model training described in Eq.~\eqref{ce:am}.
End-to-end joint training thus allows training the enhancement parameters with real far-field data, for which clean reference signals are usually not available. The only requirement is the availability of the transcription of the far-field data, which is always required for ASR training based on supervised learning.

There are several variants and extensions of multi-channel end-to-end \gls{ASR} including
\begin{itemize}
	\item Attention-based channel/array selection \cite{Braun2018,wang2019stream}
	\item Incorporation of a dereverberation component \cite{subramanian2019speech}\footnote{This is implemented based on DNN-WPE \cite{Kinoshita2017Neural} developed in \url{https://github.com/nttcslab-sp/dnn_wpe}.}
	\item Extension to multispeaker ASR \cite{chang2019mimo}
	\item Extension to target speech extraction \cite{Delcroix2019,Denisov2019}.
\end{itemize}
Although end-to-end approaches are promising, they do not reach the performance of current state-of-the-art far-field \gls{ASR} systems.
The main reason is that these solutions tend to require larger amounts of training data, which, in the case of multi-channel far-field recordings, may not always be available.
However, there has been a lot of progress in end-to-end \gls{ASR} including extensive investigations of training methods and architectures \cite{chiu2018state,zeyer2018improved}, robust training based on data augmentation \cite{SpecAugment}, and new architectures based on the transformer model \cite{dong2018speech,karita2019comparative}.

\section{Summary and remaining challenges}
\label{sec:summary}


\subsection{Summary}
This paper emphasizes that multi-channel speech enhancement is an essential component for far-field \gls{ASR}, and provides a comprehensive description of state-of-the-art enhancement techniques in Section \ref{sec:se}. The combination of powerful signal processing with deep learning significantly boosted the performance, compared to earlier signal processing-only solutions.
This  trend of solving a problem with signal processing supported by a neural network is not so often seen in other applications of deep learning. Consider, for example,  computer vision, where an entire signal processing pipeline has been replaced with a very deep network.
The main reason of this unique approach in speech enhancement is that well-established physical models exist, which can be viewed as regularizers when devising a deep learning solution.
We can thus minimize the size  of the neural networks  and can make multi-channel speech enhancement work robustly with a relatively small amount of training data.

The main focus of the description of the back-end \gls{ASR} system in Section \ref{sec:asr} is on how to make use of deep learning techniques in \gls{ASR} acoustic models for the far-field \gls{ASR} scenario.
This includes techniques like data augmentation, refinement of supervisions, and adaptation.
Note that, unlike speech enhancement, \gls{ASR} is not based on a solid physical model describing human speech perception and recognition, while at the same time single-channel data in the order of thousands 
of hours have become available also in an academic research setting. This is why pure deep learning based solutions excel at \gls{ASR}.
Overall, the fusion of neural network-supported signal processing in the front-end and the massive use of deep learning in the back-end has made far-field \gls{ASR} so reliable that it entered the consumer market with products like digital home assistants.

This paper also introduced the new research paradigm of jointly modeling front-end speech enhancement and back-end ASR acoustic models in Section \ref{sec:joint_training}.
Section \ref{sec:e2e} further extended this joint training scheme towards the emergent end-to-end \gls{ASR} framework.
The underlying idea of both approaches is to strictly follow the above established far-field ASR pipeline, but to represent it with a single neural network so that we can perform back propagation to train both speech enhancement and recognition jointly.
Currently, joint training and end-to-end approaches have not yet become as mainstream as the pipeline approach due to their complex network architecture and the lack of a sufficient amount of multi-channel far-field training data.
However, we believe that these approaches have a lot of potential to provide further breakthroughs in far-field \gls{ASR}, and we put emphasis on describing them as our most important on-going and future research directions.

\subsection{Remaining challenges}
The following subsections list remaining challenges in far-field \gls{ASR}.
For some of those, including voice activity detection and speaker diarization, there exist well-established solutions in a clean speech environment, while they remain to be challenging in far-field \gls{ASR} conditions.
\begin{itemize}
    \item \textbf{\gls{VAD}} (also called \gls{SAD}) is an essential technique to segment continuous audio signals in on-line streaming \gls {ASR}, or long audio recordings in off-line \gls{ASR} into utterances of  manageable length (up to, say, a dozen seconds).
    Traditionally, energy-based \gls{VAD} or likelihood based solutions \cite{sohn1999statistical} have been used.
    However, these methods face significant degradation in low SNR conditions. 
    Learning based methods, especially \gls{RNN}-based ones, combined with data augmentation techniques as described in Section~\ref{sec:mc_train} have become popular \cite{hughes2013recurrent,eyben2013real}, because they can detect speech activity regions by non-linear feature mapping even in the presence of low SNR.
    There are also several challenge activities including OpenSAD\footnote{https://www.nist.gov/itl/iad/mig/nist-open-speech-activity-detection-evaluation}. Further note that \gls{VAD}-related challenge activities are also included in the speaker diarization challenge, see next item. 
    \item \textbf{Speaker diarization}:
    Speaker diarization can be regarded as an extension of \gls{VAD} to multi-speaker recordings, which provides speaker identities or speaker cluster assignments for each utterance from unsegmented audio signals, i.e., it provides information about ``who speaks when'' \cite{anguera2012speaker}.
    Recently, speaker diarization has received increased attention because the focus of the \gls{ASR} research community is shifting more and more towards recognition of multi-speaker recordings such as conversations or meetings, The interest in diarization is boosted by several challenge activities including DIHARD\footnote{https://coml.lscp.ens.fr/dihard/2018/index.html} and CHiME-6.\footnote{https://chimechallenge.github.io/chime6/}
    There are two main technologies depending on whether single-channel or multi-channel data is available.
    When we have multi-channel audio signals, source speaker locations can be estimated based on beamforming, and this can in turn be exploited to provide diarization information \cite{hori2011low,anguera2012speaker}.
    In the single-channel case, people use speaker embeddings, such as the i-vector \cite{dehak2011front} or x-vector \cite{snyder2018x}, to map a speech utterance into a fixed dimensional vector, and then perform clustering on those obtained embedding vectors (e.g., agglomerative hierarchical clustering (AHC) \cite{wooters2007icsi,sell2014speaker}).
    VAD is used as an initial module in the speaker diarization pipeline to segment the recordings into manageable utterances.
    However, most single-channel techniques cannot explicitly handle regions of speech, where  more than one speaker is active. But such overlap regions are common in real conversations \cite{Barker2018CHiME5}.
    A combination of speech separation, speaker counting, and diarization based on neural networks \cite{von2019all} and permutation-free neural diarization based on multiple label classification \cite{fujita2019end} would be a promising direction to tackle regions of overlapped speech.
    \item \textbf{On-line processing}:
    Another challenge of far-field speech processing is on-line, low-latency processing which is mandatory when used in a spoken language interface. It  also has some benefits in dynamical environments, when, e.g., moving sources have to be tracked, see the next bullet in this list.
    Speech enhancement techniques often require to estimate signal statistics across frames, such as the spatial covariance matrix $\vect\Phi$ for beamforming used in Eq.~\eqref{eq:cov_estimate}  and the \gls{MCLP} coefficient matrix $\vect C $ for dereverberation, Eqs.~\eqref{eq:wpestep1} and \eqref{eq:wpestep1p}.
    If low latency is required, this statistics computation must be performed in an on-line manner, often based on recursive update equations, e.g., by a linear interpolation between previously estimated statistics and the current observations.
    Online processing for mask-based beamforming is discussed in \cite{higuchi2017online,boeddeker2018exploring}.
    \cite{Li2017} gives an overview of the development of the Google Home device and describes several online techniques \cite{caroselli2017adaptive}, especially for dereverberation.
    \cite{Kinoshita2017Neural, HeymannIWAENC2018}  realizes online \gls{WPE} dereverberation with the help of \gls{DNN}-based time varying variance estimation.
    \item \textbf{Dynamic environments: moving sensors and sources}: 
    Acoustic environments are changing over time  due to  nonstationary noise, moving sources or moving sensors.
    For example, the participants recorded in the CHiME-5 data set are moving from room to room \cite{Barker2018CHiME5}, and front-end processing has to track such moving sources accordingly.
    In addition, with wearable microphones and in moving robot scenarios \cite{nakadai2010design}, we should also take moving microphones into consideration.
    In these situations, on-line processing as discussed above is necessary to deal with adaptive estimation of enhancement filters (beamforming, dereverberation).
    Recently, there has been a challenge activity, the LOCATA Challenge\cite{evers2019locata}, on locating and tracking moving sources.
    Although this challenge mainly focuses on acoustic source localization and not on speech enhancement and recognition, their designs of dynamic environments and the defined evaluation metrics for source tracking would be a good reference for tackling far-field speech recognition in dynamic environments.
    \item \textbf{More natural conversations and spontaneous speech}.
    Our conversations are often spontaneous, and speech characteristics are quite variable and complex.
    For example, in the dinner party scenarios of CHiME~5 \cite{Barker2018CHiME5} and the Santa Barbara corpus \cite{du2000santa}, we often observe very different speaking durations, volumes,  and speaking styles during the conversation.
    Such variable speech characteristics make the statistical properties of source signals complex and renders estimation of speech statistics harder.
    In addition, the spoken contents are grammatically less regular due to  filler words, mispronunciation, stammering, etc., which makes \gls{ASR} quite challenging from both acoustic and language model perspectives.
    Finally, such conversations are challenging in terms of the data collection and annotation perspectives, because the preparation of  precise transcriptions is difficult. 
    \item \textbf{Improving signal extraction with semantic and syntactic context information}: A human's ability to track a conversation in acoustically adverse conditions (e.g., in a cocktail party) can in part be attributed to the use of context information about the discussion topic, our ``world knowledge'' and syntactic constraints we are aware of. Only few works exist towards utilizing high-level guidance for the low-level signal extraction tasks. In \cite{Takahashi2020} the speech separation is improved by feeding back deep features extracted from an end-to-end ASR system to cover the long-term dependence of phonetic aspects, while sound separation is improved in \cite{Tzinis2020} by utilizing sound classification results. Exploring ways to  support front-end processing with back-end knowledge appears to be a promising way to improve overall system performance.
    \item \textbf{Distributed microphone setup}:
    In many application scenarios, including smart homes \cite{Barker2018CHiME5,cristoforetti2014dirha}, wearable computing, and human-to-robot communication \cite{nakadai2010design}, distributed microphones can be of an advantage, compared to a single spatially concentrated microphone array.
    However, the challenge of distributed microphones is that their spatial location is often a priori unknown and may change over time. Furthermore, the  microphone characteristics can be different, e.g., if both mobile phones and desktop microphones are part of the network. Finally, and most importantly, the sampling rates of the microphones are not synchronized in general. 
    These properties often break important assumptions made in conventional front-end processing, and thus standard beamforming and dereverberation techniques cannot be straightforwardly applied.
    However, there exist several studies to tackle the distributed microphone setup including \cite{Wehr2004, ono2009blind, Cherkassky2017, araki2018meeting, afifi2018marvelo} by solving the synchronization problem to make beamforming work in this setup. There are also many works on distributed beamforming, e.g., \cite{Bertrand2011,Heusdens2012, MarkovichGolan2015}, to avoid collecting all signals at a central processing node.
    Active microphone (subset) selection instead of fusing the signals of multiple microphones is another simple yet effective approach \cite{bertrand2011applications,kumatani2011channel}.
    Also, late fusion techniques (acoustic model fusion \cite{duChime4} or hypothesis fusion \cite{kanda2018hitachi} in \gls{ASR}) instead of signal-level fusion can be a viable alternative  thanks to the relative insensitivity of acoustic models to synchronization errors.
    \item \textbf{Multimodality}: 
    A final challenges in far-field speech recognition is the use of multimodal information including videos, accelerometer, biosignals and so on.
    Such information would be complementary to audio signals, be robust against acoustic noise, and thus the fusion can bring benefits.
    In particular, audio-visual speech recognition gains a lot of attention as the video channel can provide the speaker location information for steering an acoustic beamformer. Furthermore, visual features can complement the audio features for noise robust speech recognition \cite{noda2015audio}.
    However, the visual or other multimodal data have their own distortions (e.g., brightness and frame-out issues of the image), and synchronization across different modalities is also another challenge.
\end{itemize}

%% file: chapters/probe_further.tex
\section{To probe further}
\label{sec:probe_further}





Open-source implementations are available for most of the described techniques and provide a good starting point for a more hands-on experience.

A Python implementation of the \gls{WPE} algorithm described in Sec.~\ref{subsec:dereverb} based on Numpy and Tensorflow is provided by NaraWPE~\cite{DrudeITG2018}.\footnote{\url{https://github.com/fgnt/nara_wpe}}
The Matlab implementation originally used in~\cite{Nakatani2010WPE,Yoshioka2012GeneralWPE} is available as pcode\footnote{\url{http://www.kecl.ntt.co.jp/icl/signal/wpe/index.html}}.
For beamforming as described in Sec.~\ref{ssec:Beamforming}, two different Python implementations are provided.
NN-GEV\footnote{\url{https://github.com/fgnt/nn-gev}} focuses on neural network-based mask estimation and subsequent beamforming while PB-BSS\footnote{\url{https://github.com/fgnt/pb_bss}} focuses on spatial clustering-based \gls{SPP} estimation.
Other useful toolkits implementing dereverberation and beamforming techniques include the BTK toolkit\footnote{\url{https://distantspeechrecognition.sourceforge.io/}} and Pyroomacoustics~\cite{8461310}.\footnote{\url{https://github.com/LCAV/pyroomacoustics}}
The latter one also allows to simulate acoustic scenarios to generate data.

\begin{figure}[t]
	\centering
	\hspace{-1.5em}
	\includegraphics[width=0.5\textwidth]{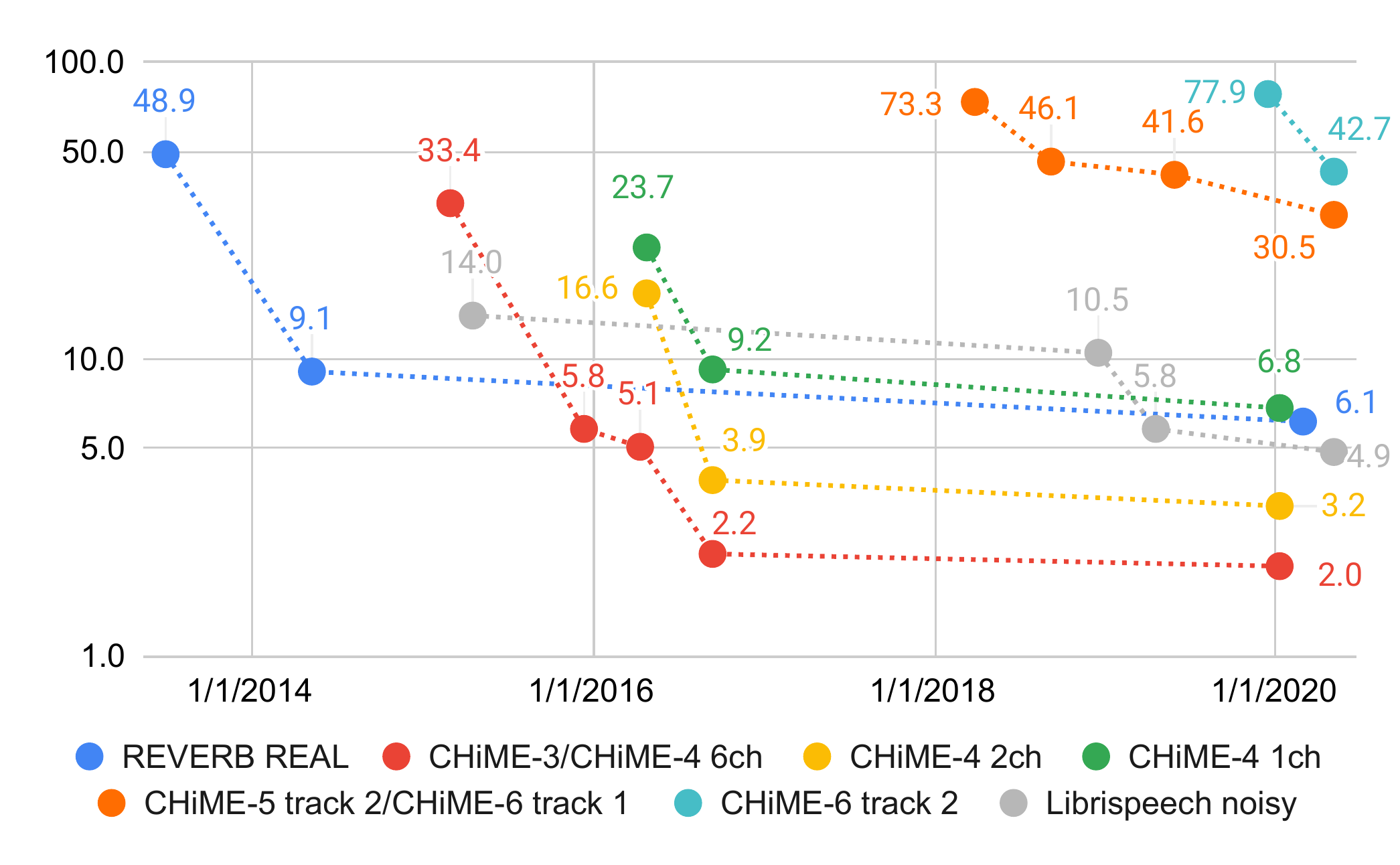}
	\caption{The WER transitions of far-field ASR systems based on the REVERB and CHiME-3/4/5/6 challenge results from their baseline systems, challenge best systems, and the follow-up studies.}
	\label{fig:asr_wers}
\end{figure}

An overview of selected implementations is given in Table~\ref{watanabe2_asr} while databases are listed in Table~\ref{chap17-tab-data-summary}.
To visualize the comprehension of the effect of far-field speech and prospective improvements for several acoustic scenarios, Fig.~\ref{fig:asr_wers}  depicts the ASR performance transition of the CHiME and REVERB challenges from the challenge baseline at the challenge release period, the challenge best system, and the challenge follow up studies. 
By referring to Fig.~\ref{fig:asr_wers} and corresponding acoustic scenarios in Table~\ref{chap17-tab-data-summary}., we can monitor the prospective improvement of various far-field ASR problems.

Note that many of these \gls{ASR} results can be reproduced by using publicly available toolkits.
For a head-start on \gls{ASR} tasks, the Kaldi toolkit~\cite{povey2011kaldi} provides several recipes for the listed databases which include some of the tools discussed above. The CHiME-6 recipe\footnote{\url{https://github.com/kaldi-asr/kaldi/tree/master/egs/chime6}} for example uses NaraWPE and PB-GSS\footnote{\url{https://github.com/fgnt/pb_chime5}} while the CHiME-3/4 recipe\footnote{\url{https://github.com/kaldi-asr/kaldi/tree/master/egs/chime4}} includes BeamformIt and NN-GEV.
ESPnet~\cite{watanabe2018espnet} also provides multichannel end-to-end ASR for the REVERB\footnote{\url{https://github.com/espnet/espnet/tree/master/egs/reverb/asr1_multich}} and CHiME-4\footnote{\url{https://github.com/espnet/espnet/tree/master/egs/chime4/asr1_multich}} data with the help of DNN-WPE.





\begin{table*}[tb]
\centering
\caption{Recent noise robust speech recognition tasks.
}
\label{chap17-tab-data-summary}
\rarray{1.3}
\input{tables/database.tex}
\end{table*}

\begin{table*}
\centering
\caption{Toolkits}
\label{watanabe2_asr}
\rarray{1.3}
\input{tables/tool.tex}
\end{table*}

%% file: tables/database.tex
\begin{tabular}{lccclccc}
\toprule
Task &
\watanabeStyleC{\!\!Vocabulary\!\! \\ size} &
\watanabeStyleC{Amount of\\ training data} &
\watanabeStyleC{Realism} &
Type of distortions &
\watanabeStyleC{Number \\ of mics} & \watanabeStyleC{Mic-speaker\\ distance} &
\watanabeStyleC{Ground\\ truth} \\
\midrule
ASpIRE~\cite{harper2015automatic} &
100k &
N/A &
Real &
Reverberation &
8/1 &
N/A &
N/A \\
\midrule
AMI~\cite{carletta2005ami} &
11k &
\watanabeStyleC{$\sim$ 107k\,utt. \\ ($\sim$ 75 h)} &
Real &
\watanabeStyleL{Multi-speaker conversations\\ Reverberation and noise} &
8 &
N/A &
Headset \\
\midrule
CHiME-3/4~\cite{chime3CS, Vincent2016Chime4} &
5k &
\watanabeStyleC{8738\,utt. \\ ($\sim$ \SI{18}{h})} &
Simu+Real &
\watanabeStyleL{Nonstationary noise in \\ four public environments} &
6/2/1 &
\SI{0.5}{m} &
\watanabeStyleC{Clean/ \\ close talk} \\
\midrule
CHiME-5/6~\cite{Barker2018CHiME5} &
100k &
\watanabeStyleC{$\sim$ 80k\,utt. \\ ($\sim$ \SI{40}{h})} &
Real &
\watanabeStyleL{Nonstationary noise, \\ multi-speaker conversations, \\ reverberation} &
32 &
\watanabeStyleC{\SI{0.5}{m} \\ to \SI{2}{m}} &
\watanabeStyleC{Binaural \\ headset} \\
\midrule
REVERB~\cite{REVERB2016} &
5k &
\watanabeStyleC{7861\,utt. \\ ($\sim$ \SI{15}{h})} & Simu+Real &
\watanabeStyleL{Reverberation in different \\ living rooms} &
8/2/1 &
\watanabeStyleC{\SI{0.5}{m} \\ to \SI{2}{m}} &
\watanabeStyleC{Clean/ \\ headset} \\
\bottomrule
\end{tabular}

%% file: tables/tool.tex
\newcommand{\zeroWidthFootnotemark}[1]{%
  \makebox[0pt][l]{\footnotemark}%
  \footnotetext{#1}%
}
\begin{tabular}{lclllccc}
\toprule
Name &
Affiliation &
Function &
Interface &
Back-end &
License &
Ref. \\
\midrule
NaraWPE & UPB\zeroWidthFootnotemark{University Paderborn} & WPE dereverberation & Python & NumPy, TensorFlow\!\!\!\! & MIT & \cite{DrudeITG2018}\zeroWidthFootnotemark{\url{https://github.com/fgnt/nara_wpe}} \\
WPE & NTT\zeroWidthFootnotemark{Nippon Telephone \& Telegraph, Japan} & WPE dereverberation & Matlab & Matlab & Custom & \cite{Nakatani2010WPE,Yoshioka2012GeneralWPE}\zeroWidthFootnotemark{\url{http://www.kecl.ntt.co.jp/icl/signal/wpe/index.html}} \\
DNN-WPE & NTT & DNN-based WPE dereverberation\!\! & Python & NumPy, PyTorch & Custom & \cite{Kinoshita2017Neural}\zeroWidthFootnotemark{\url{https://github.com/nttcslab-sp/dnn_wpe}} \\
NN-GEV & UPB & Neural mask-based beamforming & Python & NumPy, Chainer & Custom & \cite{hey_asru_2015}\zeroWidthFootnotemark{\url{https://github.com/fgnt/nn-gev}} \\
PB-BSS & UPB & Spatial clustering, beamforming & Python & NumPy & MIT & \cite{Drude2017Integration}\zeroWidthFootnotemark{\url{https://github.com/fgnt/pb_bss}}\\
BTK & CMU\zeroWidthFootnotemark{Carnegie-Mellon University, USA} & Beamforming, dereverberation & Python, C++ & C++ & MIT & N/A\zeroWidthFootnotemark{\url{https://distantspeechrecognition.sourceforge.io/}} \\
PyRoomAcoustics\!\!\!\!\!\!\!\! & EPFL\zeroWidthFootnotemark{{\'E}cole polytechnique f{\'e}d{\'e}rale de Lausanne, Switzerland} & Beamforming, RIR generation & Python & NumPy/C & MIT & \cite{8461310}\zeroWidthFootnotemark{\url{https://github.com/LCAV/pyroomacoustics}} \\
BeamformIt & ICSI & Delay-and-sum beamforming & CLI, C++ & C/C++ & N/A & \cite{beamformit}\zeroWidthFootnotemark{\url{https://github.com/xanguera/BeamformIt}} \\
HARK & HRI & Source localization, separation & CLI, Python & C++ & Custom & hark.jp \\
\bottomrule
\end{tabular}

%% file: bare_jrnl.bbl
\begin{thebibliography}{100}
\providecommand{\url}[1]{#1}
\csname url@samestyle\endcsname
\providecommand{\newblock}{\relax}
\providecommand{\bibinfo}[2]{#2}
\providecommand{\BIBentrySTDinterwordspacing}{\spaceskip=0pt\relax}
\providecommand{\BIBentryALTinterwordstretchfactor}{4}
\providecommand{\BIBentryALTinterwordspacing}{\spaceskip=\fontdimen2\font plus
\BIBentryALTinterwordstretchfactor\fontdimen3\font minus
  \fontdimen4\font\relax}
\providecommand{\BIBforeignlanguage}[2]{{%
\expandafter\ifx\csname l@#1\endcsname\relax
\typeout{** WARNING: IEEEtran.bst: No hyphenation pattern has been}%
\typeout{** loaded for the language `#1'. Using the pattern for}%
\typeout{** the default language instead.}%
\else
\language=\csname l@#1\endcsname
\fi
#2}}
\providecommand{\BIBdecl}{\relax}
\BIBdecl

\bibitem{REVERB2016}
K.~Kinoshita, M.~Delcroix, S.~Gannot, E.~Habets, R.~Haeb-Umbach, W.~Kellermann,
  V.~Leutnant, R.~Maas, T.~Nakatani, B.~Raj, A.~Sehr, and T.~Yoshioka, ``A
  summary of the {REVERB} challenge: state-of-the-art and remaining challenges
  in reverberant speech processing research,'' \emph{EURASIP Journal on
  Advances in Signal Processing}, 2016.

\bibitem{chime3CS}
J.~Barker, R.~Marxer, E.~Vincent, and S.~Watanabe, ``The third "{CH}i{ME}"
  speech separation and recognition challenge: {A}nalysis and outcomes,''
  \emph{Computer Speech and Language}, vol.~46, pp. 605--626, Nov. 2017.

\bibitem{Vincent2016Chime4}
E.~Vincent, S.~Watanabe, A.~A. Nugraha, J.~Barker, and R.~Marxer, ``An analysis
  of environment, microphone and data simulation mismatches in robust speech
  recognition,'' \emph{Computer Speech and Language}, 2016.

\bibitem{Barker2018CHiME5}
J.~Barker, S.~Watanabe, E.~Vincent, and J.~Trmal, ``The fifth {'CHiME'} speech
  separation and recognition challenge: Dataset, task and baselines,'' in
  \emph{Proc. of Annual Conference of the International Speech Communication
  Association \mbox{(Interspeech)}}, 2018.

\bibitem{Watanabe2020}
S.~Watanabe, M.~Mandel, J.~Barker, E.~Vincent, A.~Arora, X.~Chang,
  S.~Khudanpur, V.~Manohar, D.~Povey, D.~Raj, D.~Snyder, A.~S. Subramanian,
  J.~Trmal, B.~B. Yair, C.~Boeddeker, Z.~Ni, Y.~Fujita, S.~Horiguchi, N.~Kanda,
  T.~Yoshioka, and N.~Ryant, ``{CHiME-6} challenge: Tackling multispeaker
  speech recognition for unsegmented recordings,'' \emph{CoRR}, 2020.

\bibitem{harper2015automatic}
M.~{Harper}, ``The automatic speech recogition in reverberant environments
  ({ASpIRE}) challenge,'' in \emph{Proc. of \mbox{IEEE} Workshop on Automatic
  Speech Recognition and Understanding \mbox{(ASRU)}}, Dec 2015, pp. 547--554.

\bibitem{Delcroix2015Strategies}
M.~Delcroix, T.~Yoshioka, A.~Ogawa, Y.~Kubo, M.~Fujimoto, N.~Ito, K.~Kinoshita,
  M.~Espi, S.~Araki, T.~Hori \emph{et~al.}, ``Strategies for distant speech
  recognition in reverberant environments,'' \emph{EURASIP Journal on Advances
  in Signal Processing}, vol. 2015, no.~1, p.~60, 2015.

\bibitem{McDonough2014REVERB}
X.~Feng, K.~Kumatani, and J.~McDonough, ``The {CMU-MIT} {REVERB} challenge 2014
  system: description and results,'' in \emph{REVERB Challenge Workshop}, 2014.

\bibitem{Weninger2014REVERB}
F.~Weninger, S.~Watanabe, J.~L. Roux, J.~R. Hershey, Y.~Tachioka, and
  G.~Rigoll, ``The {MERL/MELCO/TUM} system for the {REVERB} challenge using
  deep recurrent neural network feature enhancement,'' in \emph{REVERB
  Challenge Workshop}, 2014.

\bibitem{Wang2020}
Z.-Q. Wang and D.~Wang, ``Deep learning based target cancellation for speech
  dereverberation,'' \emph{{IEEE} Transactions on Audio, Speech, and Language
  Processing}, vol.~28, pp. 941--950, 2020.

\bibitem{spcom2017_CHiME3}
J.Barker, R.~Marxer, E.~Vincent, and S.~Watanabe, ``Multi-microphone speech
  recognition in everyday environments,'' \emph{Computer Speech and Language},
  vol.~46, pp. 386--387, June 2017.

\bibitem{Chen2018CHiME4}
S.-J. Chen, A.~S. Subramanian, H.~Xu, and S.~Watanabe, ``Building
  state-of-the-art distant speech recognition using the {CHiME-4} challenge
  with a setup of speech enhancement baseline,'' in \emph{Proc. of Annual
  Conference of the International Speech Communication Association
  \mbox{(Interspeech)}}, 2018, pp. 1571--1575.

\bibitem{Catalin2019Single}
C.~Zoril\u{a}, C.~Boeddeker, R.~Doddipatla, and R.~Haeb-Umbach, ``An
  investigation into the effectiveness of enhancement in {ASR} training and
  test for {CHiME-5} dinner party transcription,'' in \emph{Proc. of
  \mbox{IEEE} Workshop on Automatic Speech Recognition and Understanding
  \mbox{(ASRU)}}, 2019.

\bibitem{USTC_chime6}
J.~Du, Y.-H. Tu, L.~Sun, L.~Chai, X.~Tang, M.-K. He, F.~Ma, J.~Pan, J.-Q. Gao,
  D.~Liu, C.-H. Lee, and J.-D. Chen, ``The {USTC}-{NELSLIP} systems for
  {CHiME-6} challenge,'' in \emph{6th CHiME Speech Separation and Recognition
  Challenge (CHiME-6)}, 2020.

\bibitem{6296526}
G.~Hinton, L.~Deng, D.~Yu, G.~E. Dahl, A.~r.~Mohamed, N.~Jaitly, A.~Senior,
  V.~Vanhoucke, P.~Nguyen, T.~N. Sainath, and B.~Kingsbury, ``Deep neural
  networks for acoustic modeling in speech recognition: The shared views of
  four research groups,'' \emph{{IEEE} Signal Processing Magazine}, vol.~29,
  no.~6, pp. 82--97, Nov 2012.

\bibitem{Yu2015}
D.~Yu and L.~Deng, \emph{Automatic Speech Recognition -- A Deep Learning
  Approach}.\hskip 1em plus 0.5em minus 0.4em\relax Springer, 2015.

\bibitem{LiDeHaGo2015}
J.~Li, L.~Deng, R.~Haeb-Umbach, and Y.~Gong, \emph{Robust Automatic Speech
  Recognition}.\hskip 1em plus 0.5em minus 0.4em\relax Elsevier, Oct 2015.

\bibitem{Souden2010MVDR}
M.~Souden, J.~Benesty, and S.~Affes, ``On optimal frequency-domain multichannel
  linear filtering for noise reduction,'' \emph{{IEEE} Transactions on Audio,
  Speech, and Language Processing}, vol.~18, no.~2, pp. 260--276, 2010.

\bibitem{VanCompernolle1990}
D.~Van~Compernolle, W.~Ma, F.~Xie, and M.~Van~Diest, ``Speech recognition in
  noisy environments with the aid of microphone arrays,'' \emph{Speech
  Communication}, vol.~9, no. 5-6, pp. 433--442, 1990.

\bibitem{Naylor2010}
P.~Naylor and N.~Gaubitch, Eds., \emph{Speech Dereverberation}.\hskip 1em plus
  0.5em minus 0.4em\relax Springer, 2010.

\bibitem{Nakatani2010WPE}
T.~Nakatani, T.~Yoshioka, K.~Kinoshita, M.~Miyoshi, and B.-H. Juang, ``Speech
  dereverberation based on variance-normalized delayed linear prediction,''
  \emph{IEEE Transactions on Audio, Speech, and Language Processing}, vol.~18,
  no.~7, pp. 1717--1731, 2010.

\bibitem{Araki2016Spatial}
S.~Araki, M.~Okada, T.~Higuchi, A.~Ogawa, and T.~Nakatani, ``Spatial
  correlation model based observation vector clustering and {MVDR} beamforming
  for meeting recognition,'' in \emph{Proc. of \mbox{IEEE} International
  Conference on Acoustics, Speech and Signal Processing \mbox{(ICASSP)}}.\hskip
  1em plus 0.5em minus 0.4em\relax IEEE, 2016, pp. 385--389.

\bibitem{Wu:2017:RAS:3068681.3068688}
B.~Wu, K.~Li, M.~Yang, C.-H. Lee, B.~Wu, K.~Li, M.~Yang, C.-H. Lee, B.~Wu,
  M.~Yang, C.-H. Lee, and K.~Li, ``A reverberation-time-aware approach to
  speech dereverberation based on deep neural networks,'' \emph{{IEEE}
  Transactions on Audio, Speech, and Language Processing}, vol.~25, no.~1, pp.
  102--111, Jan. 2017.

\bibitem{Hershey2016DeepClustering}
J.~R. Hershey, Z.~Chen, J.~Le~Roux, and S.~Watanabe, ``Deep clustering:
  discriminative embeddings for segmentation and separation,'' in \emph{Proc.
  of \mbox{IEEE} International Conference on Acoustics, Speech and Signal
  Processing \mbox{(ICASSP)}}.\hskip 1em plus 0.5em minus 0.4em\relax IEEE,
  2016, pp. 31--35.

\bibitem{7952155}
Z.~Chen, Y.~Luo, and N.~Mesgarani, ``Deep attractor network for
  single-microphone speaker separation,'' in \emph{Proc. of \mbox{IEEE}
  International Conference on Acoustics, Speech and Signal Processing
  \mbox{(ICASSP)}}, March 2017, pp. 246--250.

\bibitem{Kolbaek2017PIT}
M.~Kolb{\ae}k, D.~Yu, Z.-H. Tan, and J.~Jensen, ``Multitalker speech separation
  with utterance-level permutation invariant training of deep recurrent neural
  networks,'' \emph{IEEE/ACM Transactions on Audio, Speech, and Language
  Processing}, vol.~25, no.~10, pp. 1901--1913, 2017.

\bibitem{Yi19ConvTasNet}
Y.~Luo and N.~Mesgarani, ``{Conv-TasNet}: Surpassing ideal time-frequency
  magnitude masking for speech separation,'' \emph{IEEE/ACM Transactions on
  Audio, Speech, and Language Processing}, vol.~PP, pp. 1--1, May 2019.

\bibitem{Avargel2007OnMT}
Y.~Avargel and I.~Cohen, ``On multiplicative transfer function approximation in
  the short-time fourier transform domain,'' \emph{{IEEE} Signal Processing
  Letters}, vol.~14, pp. 337--340, 2007.

\bibitem{Gilloire1992}
A.~Gilloire and M.~Vetterli, ``Adaptive filtering in sub-bands with critical
  sampling: analysis, experiments, and application to acoustic echo
  cancellation,'' \emph{IEEE Transactions on Signal Processing}, vol.~40,
  no.~8, pp. 1862--1875, 1992.

\bibitem{Talmon2009}
R.~Talmon, I.~Cohen, and S.~Gannot, ``Convolutive transfer function generalized
  sidelobe canceler,'' \emph{{IEEE} Transactions on Audio, Speech, and Language
  Processing}, vol.~17, no.~7, pp. 1420--1434, 2009.

\bibitem{Li2017a}
X.~{Li}, L.~{Girin}, S.~{Gannot}, and R.~{Horaud}, ``Multichannel speech
  separation and enhancement using the convolutive transfer function,''
  \emph{IEEE/ACM Transactions on Audio, Speech, and Language Processing},
  vol.~27, no.~3, pp. 645--659, 2019.

\bibitem{PESQ}
``{ITU-T} recommendation p.862: Perceptual evaluation of speech quality
  ({PESQ}): An objective method for end-to-end speech quality assessment of
  narrow-band telephone networks and speech codecs,''
  http://www.itu.int/rec/T-REC-P.862/en, 2008.

\bibitem{Taal2010}
C.~H. {Taal}, R.~C. {Hendriks}, R.~{Heusdens}, and J.~{Jensen}, ``A short-time
  objective intelligibility measure for time-frequency weighted noisy speech,''
  in \emph{Proc. of \mbox{IEEE} International Conference on Acoustics, Speech
  and Signal Processing \mbox{(ICASSP)}}, March 2010, pp. 4214--4217.

\bibitem{Vincent2006Performance}
E.~Vincent, R.~Gribonval, and C.~F{\'e}votte, ``Performance measurement in
  blind audio source separation,'' \emph{IEEE Transactions on Audio, Speech,
  and Language Processing}, vol.~14, no.~4, pp. 1462--1469, 2006.

\bibitem{DeliangWang2018}
D.~Wang and J.~Chen, ``Supervised speech separation based on deep learning: An
  overview,'' \emph{IEEE/ACM Transactions on Audio, Speech, and Language
  Processing}, vol.~26, no.~10, pp. 1702--1726, 2018.

\bibitem{Buchner2004icassp}
H.~Buchner, R.~Aichner, and W.~Kellermann, ``{TRINICON}: a versatile framework
  for multichannel blind signal processing,'' in \emph{Proc. of \mbox{IEEE}
  International Conference on Acoustics, Speech and Signal Processing
  \mbox{(ICASSP)}}, vol. III, 2004, pp. 889--892.

\bibitem{Nakatani2019ML}
T.~Nakatani and K.~Kinoshita, ``Maximum likelihood convolutional beamformer for
  simultaneous denoising and dereverberation,'' in \emph{27th European Signal
  Processing Conference (EUSIPCO)}, 2019.

\bibitem{Slock1994mclp}
D.~T.~M. Slock, ``Blind fractionally-spaced equalization, perfectre
  construction filter-banks and multichannel linear prediction,'' in
  \emph{Proc. of \mbox{IEEE} International Conference on Acoustics, Speech and
  Signal Processing \mbox{(ICASSP)}}, vol.~4, 1994, pp. 585--588.

\bibitem{VanVeen1988Beamforming}
B.~D. {Van Veen} and K.~M. {Buckley}, ``Beamforming: A versatile approach to
  spatial filtering,'' \emph{IEEE ASSP Magazine}, vol.~5, no.~2, pp. 4--24,
  April 1988.

\bibitem{Sharon2017taslp}
S.~Gannot, E.~Vincent, S.~Markovich-Golan, and A.~Ozerov, ``A consolidated
  perspective on multimicrophone speech enhancement and source separation,''
  \emph{{IEEE} Transactions on Audio, Speech, and Language Processing},
  vol.~25, no.~4, 2017.

\bibitem{boeddeker2020icassp}
\BIBentryALTinterwordspacing
C.~Boeddeker, T.~Nakatani, K.~Kinoshita, and R.~Haeb-Umbach, ``Jointly optimal
  dereverberation and beamforming,'' Submitted to ICASSP, 2020. [Online].
  Available: \url{http://arxiv.org/abs/1910.13707}
\BIBentrySTDinterwordspacing

\bibitem{Abed-Meraim1997pe}
K.~Abed-Meraim and P.~Loubaton, ``Prediction error method for second-order
  blind identification,'' \emph{{IEEE} Transactions on Signal Processing},
  vol.~45, no.~3, pp. 694--705, 1997.

\bibitem{Vincent2018}
E.~Vincent, T.~Virtanen, and S.~Gannot, \emph{Audio source separation and
  speech enhancement}.\hskip 1em plus 0.5em minus 0.4em\relax John Wiley \&
  Sons, 2018.

\bibitem{Makino2018}
S.~Makino, Ed., \emph{Audio Source Separation}.\hskip 1em plus 0.5em minus
  0.4em\relax Springer, 2018.

\bibitem{Xiong2015}
F.~Xiong, B.~T. Meyer, N.~Moritz, R.~Rehr, J.~Anem{\"u}ller, T.~Gerkmann,
  S.~Doclo, and S.~Goetze, ``Front-end technologies for robust asr in
  reverberant environments—spectral enhancement-based dereverberation and
  auditory modulation filterbank features,'' \emph{EURASIP Journal on Advances
  in Signal Processing}, vol. 2015, no.~1, p.~70, 2015.

\bibitem{Yoshioka2012GeneralWPE}
T.~Yoshioka and T.~Nakatani, ``Generalization of multi-channel linear
  prediction methods for blind {MIMO} impulse response shortening,''
  \emph{{IEEE} Transactions on Audio, Speech, and Language Processing}, 2012.

\bibitem{caroselli2017adaptive}
J.~Caroselli, I.~Shafran, A.~Narayanan, and R.~Rose, ``Adaptive multichannel
  dereverberation for automatic speech recognition,'' in \emph{Proc. of Annual
  Conference of the International Speech Communication Association
  \mbox{(Interspeech)}}, 2017.

\bibitem{7472671}
T.~Higuchi, N.~Ito, T.~Yoshioka, and T.~Nakatani, ``Robust {MVDR} beamforming
  using time-frequency masks for online/offline asr in noise,'' in \emph{Proc.
  of \mbox{IEEE} International Conference on Acoustics, Speech and Signal
  Processing \mbox{(ICASSP)}}, March 2016, pp. 5210--5214.

\bibitem{Isik2016DeepClustering}
Y.~Isik, J.~Le~Roux, Z.~Chen, S.~Watanabe, and J.~Hershey, ``Single-channel
  multi-speaker separation using deep clustering,'' in \emph{Proc. of Annual
  Conference of the International Speech Communication Association
  \mbox{(Interspeech)}}, 2016.

\bibitem{1142739}
L.~Griffiths and C.~Jim, ``An alternative approach to linearly constrained
  adaptive beamforming,'' \emph{IEEE Transactions on Antennas and Propagation},
  vol.~30, no.~1, pp. 27--34, January 1982.

\bibitem{Makino2007BSS}
S.~Makino, T.~Lee, and H.~Sawada, \emph{Blind speech separation}.\hskip 1em
  plus 0.5em minus 0.4em\relax Springer, 2007, vol. 615.

\bibitem{Naik2014}
G.~Naik and W.~Wang, Eds., \emph{Blind Source Separation}.\hskip 1em plus 0.5em
  minus 0.4em\relax Springer, 2014.

\bibitem{Hey2016}
J.~Heymann, L.~Drude, and R.~Haeb-Umbach, ``Neural network based spectral mask
  estimation for acoustic beamforming,'' in \emph{Proc. of \mbox{IEEE}
  International Conference on Acoustics, Speech and Signal Processing
  \mbox{(ICASSP)}}, 2016.

\bibitem{7472778}
X.~Xiao, S.~Watanabe, H.~Erdogan, L.~Lu, J.~Hershey, M.~L. Seltzer, G.~Chen,
  Y.~Zhang, M.~Mandel, and D.~Yu, ``Deep beamforming networks for multi-channel
  speech recognition,'' in \emph{Proc. of \mbox{IEEE} International Conference
  on Acoustics, Speech and Signal Processing \mbox{(ICASSP)}}, March 2016, pp.
  5745--5749.

\bibitem{Drude2017Integration}
L.~Drude and R.~Haeb-Umbach, ``Tight integration of spatial and spectral
  features for {BSS} with deep clustering embeddings,'' in \emph{Proc. of
  Annual Conference of the International Speech Communication Association
  \mbox{(Interspeech)}}, 2017.

\bibitem{Kinoshita2017Neural}
K.~Kinoshita, M.~Delcroix, H.~Kwon, T.~Mori, and T.~Nakatani, ``Neural
  network-based spectrum estimation for online {WPE} dereverberation,'' in
  \emph{Proc. of Annual Conference of the International Speech Communication
  Association \mbox{(Interspeech)}}, 2017, pp. 384--388.

\bibitem{Heymann2019Joint}
J.~Heymann, L.~Drude, R.~Haeb-Umbach, K.~Kinoshita, and T.~Nakatani, ``Joint
  optimization of neural network-based {WPE} dereverberation and acoustic model
  for robust online {ASR},'' in \emph{Proc. of \mbox{IEEE} International
  Conference on Acoustics, Speech and Signal Processing \mbox{(ICASSP)}}.\hskip
  1em plus 0.5em minus 0.4em\relax IEEE, 2019, pp. 6655--6659.

\bibitem{HeymannIWAENC2018}
------, ``Frame-online {DNN}-{WPE} dereverberation,'' in \emph{Proc. IWAENC},
  Tokyo, Japan, September 2018.

\bibitem{Heymann2017Beamnet}
J.~Heymann, L.~Drude, C.~Boeddeker, P.~Hanebrink, and R.~Haeb-Umbach,
  ``{BEAMNET}: End-to-end training of a beamformer-supported multi-channel
  {ASR} system,'' in \emph{Proc. of \mbox{IEEE} International Conference on
  Acoustics, Speech and Signal Processing \mbox{(ICASSP)}}, 2017.

\bibitem{Ante2015taslp}
A.~Juki\'c, T.~van Waterschoot, T.~Gerkmann, and S.~Doclo, ``Multi-channel
  linear prediction-based speech dereverberation with sparse priors,''
  \emph{{IEEE} Transactions on Audio, Speech, and Language Processing},
  vol.~23, no.~9, pp. 1509--1520, 2015.

\bibitem{Chetupalli2019taslp}
S.~R. Chetupalli and T.~V. Sreenivas, ``Late reverberation cancellation using
  {B}ayesian estimation of multi-channel linear predictors and student's
  t-source prior,'' \emph{{IEEE} Transactions on Audio, Speech, and Language
  Processing}, vol.~27, no.~6, 2019.

\bibitem{miyoshi1988mint}
M.~Miyoshi and Y.~Kaneda, ``Inverse filtering of room acoustics,'' \emph{IEEE
  Trans. Acoustics, Speech, and Signal Processing}, vol.~36, no.~2, pp.
  145--152, 1988.

\bibitem{Braun2018a}
S.~Braun, A.~Kuklasi{\'n}ski, O.~Schwartz, O.~Thiergart, E.~A. Habets,
  S.~Gannot, S.~Doclo, and J.~Jensen, ``Evaluation and comparison of late
  reverberation power spectral density estimators,'' \emph{{IEEE} Transactions
  on Audio, Speech, and Language Processing}, vol.~26, no.~6, pp. 1056--1071,
  2018.

\bibitem{Gannot2001SP}
S.~Gannot, D.~Burshtein, and E.~Weinstein, ``Signal enhancement using
  beamforming and nonstationarity with applications to speech,'' \emph{IEEE
  Transactions on Signal Processing}, vol.~49, no.~8, pp. 1614--1626, 2001.

\bibitem{Schwarts2015TASLP}
O.~Schwartz, S.~Gannot, and E.~Habets, ``Multi-microphone speech
  dereverberation and noise reduction using relative early transfer
  functions,'' \emph{IEEE/ACM Transactions on Audio, Speech, and Language
  Processing}, vol.~23, no.~2, pp. 240--251, 2015.

\bibitem{WaKrHa08}
E.~Warsitz, A.~Krueger, and R.~Haeb-Umbach, ``Speech enhancement with a new
  generalized eigenvector blocking matrix for application in a generalized
  sidelobe canceller,'' in \emph{Proc. of \mbox{IEEE} International Conference
  on Acoustics, Speech and Signal Processing \mbox{(ICASSP)}}, 2008, pp.
  73--76.

\bibitem{Mandel2007EM}
M.~I. Mandel, D.~P. Ellis, and T.~Jebara, ``An {EM} algorithm for localizing
  multiple sound sources in reverberant environments,'' in \emph{Advances in
  neural information processing systems}, 2007, pp. 953--960.

\bibitem{TranVu2010cW}
D.~H. Tran~Vu and R.~Haeb-Umbach, ``Blind speech separation employing
  directional statistics in an expectation maximization framework,'' in
  \emph{Proc. of \mbox{IEEE} International Conference on Acoustics, Speech and
  Signal Processing \mbox{(ICASSP)}}.\hskip 1em plus 0.5em minus 0.4em\relax
  IEEE, 2010, pp. 241--244.

\bibitem{Ito2016cBMM}
N.~Ito, S.~Araki, and T.~Nakatani, ``Modeling audio directional statistics
  using a complex {Bingham} mixture model for blind source extraction from
  diffuse noise,'' in \emph{Proc. of \mbox{IEEE} International Conference on
  Acoustics, Speech and Signal Processing \mbox{(ICASSP)}}.\hskip 1em plus
  0.5em minus 0.4em\relax IEEE, 2016, pp. 465--468.

\bibitem{Ito2016cACGMM}
------, ``Complex angular central {Gaussian} mixture model for directional
  statistics in mask-based microphone array signal processing,'' in
  \emph{European Signal Processing Conference (EUSIPCO)}.\hskip 1em plus 0.5em
  minus 0.4em\relax IEEE, 2016, pp. 1153--1157.

\bibitem{Yoshioka2015CHiME3}
T.~Yoshioka, N.~Ito, M.~Delcroix, A.~Ogawa, K.~Kinoshita, M.~Fujimoto, C.~Yu,
  W.~J. Fabian, M.~Espi, T.~Higuchi \emph{et~al.}, ``The {NTT} {CHiME-3}
  system: Advances in speech enhancement and recognition for mobile
  multi-microphone devices,'' in \emph{Proc. of \mbox{IEEE} Workshop on
  Automatic Speech Recognition and Understanding \mbox{(ASRU)}}.\hskip 1em plus
  0.5em minus 0.4em\relax IEEE, 2015, pp. 436--443.

\bibitem{Li2009idealBinaryMask}
Y.~Li and D.~Wang, ``On the optimality of ideal binary time--frequency masks,''
  \emph{Speech Communication}, vol.~51, no.~3, pp. 230--239, 2009.

\bibitem{Erdogan2015Masks}
H.~Erdogan, J.~R. Hershey, S.~Watanabe, and J.~Le~Roux, ``Phase-sensitive and
  recognition-boosted speech separation using deep recurrent neural networks,''
  in \emph{Proc. of \mbox{IEEE} International Conference on Acoustics, Speech
  and Signal Processing \mbox{(ICASSP)}}.\hskip 1em plus 0.5em minus
  0.4em\relax IEEE, 2015.

\bibitem{hey_asru_2015}
J.~Heymann, L.~Drude, A.~Chinaev, and R.~Haeb-Umbach, ``{BLSTM} supported {GEV}
  beamformer front-end for the 3rd {CHiME} challenge,'' in \emph{Proc. of
  \mbox{IEEE} Workshop on Automatic Speech Recognition and Understanding
  \mbox{(ASRU)}}, December 2015.

\bibitem{Erdogan2016MVDR}
H.~Erdogan, J.~R. Hershey, S.~Watanabe, M.~I. Mandel, and J.~Le~Roux,
  ``Improved {MVDR} beamforming using single-channel mask prediction
  networks,'' in \emph{Proc. of Annual Conference of the International Speech
  Communication Association \mbox{(Interspeech)}}, 2016, pp. 1981--1985.

\bibitem{Zmolikova2017Speaker}
K.~{\v{Z}}mol{\'\i}kov{\'a}, M.~Delcroix, K.~Kinoshita, T.~Higuchi, A.~Ogawa,
  and T.~Nakatani, ``Speaker-aware neural network based beamformer for speaker
  extraction in speech mixtures,'' in \emph{Proc. of Annual Conference of the
  International Speech Communication Association \mbox{(Interspeech)}}, 2017,
  pp. 2655--2659.

\bibitem{Zmolikova2019SpeakerBeam}
K.~{\v{Z}}mol{\'\i}kov{\'a}, M.~Delcroix, K.~Kinoshita, T.~Ochiai, T.~Nakatani,
  L.~Burget, and J.~{\v{C}}ernock{\`y}, ``Speaker{B}eam: Speaker aware neural
  network for target speaker extraction in speech mixtures,'' \emph{IEEE
  Journal of Selected Topics in Signal Processing}, vol.~13, no.~4, pp.
  800--814, 2019.

\bibitem{Vesely2016Summarizing}
K.~Vesel{\`y}, S.~Watanabe, K.~{\v{Z}}mol{\'\i}kov{\'a}, M.~Karafi{\'a}t,
  L.~Burget, and J.~H. {\v{C}}ernock{\`y}, ``Sequence summarizing neural
  network for speaker adaptation,'' in \emph{Proc. of \mbox{IEEE} International
  Conference on Acoustics, Speech and Signal Processing \mbox{(ICASSP)}}.\hskip
  1em plus 0.5em minus 0.4em\relax IEEE, 2016, pp. 5315--5319.

\bibitem{Wang2018VoiceFilter}
Q.~Wang, H.~Muckenhirn, K.~Wilson, P.~Sridhar, Z.~Wu, J.~Hershey, R.~A.
  Saurous, R.~J. Weiss, Y.~Jia, and I.~L. Moreno, ``Voicefilter: Targeted voice
  separation by speaker-conditioned spectrogram masking,'' \emph{arXiv preprint
  arXiv:1810.04826}, 2018.

\bibitem{Boeddeker2018CHiME5}
C.~Boeddeker, J.~Heitkaemper, J.~Schmalenstroeer, L.~Drude, J.~Heymann, and
  R.~Haeb-Umbach, ``Front-end processing for the {CHiME-5} dinner party
  scenario,'' in \emph{CHiME5 Workshop}, 2018.

\bibitem{Sawada2004Robust}
H.~Sawada, R.~Mukai, S.~Araki, and S.~Makino, ``A robust and precise method for
  solving the permutation problem of frequency-domain blind source
  separation,'' \emph{IEEE Transactions on Speech and Audio Processing},
  vol.~12, no.~5, pp. 530--538, 2004.

\bibitem{Yu2016Invariant}
D.~Yu, M.~Kolb{\ae}k, Z.-H. Tan, and J.~Jensen, ``Permutation invariant
  training of deep models for speaker-independent multi-talker speech
  separation,'' \emph{arXiv preprint arXiv:1607.00325}, 2016.

\bibitem{Drude2019Unsupervised}
L.~Drude, D.~Hasenklever, and R.~Haeb-Umbach, ``Unsupervised training of a deep
  clustering model for multichannel blind source separation,'' in \emph{Proc.
  of \mbox{IEEE} International Conference on Acoustics, Speech and Signal
  Processing \mbox{(ICASSP)}}.\hskip 1em plus 0.5em minus 0.4em\relax IEEE,
  2019.

\bibitem{Seetharaman2018Bootstrapping}
P.~Seetharaman, G.~Wichern, J.~Le~Roux, and B.~Pardo, ``Bootstrapping
  single-channel source separation via unsupervised spatial clustering on
  stereo mixtures,'' in \emph{Proc. of \mbox{IEEE} International Conference on
  Acoustics, Speech and Signal Processing \mbox{(ICASSP)}}.\hskip 1em plus
  0.5em minus 0.4em\relax IEEE, 2019.

\bibitem{Tzinis2018Unsupervised}
E.~Tzinis, S.~Venkataramani, and P.~Smaragdis, ``Unsupervised deep clustering
  for source separation: Direct learning from mixtures using spatial
  information,'' in \emph{Proc. of \mbox{IEEE} International Conference on
  Acoustics, Speech and Signal Processing \mbox{(ICASSP)}}.\hskip 1em plus
  0.5em minus 0.4em\relax IEEE, 2019.

\bibitem{Wang2018MCDC}
Z.-Q. Wang, J.~Le~Roux, and J.~R. Hershey, ``Multi-channel deep clustering:
  Discriminative spectral and spatial embeddings for speaker-independent speech
  separation,'' in \emph{Proc. of \mbox{IEEE} International Conference on
  Acoustics, Speech and Signal Processing \mbox{(ICASSP)}}.\hskip 1em plus
  0.5em minus 0.4em\relax IEEE, 2018.

\bibitem{Nakatani2017Integrating}
T.~Nakatani, N.~Ito, T.~Higuchi, S.~Araki, and K.~Kinoshita, ``Integrating
  {DNN}-based and spatial clustering-based mask estimation for robust {MVDR}
  beamforming,'' in \emph{Proc. of \mbox{IEEE} International Conference on
  Acoustics, Speech and Signal Processing \mbox{(ICASSP)}}.\hskip 1em plus
  0.5em minus 0.4em\relax IEEE, 2017.

\bibitem{Drude2019Integration}
L.~Drude and R.~Haeb-Umbach, ``Integration of neural networks and probabilistic
  spatial models for acoustic blind source separation,'' \emph{IEEE Journal of
  Selected Topics in Signal Processing}, 2019.

\bibitem{Drude2018WPE}
L.~Drude, C.~Boeddeker, J.~Heymann, R.~Haeb-Umbach, K.~Kinoshita, M.~Delcroix,
  and T.~Nakatani, ``Integrating neural network based beamforming and weighted
  prediction error dereverberation,'' in \emph{Proc. of Annual Conference of
  the International Speech Communication Association \mbox{(Interspeech)}},
  2018, pp. 3043--3047.

\bibitem{Yoshioka2018Recognizing}
T.~Yoshioka, H.~Erdogan, Z.~Chen, X.~Xiao, and F.~Alleva, ``Recognizing
  overlapped speech in meetings: A multichannel separation approach using
  neural networks,'' \emph{Proc. of Annual Conference of the International
  Speech Communication Association \mbox{(Interspeech)}}, pp. 3038--3042, 2018.

\bibitem{Rabiner2008}
L.~Rabiner and B.-H. Juang, ``Historical perspective of the field of
  {ASR/NLU},'' in \emph{Springer handbook of speech processing}.\hskip 1em plus
  0.5em minus 0.4em\relax Springer, 2008, pp. 521--538.

\bibitem{rabiner1989tutorial}
L.~R. Rabiner, ``A tutorial on hidden {Markov} models and selected applications
  in speech recognition,'' \emph{Proceedings of the {IEEE}}, vol.~77, no.~2,
  pp. 257--286, 1989.

\bibitem{Huang:2001}
X.~Huang, A.~Acero, and H.-W. Hon, \emph{Spoken Language Processing: A Guide to
  Theory, Algorithm, and System Development}, 1st~ed.\hskip 1em plus 0.5em
  minus 0.4em\relax Upper Saddle River, NJ, USA: Prentice Hall PTR, 2001.

\bibitem{graves2006connectionist}
A.~Graves, S.~Fern{\'a}ndez, F.~Gomez, and J.~Schmidhuber, ``Connectionist
  temporal classification: labelling unsegmented sequence data with recurrent
  neural networks,'' in \emph{Proc. of International Conference on Machine
  Learning \mbox{(ICML)}}, 2006, pp. 369--376.

\bibitem{Li2017}
\BIBentryALTinterwordspacing
B.~Li, T.~Sainath, A.~Narayanan, J.~Caroselli, M.~Bacchiani, A.~Misra,
  I.~Shafran, H.~Sak, G.~Pundak, K.~Chin, K.~C. Sim, R.~J. Weiss, K.~Wilson,
  E.~Variani, C.~Kim, O.~Siohan, M.~Weintraub, E.~McDermott, R.~Rose, and
  M.~Shannon, ``Acoustic modeling for {G}oogle {H}ome,'' in \emph{Proc. of
  Annual Conference of the International Speech Communication Association
  \mbox{(Interspeech)}}, 2017. [Online]. Available:
  \url{http://www.cs.cmu.edu/~chanwook/MyPapers/b_li_interspeech_2017.pdf}
\BIBentrySTDinterwordspacing

\bibitem{21701}
A.~{Waibel}, T.~{Hanazawa}, G.~{Hinton}, K.~{Shikano}, and K.~J. {Lang},
  ``Phoneme recognition using time-delay neural networks,'' \emph{{IEEE}
  Transactions on Acoustics, Speech, and Signal Processing}, vol.~37, no.~3,
  pp. 328--339, 1989.

\bibitem{7178920}
J.~{Huang}, J.~{Li}, and Y.~{Gong}, ``An analysis of convolutional neural
  networks for speech recognition,'' in \emph{Proc. of \mbox{IEEE}
  International Conference on Acoustics, Speech and Signal Processing
  \mbox{(ICASSP)}}, 2015, pp. 4989--4993.

\bibitem{7178838}
T.~N. {Sainath}, O.~{Vinyals}, A.~{Senior}, and H.~{Sak}, ``Convolutional, long
  short-term memory, fully connected deep neural networks,'' in \emph{Proc. of
  \mbox{IEEE} International Conference on Acoustics, Speech and Signal
  Processing \mbox{(ICASSP)}}, 2015, pp. 4580--4584.

\bibitem{vaswani2017attention}
A.~Vaswani, N.~Shazeer, N.~Parmar, J.~Uszkoreit, L.~Jones, A.~N. Gomez,
  {\L}.~Kaiser, and I.~Polosukhin, ``Attention is all you need,'' in
  \emph{Advances in neural information processing systems}, 2017, pp.
  5998--6008.

\bibitem{karita2019comparative}
S.~Karita, N.~Chen, T.~Hayashi, T.~Hori, H.~Inaguma, Z.~Jiang, M.~Someki,
  N.~E.~Y. Soplin, R.~Yamamoto, X.~Wang, S.~Watanabe, T.~Yoshimura, and
  W.~Zhang, ``A comparative study on transformer vs {RNN} in speech
  applications,'' in \emph{Proc. of \mbox{IEEE} Workshop on Automatic Speech
  Recognition and Understanding \mbox{(ASRU)}}, 2019.

\bibitem{9054345}
Y.~{Wang}, A.~{Mohamed}, D.~{Le}, C.~{Liu}, A.~{Xiao}, J.~{Mahadeokar},
  H.~{Huang}, A.~{Tjandra}, X.~{Zhang}, F.~{Zhang}, C.~{Fuegen}, G.~{Zweig},
  and M.~L. {Seltzer}, ``Transformer-based acoustic modeling for hybrid speech
  recognition,'' in \emph{Proc. of \mbox{IEEE} International Conference on
  Acoustics, Speech and Signal Processing \mbox{(ICASSP)}}, 2020, pp.
  6874--6878.

\bibitem{Mikolov}
T.~Mikolov, M.~Karafi{\'{a}}t, L.~Burget, J.~Cernock{\'{y}}, and S.~Khudanpur,
  ``Recurrent neural network based language model,'' in \emph{Proc. of Annual
  Conference of the International Speech Communication Association
  \mbox{(Interspeech)}}, 2010, pp. 1045--1048.

\bibitem{duChime4}
J.~Du, Y.-H. Tu, L.~Sun, F.~Ma, H.-K. Wang, J.~Pan, C.~Liu, J.-D. Chen, and
  C.-H. Lee, ``{The {USTC}-{iFlytek} System for {CHiME-4} Challenge},'' in
  \emph{CHiME4 Workshop}, 2016.

\bibitem{DuChime52018}
\BIBentryALTinterwordspacing
J.~Du, T.~Gao, L.~Sun, F.~Ma, Y.~Fang, D.-Y. Liu, Q.~Zhang, X.~Zhang, H.-K.
  Wang, J.~Pan, J.-Q. Gao, C.-H. Lee, and J.-D. Chen, ``{The {USTC}-{iFlytek}
  systems for {CHiME-5} Challenge},'' in \emph{Proc. CHiME 2018 Workshop on
  Speech Processing in Everyday Environments}, 2018, pp. 11--15. [Online].
  Available: \url{http://dx.doi.org/10.21437/CHiME.2018-3}
\BIBentrySTDinterwordspacing

\bibitem{vesely2013sequence}
K.~Vesel{\`y}, A.~Ghoshal, L.~Burget, and D.~Povey, ``Sequence-discriminative
  training of deep neural networks,'' in \emph{Proc. of Annual Conference of
  the International Speech Communication Association \mbox{(Interspeech)}},
  2013, pp. 2345--2349.

\bibitem{DBLP:conf/interspeech/PoveyPGGMNWK16}
D.~Povey, V.~Peddinti, D.~Galvez, P.~Ghahremani, V.~Manohar, X.~Na, Y.~Wang,
  and S.~Khudanpur, ``Purely sequence-trained neural networks for {ASR} based
  on lattice-free {MMI},'' in \emph{Proc. of Annual Conference of the
  International Speech Communication Association \mbox{(Interspeech)}}.\hskip
  1em plus 0.5em minus 0.4em\relax {ISCA}, 2016, pp. 2751--2755.

\bibitem{kanda2018hitachi}
N.~Kanda, R.~Ikeshita, S.~Horiguchi, Y.~Fujita, K.~Nagamatsu, X.~Wang,
  V.~Manohar, N.~E.~Y. Soplin, M.~Maciejewski, S.-J. Chen \emph{et~al.}, ``The
  {Hitachi/JHU CHiME-5} system: Advances in speech recognition for everyday
  home environments using multiple microphone arrays,'' \emph{Proc. CHiME-5},
  pp. 6--10, 2018.

\bibitem{nakamura-etal-2000-acoustical}
S.~Nakamura, K.~Hiyane, F.~Asano, T.~Nishiura, and T.~Yamada, ``Acoustical
  sound database in real environments for sound scene understanding and
  hands-free speech recognition,'' in \emph{Proceedings of the Second
  International Conference on Language Resources and Evaluation ({LREC}{'}00)},
  Athens, Greece, May 2000.

\bibitem{jeub2009}
M.~{Jeub}, M.~{Schafer}, and P.~{Vary}, ``A binaural room impulse response
  database for the evaluation of dereverberation algorithms,'' in \emph{2009
  16th International Conference on Digital Signal Processing}, July 2009, pp.
  1--5.

\bibitem{Szoke2019}
I.~{Szöke}, M.~{Skácel}, L.~{Mošner}, J.~{Paliesek}, and J.~{Černocký},
  ``Building and evaluation of a real room impulse response dataset,''
  \emph{IEEE Journal of Selected Topics in Signal Processing}, vol.~13, no.~4,
  pp. 863--876, Aug 2019.

\bibitem{Allen1979Image}
J.~Allen and D.~Berkley, ``Image method for efficiently simulating small-room
  acoustics,'' \emph{The Journal of the Acoustical Society of America},
  vol.~65, pp. 943--950, 1979.

\bibitem{Habets2010}
\BIBentryALTinterwordspacing
E.~A.~P. Habets, ``{Room impulse response generator},'' Technische Universiteit
  Eindhoven, Tech. Rep., 2010. [Online]. Available:
  \url{http://home.tiscali.nl/ehabets/rir\_generator/rir\_generator.pdf}
\BIBentrySTDinterwordspacing

\bibitem{musan2015}
D.~Snyder, G.~Chen, and D.~Povey, ``{MUSAN}: {A} {M}usic, {S}peech, and {N}oise
  {C}orpus,'' 2015, arXiv:1510.08484v1.

\bibitem{KoIS15}
T.~Ko, V.~Peddinti, D.~Povey, and S.~Khudanpur, ``Audio augmentation for speech
  recognition.'' in \emph{Proc. of Annual Conference of the International
  Speech Communication Association \mbox{(Interspeech)}}.\hskip 1em plus 0.5em
  minus 0.4em\relax ISCA, 2015, pp. 3586--3589.

\bibitem{Hsu_2017}
W.~{Hsu}, Y.~{Zhang}, and J.~{Glass}, ``Unsupervised domain adaptation for
  robust speech recognition via variational autoencoder-based data
  augmentation,'' in \emph{2017 IEEE Automatic Speech Recognition and
  Understanding Workshop (ASRU)}, 2017, pp. 16--23.

\bibitem{adversarial_examples}
\BIBentryALTinterwordspacing
S.~Sun, C.~Yeh, M.~Ostendorf, M.~Hwang, and L.~Xie, ``Training augmentation
  with adversarial examples for robust speech recognition,'' \emph{CoRR}, vol.
  abs/1806.02782, 2018. [Online]. Available:
  \url{http://arxiv.org/abs/1806.02782}
\BIBentrySTDinterwordspacing

\bibitem{Hu_2018}
H.~{Hu}, T.~{Tan}, and Y.~{Qian}, ``Generative adversarial networks based data
  augmentation for noise robust speech recognition,'' in \emph{Proc. of
  \mbox{IEEE} International Conference on Acoustics, Speech and Signal
  Processing \mbox{(ICASSP)}}, 2018, pp. 5044--5048.

\bibitem{Qian_2019}
Y.~Qian, H.~Hu, and T.~Tan, ``Data augmentation using generative adversarial
  networks for robust speech recognition,'' \emph{Speech Communication}, vol.
  114, 08 2019.

\bibitem{SpecAugment}
D.~S. Park, W.~Chan, Y.~Zhang, C.-C. Chiu, B.~Zoph, E.~D. Cubuk, and Q.~V. Le,
  ``Spec{A}ugment: A simple augmentation method for automatic speech
  recognition,'' in \emph{Proc. of Annual Conference of the International
  Speech Communication Association \mbox{(Interspeech)}}, 2019.

\bibitem{STC_chime6}
I.~Medennikov, M.~Korenevsky, T.~Prisyach, Y.~Khokhlov, M.~Korenevskaya,
  I.~Sorokin, T.~Timofeeva, A.~Mitrofanov, A.~Andrusenko, I.~Podluzhny,
  A.~Laptev, and A.~Romanenko, ``The {STC} system for the {CHiME-6}
  challenge,'' in \emph{6th CHiME Speech Separation and Recognition Challenge
  (CHiME-6)}, 2020.

\bibitem{Xue_2018}
S.~{Xue}, Z.~{Yan}, T.~{Yu}, and Z.~{Liu}, ``A study on improving acoustic
  model for robust and far-field speech recognition,'' in \emph{2018 IEEE 23rd
  International Conference on Digital Signal Processing (DSP)}, 2018, pp. 1--5.

\bibitem{Tang_2018}
\BIBentryALTinterwordspacing
H.~Tang, W.-N. Hsu, F.~Grondin, and J.~Glass, ``A study of enhancement,
  augmentation and autoencoder methods for domain adaptation in distant speech
  recognition,'' \emph{Interspeech 2018}, Sep 2018. [Online]. Available:
  \url{http://dx.doi.org/10.21437/Interspeech.2018-2030}
\BIBentrySTDinterwordspacing

\bibitem{Manohar_2019}
V.~{Manohar}, S.~{Chen}, Z.~{Wang}, Y.~{Fujita}, S.~{Watanabe}, and
  S.~{Khudanpur}, ``Acoustic modeling for overlapping speech recognition: Jhu
  chime-5 challenge system,'' in \emph{Proc. of \mbox{IEEE} International
  Conference on Acoustics, Speech and Signal Processing \mbox{(ICASSP)}}, 2019,
  pp. 6665--6669.

\bibitem{Kanda_2019}
N.~{Kanda}, Y.~{Fujita}, S.~{Horiguchi}, R.~{Ikeshita}, K.~{Nagamatsu}, and
  S.~{Watanabe}, ``Acoustic modeling for distant multi-talker speech
  recognition with single- and multi-channel branches,'' in \emph{Proc. of
  \mbox{IEEE} International Conference on Acoustics, Speech and Signal
  Processing \mbox{(ICASSP)}}, 2019, pp. 6630--6634.

\bibitem{Peddinti+2016}
\BIBentryALTinterwordspacing
V.~Peddinti, V.~Manohar, Y.~Wang, D.~Povey, and S.~Khudanpur, ``Far-field {ASR}
  without parallel data,'' in \emph{Proc. of Annual Conference of the
  International Speech Communication Association \mbox{(Interspeech)}}, 2016,
  pp. 1996--2000. [Online]. Available:
  \url{http://dx.doi.org/10.21437/Interspeech.2016-1475}
\BIBentrySTDinterwordspacing

\bibitem{decroix_IS13}
M.~Delcroix, Y.~Kubo, T.~Nakatani, and A.~Nakamura, ``Is speech enhancement
  pre-processing still relevant when using deep neural networks for acoustic
  modeling?'' in \emph{Proc. of Annual Conference of the International Speech
  Communication Association \mbox{(Interspeech)}}.\hskip 1em plus 0.5em minus
  0.4em\relax ISCA, 2013, pp. 2992--2996.

\bibitem{vincent:hal-01588876}
\BIBentryALTinterwordspacing
E.~Vincent, ``{When mismatched training data outperform matched data},'' in
  \emph{{Systematic approaches to deep learning methods for audio}}, Vienna,
  Austria, Sep. 2017. [Online]. Available:
  \url{https://hal.inria.fr/hal-01588876}
\BIBentrySTDinterwordspacing

\bibitem{Gonzalez:2015:MTT:2812295.2812300}
C.~R. Gonz\'{a}lez and Y.~S. Abu-Mostafa, ``Mismatched training and test
  distributions can outperform matched ones,'' \emph{Neural Comput.}, vol.~27,
  no.~2, pp. 365--387, Feb. 2015.

\bibitem{liao2013speaker}
H.~{Liao}, ``Speaker adaptation of context dependent deep neural networks,'' in
  \emph{Proc. of \mbox{IEEE} International Conference on Acoustics, Speech and
  Signal Processing \mbox{(ICASSP)}}, May 2013, pp. 7947--7951.

\bibitem{yu2013kl}
D.~{Yu}, K.~{Yao}, H.~{Su}, G.~{Li}, and F.~{Seide}, ``Kl-divergence
  regularized deep neural network adaptation for improved large vocabulary
  speech recognition,'' in \emph{Proc. of \mbox{IEEE} International Conference
  on Acoustics, Speech and Signal Processing \mbox{(ICASSP)}}, May 2013, pp.
  7893--7897.

\bibitem{erdogan2016multi}
H.~Erdogan, T.~Hayashi, J.~R. Hershey, T.~Hori, C.~Hori, W.-N. Hsu, S.~Kim,
  J.~Le~Roux, Z.~Meng, and S.~Watanabe, ``Multi-channel speech recognition:
  {LSTM}s all the way through,'' in \emph{CHiME-4 workshop}, 2016, pp. 1--4.

\bibitem{xiao2017time}
X.~Xiao, S.~Zhao, D.~L. Jones, E.~S. Chng, and H.~Li, ``On time-frequency mask
  estimation for {MVDR} beamforming with application in robust speech
  recognition,'' in \emph{Proc. of \mbox{IEEE} International Conference on
  Acoustics, Speech and Signal Processing \mbox{(ICASSP)}}.\hskip 1em plus
  0.5em minus 0.4em\relax IEEE, 2017, pp. 3246--3250.

\bibitem{Heymann_ICASSP18}
J.~{Heymann}, M.~{Bacchiani}, and T.~N. {Sainath}, ``Performance of mask based
  statistical beamforming in a smart home scenario,'' in \emph{Proc. of
  \mbox{IEEE} International Conference on Acoustics, Speech and Signal
  Processing \mbox{(ICASSP)}}, April 2018, pp. 6722--6726.

\bibitem{chorowski2015attention}
J.~K. Chorowski, D.~Bahdanau, D.~Serdyuk, K.~Cho, and Y.~Bengio,
  ``Attention-based models for speech recognition,'' in \emph{Proc. of Advances
  in neural information processing systems \mbox{(NeurIPS)}}, 2015, pp.
  577--585.

\bibitem{chan2016listen}
W.~Chan, N.~Jaitly, Q.~Le, and O.~Vinyals, ``Listen, attend and spell: A neural
  network for large vocabulary conversational speech recognition,'' in
  \emph{Proc. of \mbox{IEEE} International Conference on Acoustics, Speech and
  Signal Processing \mbox{(ICASSP)}}, 2016, pp. 4960--4964.

\bibitem{graves2013speech}
A.~Graves, A.-R. Mohamed, and G.~Hinton, ``Speech recognition with deep
  recurrent neural networks,'' in \emph{Proc. of \mbox{IEEE} International
  Conference on Acoustics, Speech and Signal Processing \mbox{(ICASSP)}}, 2013,
  pp. 6645--6649.

\bibitem{kim2017joint}
S.~Kim, T.~Hori, and S.~Watanabe, ``Joint {CTC}-attention based end-to-end
  speech recognition using multi-task learning,'' in \emph{Proc. of \mbox{IEEE}
  International Conference on Acoustics, Speech and Signal Processing
  \mbox{(ICASSP)}}, 2017, pp. 4835--4839.

\bibitem{Ochiai2017MultichannelES}
T.~Ochiai, S.~Watanabe, T.~Hori, and J.~R. Hershey, ``Multichannel end-to-end
  speech recognition,'' in \emph{Proc. of International Conference on Machine
  Learning \mbox{(ICML)}}, 2017.

\bibitem{ochiai2017unified}
T.~Ochiai, S.~Watanabe, T.~Hori, J.~R. Hershey, and X.~Xiao, ``Unified
  architecture for multichannel end-to-end speech recognition with neural
  beamforming,'' \emph{{IEEE} Journal of Selected Topics in Signal Processing},
  vol.~11, no.~8, pp. 1274--1288, 2017.

\bibitem{Braun2018}
S.~Braun, D.~Neil, J.~Anumula, E.~Ceolini, and S.-C. Liu, ``Multi-channel
  attention for end-to-end speech recognition,'' in \emph{Proc. of Annual
  Conference of the International Speech Communication Association
  \mbox{(Interspeech)}}, 2018, pp. 17--21.

\bibitem{wang2019stream}
X.~Wang, R.~Li, S.~H. Mallidi, T.~Hori, S.~Watanabe, and H.~Hermansky, ``Stream
  attention-based multi-array end-to-end speech recognition,'' in \emph{Proc.
  of \mbox{IEEE} International Conference on Acoustics, Speech and Signal
  Processing \mbox{(ICASSP)}}.\hskip 1em plus 0.5em minus 0.4em\relax IEEE,
  2019, pp. 7105--7109.

\bibitem{subramanian2019speech}
A.~S. Subramanian, X.~Wang, M.~K. Baskar, S.~Watanabe, T.~Taniguchi, D.~Tran,
  and Y.~Fujita, ``Speech enhancement using end-to-end speech recognition
  objectives,'' in \emph{Proc. of \mbox{IEEE} \mbox{ASSP} Workshop on
  Applications of Signal Processing to Audio and Acoustics}.\hskip 1em plus
  0.5em minus 0.4em\relax IEEE, 2019.

\bibitem{chang2019mimo}
X.~Chang, W.~Zhang, Y.~Qian, J.~L. Roux, and S.~Watanabe, ``{MIMO-SPEECH}:
  End-to-end multi-channel multi-speaker speech recognition,'' in \emph{Proc.
  of \mbox{IEEE} Workshop on Automatic Speech Recognition and Understanding
  \mbox{(ASRU)}}, 2019.

\bibitem{Delcroix2019}
M.~Delcroix, S.~Watanabe, T.~Ochiai, K.~Kinoshita, S.~Karita, A.~Ogawa, and
  T.~Nakatani, ``{End-to-End SpeakerBeam for Single Channel Target Speech
  Recognition},'' in \emph{Proc. of Annual Conference of the International
  Speech Communication Association \mbox{(Interspeech)}}, 2019, pp. 451--455.

\bibitem{Denisov2019}
P.~Denisov and N.~T. Vu, ``{End-to-End Multi-Speaker Speech Recognition Using
  Speaker Embeddings and Transfer Learning},'' in \emph{Proc. Interspeech
  2019}, 2019, pp. 4425--4429.

\bibitem{chiu2018state}
C.-C. Chiu, T.~N. Sainath, Y.~Wu, R.~Prabhavalkar, P.~Nguyen, Z.~Chen,
  A.~Kannan, R.~J. Weiss, K.~Rao, E.~Gonina \emph{et~al.}, ``State-of-the-art
  speech recognition with sequence-to-sequence models,'' in \emph{Proc. of
  \mbox{IEEE} International Conference on Acoustics, Speech and Signal
  Processing \mbox{(ICASSP)}}.\hskip 1em plus 0.5em minus 0.4em\relax IEEE,
  2018, pp. 4774--4778.

\bibitem{zeyer2018improved}
A.~Zeyer, K.~Irie, R.~Schl{\"u}ter, and H.~Ney, ``Improved training of
  end-to-end attention models for speech recognition,'' \emph{Proc. of Annual
  Conference of the International Speech Communication Association
  \mbox{(Interspeech)}}, pp. 7--11, 2018.

\bibitem{dong2018speech}
L.~Dong, S.~Xu, and B.~Xu, ``Speech-transformer: a no-recurrence
  sequence-to-sequence model for speech recognition,'' in \emph{Proc. of
  \mbox{IEEE} International Conference on Acoustics, Speech and Signal
  Processing \mbox{(ICASSP)}}.\hskip 1em plus 0.5em minus 0.4em\relax IEEE,
  2018, pp. 5884--5888.

\bibitem{sohn1999statistical}
J.~Sohn, N.~S. Kim, and W.~Sung, ``A statistical model-based voice activity
  detection,'' \emph{{IEEE} Signal Processing Letters}, vol.~6, no.~1, pp.
  1--3, 1999.

\bibitem{hughes2013recurrent}
T.~Hughes and K.~Mierle, ``Recurrent neural networks for voice activity
  detection,'' in \emph{Proc. of \mbox{IEEE} International Conference on
  Acoustics, Speech and Signal Processing \mbox{(ICASSP)}}.\hskip 1em plus
  0.5em minus 0.4em\relax IEEE, 2013, pp. 7378--7382.

\bibitem{eyben2013real}
F.~Eyben, F.~Weninger, S.~Squartini, and B.~Schuller, ``Real-life voice
  activity detection with {LSTM} recurrent neural networks and an application
  to hollywood movies,'' in \emph{Proc. of \mbox{IEEE} International Conference
  on Acoustics, Speech and Signal Processing \mbox{(ICASSP)}}.\hskip 1em plus
  0.5em minus 0.4em\relax IEEE, 2013, pp. 483--487.

\bibitem{anguera2012speaker}
X.~Anguera, S.~Bozonnet, N.~Evans, C.~Fredouille, G.~Friedland, and O.~Vinyals,
  ``Speaker diarization: A review of recent research,'' \emph{IEEE Transactions
  on Audio, Speech, and Language Processing}, vol.~20, no.~2, pp. 356--370,
  2012.

\bibitem{hori2011low}
T.~Hori, S.~Araki, T.~Yoshioka, M.~Fujimoto, S.~Watanabe, T.~Oba, A.~Ogawa,
  K.~Otsuka, D.~Mikami, K.~Kinoshita, T.~Nakatani, A.~Nakamura, and J.~Yamato,
  ``Low-latency real-time meeting recognition and understanding using distant
  microphones and omni-directional camera,'' \emph{{IEEE} Transactions on
  Audio, Speech, and Language Processing}, vol.~20, no.~2, pp. 499--513, 2011.

\bibitem{dehak2011front}
N.~Dehak, P.~J. Kenny, R.~Dehak, P.~Dumouchel, and P.~Ouellet, ``Front-end
  factor analysis for speaker verification,'' \emph{{IEEE} Transactions on
  Audio, Speech, and Language Processing}, vol.~19, no.~4, pp. 788--798, 2011.

\bibitem{snyder2018x}
D.~Snyder, D.~Garcia-Romero, G.~Sell, D.~Povey, and S.~Khudanpur, ``X-vectors:
  Robust dnn embeddings for speaker recognition,'' in \emph{Proc. of
  \mbox{IEEE} International Conference on Acoustics, Speech and Signal
  Processing \mbox{(ICASSP)}}.\hskip 1em plus 0.5em minus 0.4em\relax IEEE,
  2018, pp. 5329--5333.

\bibitem{wooters2007icsi}
C.~Wooters and M.~Huijbregts, ``The {ICSI} {RT07s} speaker diarization
  system,'' in \emph{Multimodal Technologies for Perception of Humans}.\hskip
  1em plus 0.5em minus 0.4em\relax Springer, 2007, pp. 509--519.

\bibitem{sell2014speaker}
G.~Sell and D.~Garcia-Romero, ``Speaker diarization with {PLDA} i-vector
  scoring and unsupervised calibration,'' in \emph{2014 IEEE Spoken Language
  Technology Workshop (SLT)}.\hskip 1em plus 0.5em minus 0.4em\relax IEEE,
  2014, pp. 413--417.

\bibitem{von2019all}
T.~von Neumann, K.~Kinoshita, M.~Delcroix, S.~Araki, T.~Nakatani, and
  R.~Haeb-Umbach, ``All-neural online source separation, counting, and
  diarization for meeting analysis,'' in \emph{Proc. of \mbox{IEEE}
  International Conference on Acoustics, Speech and Signal Processing
  \mbox{(ICASSP)}}.\hskip 1em plus 0.5em minus 0.4em\relax IEEE, 2019, pp.
  91--95.

\bibitem{fujita2019end}
Y.~Fujita, N.~Kanda, S.~Horiguchi, Y.~Xue, K.~Nagamatsu, and S.~Watanabe,
  ``End-to-end neural speaker diarization with self-attention,'' in \emph{Proc.
  of \mbox{IEEE} Workshop on Automatic Speech Recognition and Understanding
  \mbox{(ASRU)}}, 2019.

\bibitem{higuchi2017online}
T.~Higuchi, N.~Ito, S.~Araki, T.~Yoshioka, M.~Delcroix, and T.~Nakatani,
  ``Online {MVDR} beamformer based on complex gaussian mixture model with
  spatial prior for noise robust asr,'' \emph{IEEE/ACM Transactions on Audio,
  Speech, and Language Processing}, vol.~25, no.~4, pp. 780--793, 2017.

\bibitem{boeddeker2018exploring}
C.~Boeddeker, H.~Erdogan, T.~Yoshioka, and R.~Haeb-Umbach, ``Exploring
  practical aspects of neural mask-based beamforming for far-field speech
  recognition,'' in \emph{Proc. of \mbox{IEEE} International Conference on
  Acoustics, Speech and Signal Processing \mbox{(ICASSP)}}.\hskip 1em plus
  0.5em minus 0.4em\relax IEEE, 2018, pp. 6697--6701.

\bibitem{nakadai2010design}
K.~Nakadai, T.~Takahashi, H.~G. Okuno, H.~Nakajima, Y.~Hasegawa, and
  H.~Tsujino, ``Design and implementation of robot audition system
  {'HARK'}—open source software for listening to three simultaneous
  speakers,'' \emph{Advanced Robotics}, vol.~24, no. 5-6, pp. 739--761, 2010.

\bibitem{evers2019locata}
C.~Evers, H.~Loellmann, H.~Mellmann, A.~Schmidt, H.~Barfuss, P.~Naylor, and
  W.~Kellermann, ``The {LOCATA} challenge: Acoustic source localization and
  tracking,'' \emph{arXiv preprint arXiv:1909.01008}, 2019.

\bibitem{du2000santa}
J.~W. Du~Bois, W.~L. Chafe, C.~Meyer, S.~A. Thompson, and N.~Martey, ``Santa
  {B}arbara corpus of spoken {A}merican {E}nglish,'' \emph{CD-ROM.
  Philadelphia: Linguistic Data Consortium}, 2000.

\bibitem{Takahashi2020}
N.~Takahashi, M.~K. Singh, S.~Basak, P.~Sudarsanam, S.~Ganapathy, and
  Y.~Mitsufuji, ``Improving voice separation by incorporating end-to-end speech
  recognition,'' in \emph{Proc. of \mbox{IEEE} International Conference on
  Acoustics, Speech and Signal Processing \mbox{(ICASSP)}}.\hskip 1em plus
  0.5em minus 0.4em\relax IEEE, 2020, pp. 41--45.

\bibitem{Tzinis2020}
E.~Tzinis, S.~Wisdom, J.~R. Hershey, A.~Jansen, and D.~P. Ellis, ``Improving
  universal sound separation using sound classification,'' in \emph{Proc. of
  \mbox{IEEE} International Conference on Acoustics, Speech and Signal
  Processing \mbox{(ICASSP)}}.\hskip 1em plus 0.5em minus 0.4em\relax IEEE,
  2020, pp. 96--100.

\bibitem{cristoforetti2014dirha}
L.~Cristoforetti, M.~Ravanelli, M.~Omologo, A.~Sosi, A.~Abad, M.~Hagm{\"u}ller,
  and P.~Maragos, ``The {DIRHA} simulated corpus.'' in \emph{LREC}, 2014, pp.
  2629--2634.

\bibitem{Wehr2004}
S.~Wehr, I.~Kozintsev, R.~Lienhart, and W.~Kellermann, ``Synchronization of
  acoustic sensors for distributed ad-hoc audio networks and its use for blind
  source separation,'' in \emph{IEEE Sixth International Symposium on
  Multimedia Software Engineering}.\hskip 1em plus 0.5em minus 0.4em\relax
  IEEE, 2004, pp. 18--25.

\bibitem{ono2009blind}
N.~Ono, H.~Kohno, N.~Ito, and S.~Sagayama, ``Blind alignment of asynchronously
  recorded signals for distributed microphone array,'' in \emph{Proc. of
  \mbox{IEEE} \mbox{ASSP} Workshop on Applications of Signal Processing to
  Audio and Acoustics}.\hskip 1em plus 0.5em minus 0.4em\relax IEEE, 2009, pp.
  161--164.

\bibitem{Cherkassky2017}
D.~Cherkassky and S.~Gannot, ``Blind synchronization in wireless acoustic
  sensor networks,'' \emph{{IEEE} Transactions on Audio, Speech, and Language
  Processing}, vol.~25, no.~3, pp. 651--661, 2017.

\bibitem{araki2018meeting}
S.~Araki, N.~Ono, K.~Kinoshita, and M.~Delcroix, ``Meeting recognition with
  asynchronous distributed microphone array using block-wise refinement of
  mask-based mvdr beamformer,'' in \emph{Proc. of \mbox{IEEE} International
  Conference on Acoustics, Speech and Signal Processing \mbox{(ICASSP)}}.\hskip
  1em plus 0.5em minus 0.4em\relax IEEE, 2018, pp. 5694--5698.

\bibitem{afifi2018marvelo}
H.~Afifi, J.~Schmalenstroeer, J.~Ullmann, R.~Haeb-Umbach, and H.~Karl,
  ``Marvelo-a framework for signal processing in wireless acoustic sensor
  networks,'' in \emph{Speech Communication; 13th ITG-Symposium}.\hskip 1em
  plus 0.5em minus 0.4em\relax VDE, 2018, pp. 1--5.

\bibitem{Bertrand2011}
A.~Bertrand and M.~Moonen, ``Distributed node-specific lcmv beamforming in
  wireless sensor networks,'' \emph{IEEE Transactions on Signal Processing},
  vol.~60, no.~1, pp. 233--246, 2011.

\bibitem{Heusdens2012}
R.~Heusdens, G.~Zhang, R.~C. Hendriks, Y.~Zeng, and W.~B. Kleijn, ``Distributed
  mvdr beamforming for (wireless) microphone networks using message passing,''
  in \emph{IWAENC 2012; International Workshop on Acoustic Signal
  Enhancement}.\hskip 1em plus 0.5em minus 0.4em\relax VDE, 2012, pp. 1--4.

\bibitem{MarkovichGolan2015}
S.~Markovich-Golan, A.~Bertrand, M.~Moonen, and S.~Gannot, ``Optimal
  distributed minimum-variance beamforming approaches for speech enhancement in
  wireless acoustic sensor networks,'' \emph{Signal Processing}, vol. 107, pp.
  4--20, 2015.

\bibitem{bertrand2011applications}
A.~Bertrand, ``Applications and trends in wireless acoustic sensor networks: A
  signal processing perspective,'' in \emph{2011 18th IEEE symposium on
  communications and vehicular technology in the Benelux (SCVT)}.\hskip 1em
  plus 0.5em minus 0.4em\relax IEEE, 2011, pp. 1--6.

\bibitem{kumatani2011channel}
K.~Kumatani, J.~McDonough, J.~F. Lehman, and B.~Raj, ``Channel selection based
  on multichannel cross-correlation coefficients for distant speech
  recognition,'' in \emph{2011 Joint Workshop on Hands-free Speech
  Communication and Microphone Arrays}.\hskip 1em plus 0.5em minus 0.4em\relax
  IEEE, 2011, pp. 1--6.

\bibitem{noda2015audio}
K.~Noda, Y.~Yamaguchi, K.~Nakadai, H.~G. Okuno, and T.~Ogata, ``Audio-visual
  speech recognition using deep learning,'' \emph{Applied Intelligence},
  vol.~42, no.~4, pp. 722--737, 2015.

\bibitem{DrudeITG2018}
L.~Drude, J.~Heymann, C.~Boeddeker, and R.~Haeb-Umbach, ``{NARA-WPE}: A
  {Python} package for weighted prediction error dereverberation in {Numpy} and
  {Tensorflow} for online and offline processing,'' in \emph{ITG 2018,
  Oldenburg, Germany}, October 2018.

\bibitem{8461310}
R.~{Scheibler}, E.~{Bezzam}, and I.~{Dokmanić}, ``Pyroomacoustics: A python
  package for audio room simulation and array processing algorithms,'' in
  \emph{Proc. of \mbox{IEEE} International Conference on Acoustics, Speech and
  Signal Processing \mbox{(ICASSP)}}, 2018, pp. 351--355.

\bibitem{povey2011kaldi}
D.~Povey, A.~Ghoshal, G.~Boulianne, L.~Burget, O.~Glembek, N.~Goel,
  M.~Hannemann, P.~Motlicek, Y.~Qian, P.~Schwarz \emph{et~al.}, ``The kaldi
  speech recognition toolkit,'' in \emph{Proc. of ASRU'11}, no.
  EPFL-CONF-192584.\hskip 1em plus 0.5em minus 0.4em\relax IEEE Signal
  Processing Society, 2011.

\bibitem{watanabe2018espnet}
S.~Watanabe, T.~Hori, S.~Karita, T.~Hayashi, J.~Nishitoba, Y.~Unno, N.-E.~Y.
  Soplin, J.~Heymann, M.~Wiesner, N.~Chen \emph{et~al.}, ``Espnet: End-to-end
  speech processing toolkit,'' \emph{Proc. of Annual Conference of the
  International Speech Communication Association \mbox{(Interspeech)}}, pp.
  2207--2211, 2018.

\bibitem{carletta2005ami}
J.~Carletta, S.~Ashby, S.~Bourban, M.~Flynn, M.~Guillemot, T.~Hain, J.~Kadlec,
  V.~Karaiskos, W.~Kraaij, M.~Kronenthal \emph{et~al.}, ``The {AMI} meeting
  corpus: A pre-announcement,'' in \emph{International workshop on machine
  learning for multimodal interaction}.\hskip 1em plus 0.5em minus 0.4em\relax
  Springer, 2005, pp. 28--39.

\bibitem{beamformit}
X.~Anguera, C.~Wooters, and J.~Hernando, ``Acoustic beamforming for speaker
  diarization of meetings,'' vol.~15, no.~7, September 2007, pp. 2011--2021.

\end{thebibliography}
